\documentclass[floatfix,aps,preprint,nofootinbib]{revtex4-2}

\usepackage[paperwidth=210mm,paperheight=297mm,centering,hmargin=2cm,vmargin=2.5cm]{geometry}

\usepackage[utf8]{inputenc}
\usepackage{amsfonts}
\usepackage{color}

\usepackage{caption}
\usepackage{booktabs}

\usepackage{amssymb}
\usepackage{amsmath}
\usepackage{stmaryrd}
\usepackage{latexsym}

\usepackage{xcolor}
\definecolor{myurlcolor}{HTML}{08457E}
\definecolor{mylinkcolor}{HTML}{2A52BE}
\definecolor{mycitecolor}{HTML}{E30022}

\usepackage{float}
\usepackage[colorlinks, linkcolor=mylinkcolor, citecolor=mycitecolor, urlcolor=myurlcolor, linktocpage=true]{hyperref}

\def\equationautorefname~#1\null{(#1)\null}
\def\tableautorefname~#1\null{(#1)\null}
\def\figureautorefname~#1\null{(#1)\null}
\def\sectionautorefname~#1\null{(#1)\null}

\let\origref\autoref
\def\autoref#1{\textbf{\origref{#1}}}

\let\origcite\cite
\def\cite#1{\textbf{\origcite{#1}}}

\usepackage{multirow}
\usepackage{tikz}
\usepackage{pgfplots}
\pgfplotsset{compat=1.15}

\usepackage{titlesec}
\titleformat*{\section}{\centering\small\bfseries\scshape}
\titleformat*{\subsection}{\small\bfseries\scshape}
\titleformat*{\subsubsection}{\small\bfseries\scshape}

\newcommand{\be}{\begin{equation}}
\newcommand{\ee}{\end{equation}}
\newcommand{\bea}{\begin{eqnarray}}
\newcommand{\eea}{\end{eqnarray}}
\newcommand{\benn}{\begin{eqnarray*}}
\newcommand{\eenn}{\end{eqnarray*}}

\def\bse{\begin{subequations}}%
\def\ese{\end{subequations}}%

\def\L{{\cal L}}      
\def\S{{\cal S}}  
\def\R{{\cal R}}     

\def\Rsq{{\mathfrak{R}^2}}
\newcommand{\Rsdl}[1]{\mathbb{R}_{#1}}

\newcommand{\sfrac}[2]{\dfrac{\,#1\,}{\,#2\,}}

\newcommand{\der}[2]{\sfrac{d #1}{d #2}}

\newcommand{\al}{\alpha}
\newcommand{\bt}{\beta}

\let\oldsqrt\sqrt
\def\sqrt{\mathpalette\DHLhksqrt}
\def\DHLhksqrt#1#2{%
	\setbox0=\hbox{$#1\oldsqrt{#2\,}$}\dimen0=\ht0
	\advance\dimen0-0.4\ht0
	\setbox2=\hbox{\vrule height\ht0 depth -\dimen0}%
	{\box0\lower0.4pt\box2}}

\def\apj{Astrophysical Journal}

\def\prd{Physical Review D}
\def\prc{Physical Review C}

\def\aap{Astronomy and Astrophysics}

\begin{document}

\title{Numerical Approach to the Exterior Solution of Spherically Symmetric and Static Configuration in Scalar-Tensor Theories \vspace{1cm}}

\author{A. Sava{\c s} Arapo{\u g}lu}
\email{arapoglu@itu.edu.tr}

\author{Sermet {\c C}a{\u g}an}
\email{cagans@itu.edu.tr}

\author{A. Emrah Y{\"u}kselci}
\email{yukselcia@itu.edu.tr}
\affiliation{Istanbul Technical University, Department of Physics, 34469 Maslak, Istanbul, Turkey \vspace{2cm}}

\begin{abstract}
We numerically examine the exterior solution of spherically symmetric and static configuration in scalar-tensor theories by using the nonminimally coupled scalar field with zero potential as our sample model. Our main purpose in this work is to fit the resulting data of the numerical solutions in the interested region by seeking for approximate analytical expressions which are weakly dependent of the parameters of a model, such as the nonminimal coupling constant in the present case. To this end, we determine the main forms of the mass and the metric functions in terms of the scalar field and their surface values. Then, we provide a function for the scalar field that contains only the mass and the radius of the configuration together with the surface and the asymptotic values of the scalar field. Therefore, we show that the exterior solution can be expressed in a form which does not depend on the parameters of a chosen model up to an order of accuracy around $10^{-5}$.
\end{abstract}

\maketitle
\clearpage
\raggedbottom

\section{INTRODUCTION}

The spherically symmetric and static solutions in general relativity (GR) are encountered in many applications for modeling the compact astrophysical objects such as black holes and neutron stars \cite{psaltis-rw-2008}. As is well known, the exterior solution for this type of configurations in vacuum is known to be the Schwarzschild solution \cite{Schwarzschild1916,Droste1916} which is characterized by the quantity named as the Schwarzschild radius that is indeed equal to $2GM/c^2$ \cite{Misner1973,Landau1975}. Hence, the only parameter which affects the exterior solution and is determined by the interior solution is the mass of the body. However, unlike vacuum solutions, in the case of scalar-tensor theories the mass function, which gives the aforementioned mass value of the body at the radius of the star, continues to change outside the configuration because of the contribution coming from the scalar field that maintains its radial evolution (e.g.,\ Refs.\ \cite{zag92,har98,salgado1998,hor11,kazanas2014,fuzfa1,sot18,Olmo2019,Arapoglu2019,Arapoglu2020}). Consequently, this has an impact on the metric functions as well. 

In spite of the fact that the numerical techniques are usually adopted for the solution in scalar-tensor theories due to the nonlinearity of the differential equations, there are some analytical solutions as well. For instance, in the context of the spontaneous scalarization it has been shown that an analytical solution can be found in Just coordinates which is not analytically invertible to the Schwarzschild coordinates \cite{Damour:1992,yunes-silva2018}. Other attempts were made in Ref.\ \cite{Yazadjiev2011} in the presence of a phantom scalar field and in Ref.\ \cite{Saffer2019} for scalar-Gauss-Bonnet gravity with the perturbative approach. The solution strategy of the numerical techniques for that kind of setting is to iterate the variables forming the numerical equations originated from the differential equations starting from the center of configuration. Radius of the star is determined at a point corresponding to the vanishing pressure controlled by the Tolman–Oppenheimer–Volkoff (TOV) equation \cite{Tolman1934,Tolman1939,Oppenheimer1939}. That point also describes the total mass of the star specified as the value of the mass function at the surface. Furthermore, one should check whether the asymptotic values of the metric functions match with the ones in Minkowski space-time. When solving the equations with this method, it is necessary to supply an equation of state (EOS), which relates the pressure and the energy density of the fluid  \cite{EoS1,EoS2,EoS3} in the form of tabulated data that is included to the system through interpolation. On the other hand, there are also some analytical expressions for EOS that approximately relate the pressure and the density \cite{Horedt2004,Potekhin2013}. (Although there is a well-known degeneracy between the theory of gravity and the EOS, we do not mention that in this paper and refer the reader to Refs.\ \cite{degeneracy1,degeneracy2,degeneracy3,dgnrcy4_Doneva:2017,dgnrcy5_Shao:2019}.)

In this work, we use the nonminimally coupled scalar field as our sample model since this is one of the well-motivated extensions of GR stemming from the quantization of the scalar field in curved space-time \cite{chernikov1968,callan-etal1970,birrell-davies1980,birrell-davies-1982}. We investigate the case with zero potential whose underlying reasons to use was shortly discussed in Ref.\ \cite{Arapoglu2019}. We implement the same numerical strategy briefly described above with the help of one particular EOS, namely SLy \cite{eos_sly}, to determine the mass and the radius of the star as well as the radial evolution of the variables. Then, by observing and comparing the behavior of the mass and the metric functions together with the scalar field outside we try to find analytical models that fit well with the data and, at the same time, do not contain the model parameters explicitly with validity in some particular range of their values that we aim to analyze based on the previous results \cite{Arapoglu2019}.

The plan of the paper is as follows : In Sec.\ \autoref{sec:setup} we give the main equations that are solved numerically. Then, we describe the method implemented throughout this work in Sec.\ \autoref{sec:method} including the numerical error and some necessary definitions. We discuss the data fitting and represent the final results in Sec.\ \autoref{sec:fitting}. We give the concluding remarks in Sec.\ \autoref{sec:conclusion}.

\section{MAIN EQUATIONS} \label{sec:setup}
We begin with the following action
\begin{equation}
	\S = \int \! d^4 x \, \sqrt{|g|} \, \bigg[ \sfrac{1}{\, 2 \kappa \,} \R + \sfrac{\,1\,}{2} \xi \phi^2 \R - \sfrac{\,1\,}{2} \nabla^c \phi \, \nabla_{\!\!c} \, \phi - V\!(\phi) + \L^m \bigg]
\label{eq:action}
\end{equation}
where $\kappa = 8\pi$ and $\xi$ is the coupling constant\footnote{We use geometrical units ($G = c = 1$) throughout this study.}. Einstein's field equations are obtained as
\begin{equation}
	 \R_{\mu \nu} - \sfrac{\,1\,}{2} \R \, g_{\mu \nu} = \, \kappa_{\rm eff} \big[ T_{\mu\nu}^{(m)} + T_{\mu\nu}^{(\phi)} \big]
\label{eq:fieldeq1}
\end{equation}
where $\kappa_{\rm eff}$, energy-momentum tensors for the fluid $T_{\mu\nu}^{(m)}$ and for the scalar field $T_{\mu\nu}^{(\phi)}$ are given in the following forms
\begin{equation}
	\kappa_{\rm eff}(\phi) = \kappa \big( 1 + \kappa \xi \phi^2 \big)^{-1} \,\, ,
\label{eq:kappa_eff}
\end{equation}
\begin{equation}
	T_{\mu\nu}^{(m)} = \big(\rho + P\big) u_\mu u_\nu + P g_{\mu\nu} \,\, ,
\label{eq:matter_tensor}
\end{equation}
\begin{equation}
	T_{\mu\nu}^{(\phi)} = \nabla_{\!\!\mu} \, \phi \, \nabla_{\!\nu} \, \phi \\[2pt] - g_{\mu \nu} \bigg[ \sfrac{1}{2} \, \nabla^c \phi \, \nabla_{\!\!c} \, \phi + \, V\!(\phi) \bigg] - \xi \big( g_{\mu \nu} \boxempty - \nabla_{\!\!\mu} \, \nabla_{\!\nu} \big) \phi^2 \: .
\label{eq:scalar_tensor}
\end{equation}

On the other hand, variation of the above action with respect to the scalar field yields the following expression: 
\begin{equation}
	\boxempty \! \phi + \xi \phi \R - \dfrac{dV\!(\phi)}{d \phi} = 0 \, .
\label{eq:fieldeq2}
\end{equation}

We use the static and spherically symmetric metric in the form of
\begin{equation}
	ds^2 = - e^{2f(r)} dt^2 + e^{2g(r)} dr^2 + r^2 \big(d\theta^2 + \sin^2\theta \, d\varphi^2\big),
\label{eq:metric}
\end{equation}
and it yields the following set of equations
\begin{subequations}
	\begin{align}
	g' &= \bigg[ \sfrac{1}{r} \!+\! \kappa_{\rm eff} \xi \phi \phi' \bigg]^{-1} \bigg[ \sfrac{1 \!-\! e^{2g}}{2r^2} \!+\! \kappa_{\rm eff} \bigg\{ \sfrac{1}{2} \big(\rho \!+\! V\big)e^{2g} + \xi \phi \bigg( \phi'' \!+\! \sfrac{2\phi'}{r} \bigg) \!+\! \bigg(\xi \!+\! \sfrac{1}{4}\bigg)\phi'^2 \bigg\} \bigg] \label{eq:components_tt} \: , \\[3mm] 
	f' &= \bigg[ \sfrac{1}{r} \!+\! \kappa_{\rm eff} \xi \phi \phi' \bigg]^{-1} \bigg[ \!-\! \sfrac{1 \!-\! e^{2g}}{2r^2} + \kappa_{\rm eff} \bigg\{ \sfrac{1}{2} \big(P \!-\! V\big)e^{2g} + \sfrac{1}{4}\phi'^2 - \sfrac{2\xi\phi\phi'}{r} \bigg\} \bigg] \label{eq:components_rr} \: , \\[3mm] 
	f'' &= -\big(f'\!-\!g'\big) \Big[ f' \!+\! \sfrac{1}{r} \Big] \!+\! \kappa_{\rm eff} \bigg\{ \! \big(P\!-\!V\big)e^{2g} \!-\! 2\xi\phi \bigg[ \phi'' \!+\! \phi' \bigg( \! f'\!-\!g'\!+\!\sfrac{1}{r} \! \bigg) \! \bigg] \!-\! \bigg( \! 2\xi \!+\! \sfrac{1}{2}\bigg) \phi'^2 \bigg\} \: , \label{eq:components_QQ} \\[3mm] 
	\phi'' &= - \bigg[f' \!-\! g' \!+\! \sfrac{2}{r}\bigg]\phi' + 2\xi\phi \bigg[f'' \!+\! \big(f'\!-\!g'\big) \Big( f' \!+\! \sfrac{2}{r} \Big) + \sfrac{1 \!-\! e^{2g}}{r^2}\bigg] + e^{2g} \der{V}{\phi} \, , \label{eq:scalar_eq}
	\end{align}
\label{eq:components_eq}%
\end{subequations}
where the first three equations are obtained from the $tt$, $rr$, and $\theta\theta$ components of Eq.\ \autoref{eq:fieldeq1}, respectively, and Eq.\ \autoref{eq:fieldeq2} becomes the last one. Here prime denotes the derivative with respect to coordinate distance $r$.

Expression for the pressure, i.e. the TOV equation, is obtained from the energy-momentum conservation of the fluid ($\nabla^{\mu} T_{\mu\nu}^{(m)} = 0$) as
\begin{equation}
	P' = - f'\big(P + \rho\big)  \,\, .
\label{eq:pressure_eq}
\end{equation}

The mass function for a system without the scalar field is defined as $m(r) = r(1 - e^{-2g})/2$. However, the presence of the scalar field causes this definition to change since it contributes to the mass of the configuration by both coupling with the density and by its own evolution. The two possible definitions in such a case, namely the ADM \cite{adm_mass} and the Komar \cite{komar_mass} masses, coincide for the static configurations in scalar-tensor theories \cite{mass_coincides}. Therefore, similar to the one given in Ref.\ \cite{salgado1998} we define the mass function in the following form,
\begin{equation}
	M(r) \equiv 4\pi \int_{0}^{r} r'^2 E(r') \, dr' ,
\label{eq:mass}
\end{equation}
where the function in the integrand is given as 
\vspace{2mm}
\begin{equation}
	E(r) = \sfrac{\kappa_{\rm eff}}{\kappa} \bigg[ \rho + \sfrac{1}{2} \big( \phi' e^{-g} \big)^2 + V(\phi) + 2\xi \bigg\{ \phi \bigg( \! \phi'' + \phi' \bigg[ \sfrac{2}{r} \!-\! g' \bigg] \bigg) + (\phi')^2 \bigg\} e^{-2g} \bigg] \: ,
\label{eq:mass_E}
\end{equation}
which is obtained from the $tt$ component of the total energy-momentum tensor. In the limit $r \rightarrow \infty$ the mass function gives the ADM mass, i.e. $M(r \rightarrow \infty) = M_{\rm{ADM}}$. As mentioned in the previous section the mass function continues its radial evolution due to the scalar field contribution at distances $r>R$ while its value at $r=R$ gives the mass of the body. The radius of the star, on the other hand, is determined at a point where the density of the fluid, or correspondingly the pressure, vanishes, i.e. $\rho(r=R)=0$.

We will numerically solve the equation set \autoref{eq:components_eq} together with Eq.\ \autoref{eq:pressure_eq} and evaluate the mass function through Eq.\ \autoref{eq:mass}. Since we will consider the zero potential case as mentioned before, we set $V(\phi)=0$ in the above expressions and $P=\rho=0$ in the exterior region as well. We set $P(0)=P_{\rm{c}}$, $M(0)=0$, $g(0)=0$, $f(0)=f_{\rm{in}}$, $\phi(0)=\phi_{\rm{c}}$ as initial values for the integration. The initial value of the scalar field, $\phi_{\rm{c}}$, should be chosen such that the scalar field value at infinity does not violate the observational constraints such as the limitations on PPN parameters \cite{ppn_gamma,ppn_beta_1,ppn_beta_2,ppn_review} and the analysis of dipole radiation in pulsar-white dwarf binary systems \cite{dipole_rad_Freire:2012,dipole_rad_Shao:2017}. The effect of the nonminimal coupling constant, $\xi$, on the choice of $\phi_{\rm{c}}$ also has to be taken into account \cite{Arapoglu2019}. On the other hand, the central pressure used throughout the work and extracted from the EOS dataset has been chosen as $P_{\rm{c}} \approx 3 \times 10^{15} \: \mbox{g/cm}^3$ corresponding to $P_{\rm{c}} \approx 2.25 \times 10^{-3} \: \mbox{km}^{-2}$ in geometrical units that we use. Finally, the initial value of $f(r)$ is determined such that its asymptotic value goes to zero so that the metric functions approach to their counterpart in Minkowski space-time at infinity.

\section{METHOD} \label{sec:method}

In this section we will explain the method followed in this paper. We will represent the error in numerical calculations, make the definitions for the mass function and the scalar field as well as the deviations in the metric functions, and briefly discuss their behavior in the light of one particular example.

\subsection{Numerical Error}

\begin{figure*}[!h]
	\centering

	\begin{tabular}{@{}c@{}}\hspace{-5mm}
		\includegraphics[width=.5\linewidth]{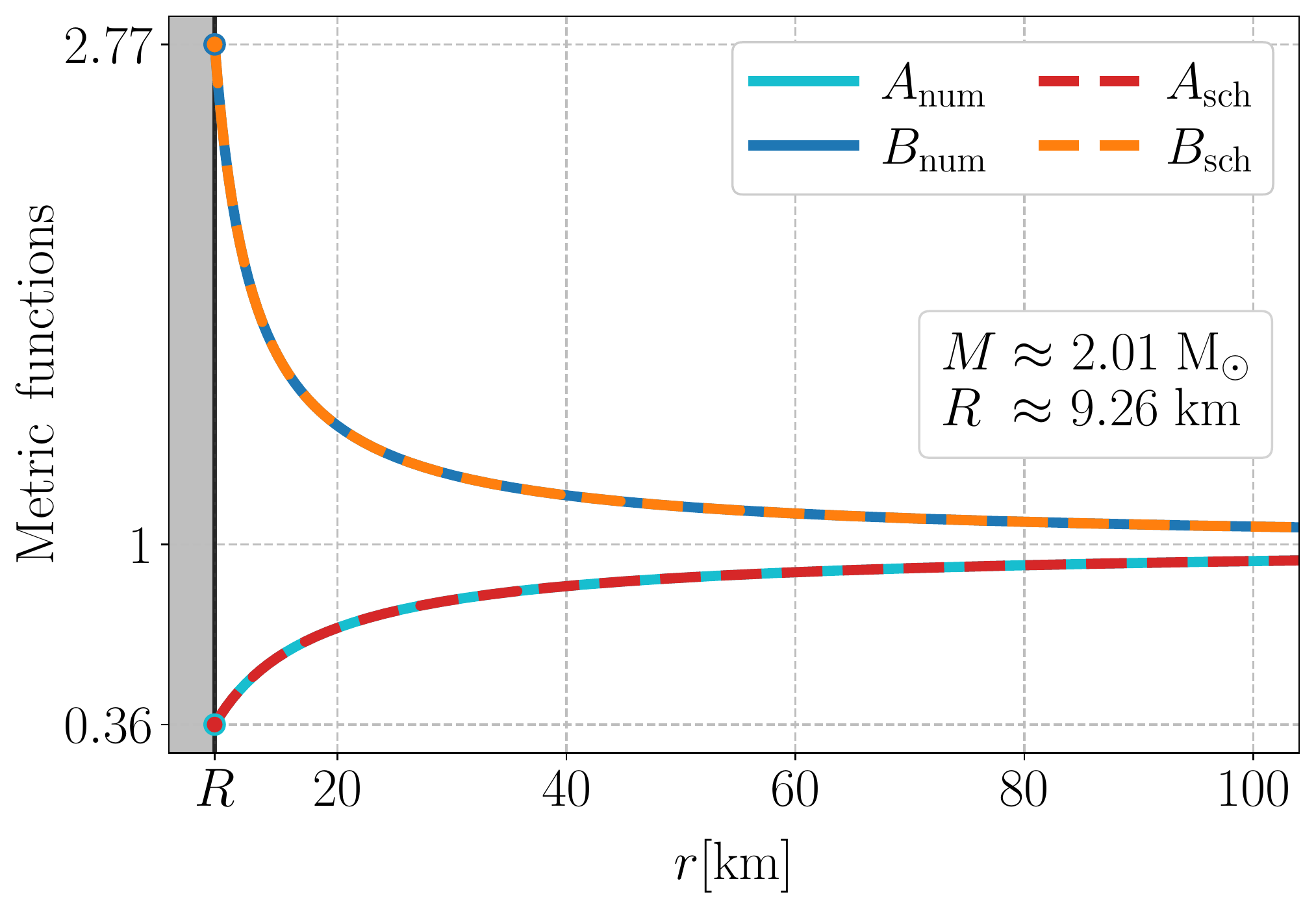}
		\label{fig:GR_solution}
	\end{tabular}\hspace{3mm}
	\begin{tabular}{@{}c@{}}
		\includegraphics[width=.47\linewidth]{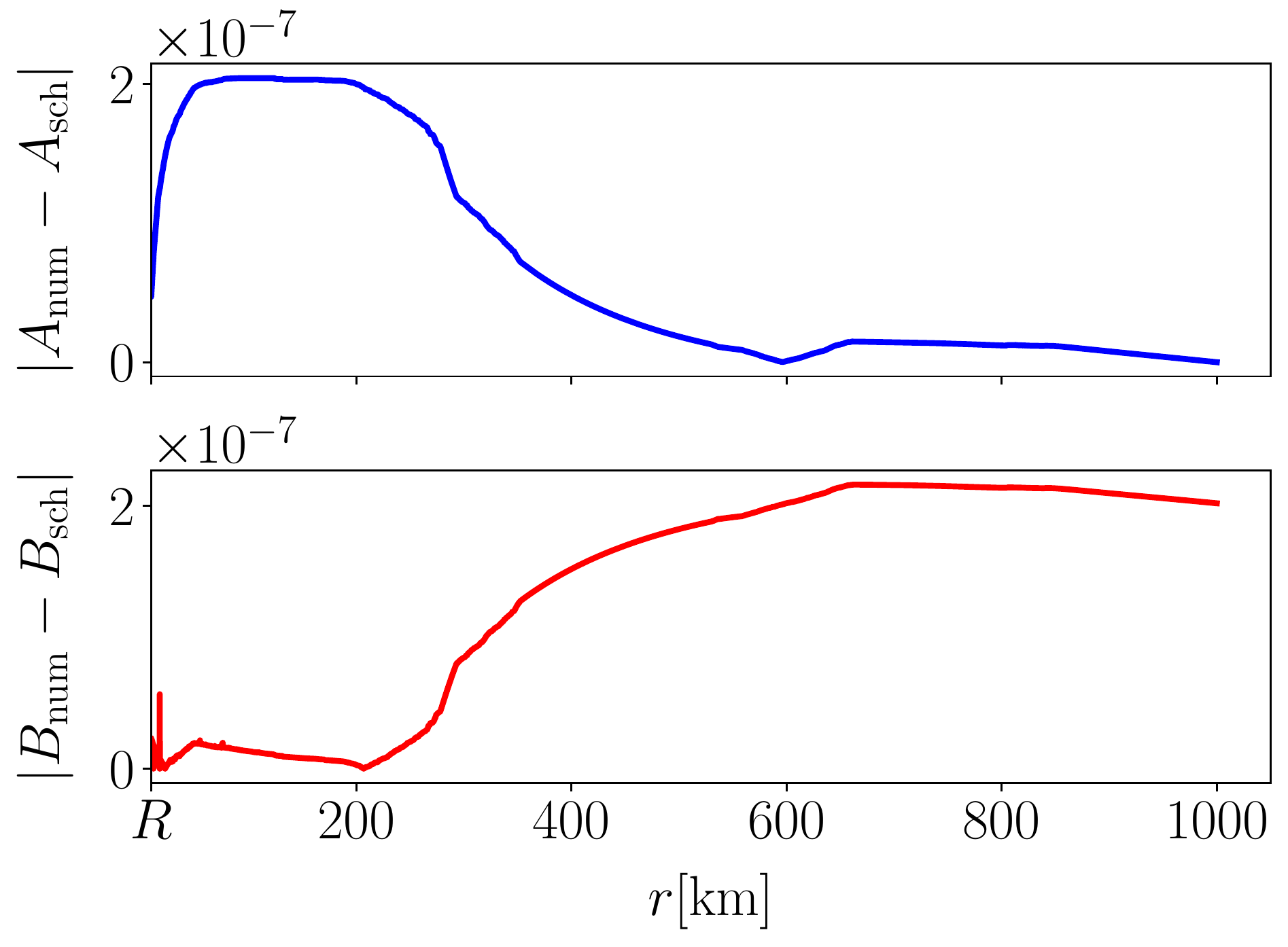}
		\label{fig:GR_abs_err}
	\end{tabular}
	
	\caption{Comparison of analytical and numerical solutions in GR for a specific configuration with $M \approx 2.01\;{\rm M}_\odot$ and $R \approx 9.26\;{\rm km}$. The central pressure value is $P_{\rm{c}}=2.25 \times 10^{-3} \: \mbox{km}^{-2}$.}
	\label{fig:GR_sol_abs_err}
\end{figure*}

Here we determine the numerical error in our code by using the well-known analytical solution of spherically symmetric and static configuration in order to show the valid range of our approach in a numerical point of view. Although we have used the metric functions given in Eq.\ \autoref{eq:metric} due to their convenience in the numerical calculations, in order to present the results we prefer the following form:
\begin{equation}
	ds^2 = - A(r) dt^2 + B(r) dr^2 + r^2 \big(d\theta^2 + \sin^2\theta \, d\varphi^2\big) \;.
\label{eq:metric2}
\end{equation}
We compute the radial evolution of the metric functions numerically thorough
\begin{equation}
	\sfrac{B'}{2B} = \sfrac{1 - B}{2r} \:, \qquad  \sfrac{A'}{2A} = - \sfrac{1 - B}{2r} \: ,
\label{eq:numerical_sch}
\end{equation}
and compare the results with the analytical one, that is the Schwarzchild solution, 
\begin{equation}
    A(r) = 1 - \sfrac{2M}{r} \qquad \mbox{and} \qquad B(r) = \bigg( 1 - \sfrac{2M}{r}\bigg)^{-1} \;
\end{equation}
to define the absolute error in our numerical calculations. Here $M$ is the mass of the configuration. For all computations we implement the same code which uses the standard \texttt{odeint} package of \texttt{SciPy} \cite{2020SciPy-NMeth} as the integrator with both modified tolerances of $10^{-10}$. As seen in Fig.\ \autoref{fig:GR_sol_abs_err} the difference between two solutions is in the order of $10^{-7}$ which determines a certain restriction when fitting an analytical model to numerical data.

\vspace{5mm}

\subsection{Definitions}

Here we define the mass function, the scalar field, and the metric functions in order to obtain appropriate forms for the data fitting.

We begin with the total mass function which can be written as
\begin{equation}
M_{\rm{total}}(r) =
\begin{cases}
   M_{\rm{in}}(r) & , \quad r < R \\
   M_{\rm{s}} & , \quad r = R \\
   M_{\rm{s}} + M_{\rm{ext}}(r) & , \quad r > R
\end{cases}
\end{equation}
where the radius, $R$, is determined via the pressure condition $P(R)=0$ as usual. The mass of the star is defined at its surface as $M_{\rm{s}} \equiv M_{\rm{total}}(R)$. Then, the total mass function outside the star is given as $M_{\rm{total}}(r) = M_{\rm{s}} + M_{\rm{ext}}(r)$. Here, $M_{\rm{in}}$ is determined by the EOS and its coupling with the scalar field while $M_{\rm{ext}}$ consists of the scalar field contribution only. We also define the asymptotic value of the total mass function as $M_\infty \equiv M_{\rm{total}}(r \rightarrow \infty)$. Regarding the scalar field, we write $\phi_{\rm{s}} \equiv \phi(R)$, $\phi_\infty \equiv \phi(r \rightarrow \infty)$, and $\phi_{\rm{ext}}(r)$ in a similar manner.

On the other hand, we describe the Schwarzschild-like solutions as
\begin{equation}
    A_{\rm{sch}}(r) = 1 - \sfrac{2M_{\rm s}}{r} \qquad \mbox{and} \qquad B_{\rm{sch}}(r) = \bigg( 1 - \sfrac{2M_{\rm s}}{r}\bigg)^{-1} \;.
    \label{eq:AB_sch_like}
\end{equation}
It is noteworthy to point out that unlike $M$ in GR, $M_{\rm{s}}$ contains the contributions from the density of the star as well as the scalar field as can be seen from Eq.\ \autoref{eq:mass}.

We assume that the exterior solutions for the metric functions are in the form of
\begin{equation}
    A(r) = 1 - \sfrac{2M_{\rm s}}{r} - \Delta A(r) \qquad \mbox{and} \qquad B(r) = \bigg( 1 - \sfrac{2M_{\rm s}}{r} - \sfrac{1}{\Delta B(r)} \bigg)^{-1}
\end{equation}
where $\Delta A(r)$ and $\Delta B(r)$ represent deviations from the Schwarzschild-like solutions and are calculated by
\begin{equation}
    \Delta A(r) = A_{\rm{sch}}(r) - A(r) \qquad \mbox{and} \qquad \sfrac{1}{\Delta B(r)} = \sfrac{1}{B_{\rm{sch}}(r)} - \sfrac{1}{B(r)} \;.
\end{equation}
Hence, we will try to find out the forms for $M_{\rm{ext}}(r)$, $\phi_{\rm{ext}}(r)$, $\Delta A$ and $\Delta B$ by modeling the data coming from the numerical solutions.

\vspace{5mm}

\begin{figure*}[!t]
	\centering

	\begin{tabular}{@{}c@{}}\hspace{-5mm}
		\includegraphics[width=.5\linewidth]{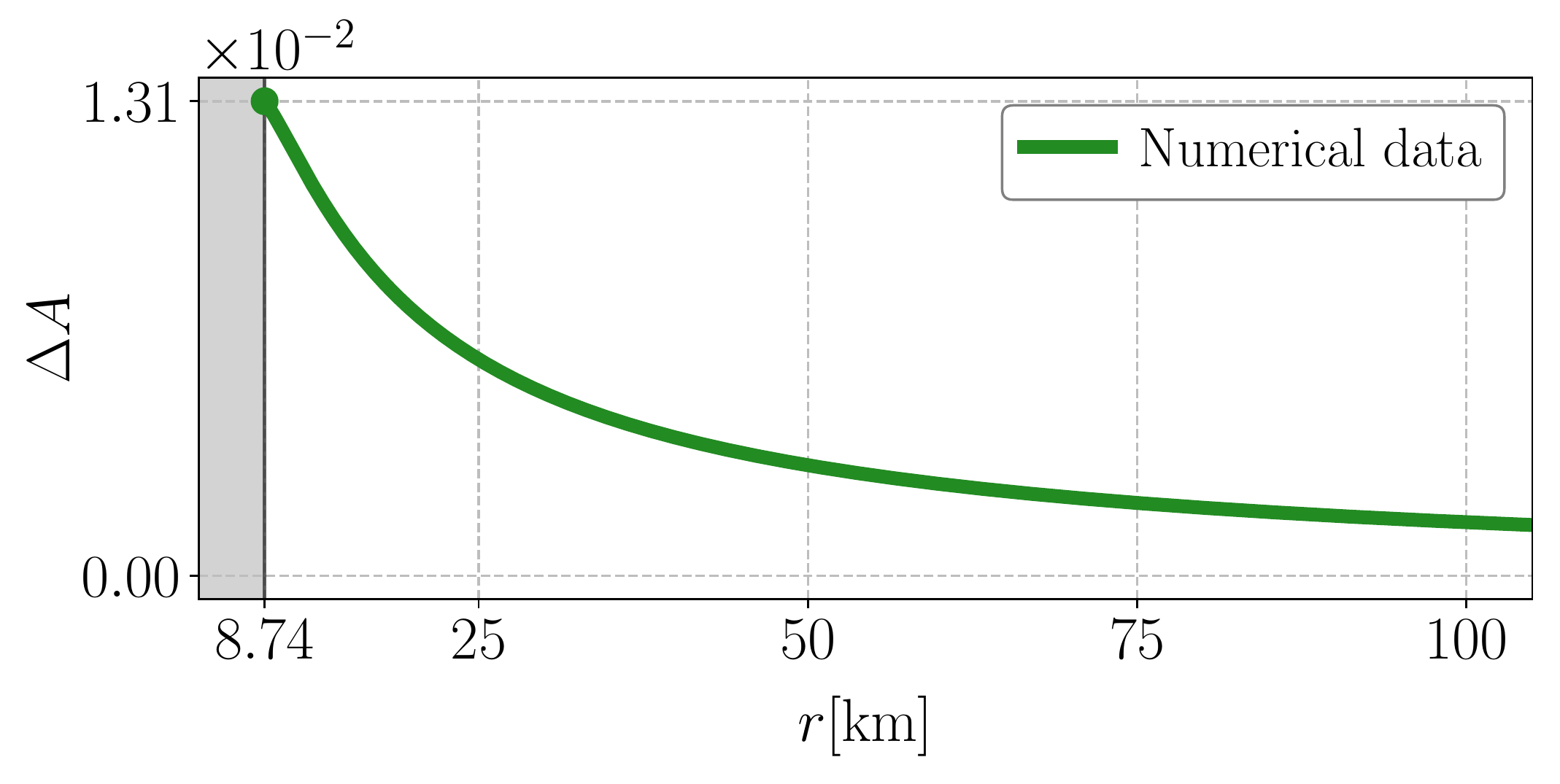}
	\end{tabular}\hspace{1mm}
	\begin{tabular}{@{}c@{}}
		\includegraphics[width=.5\linewidth]{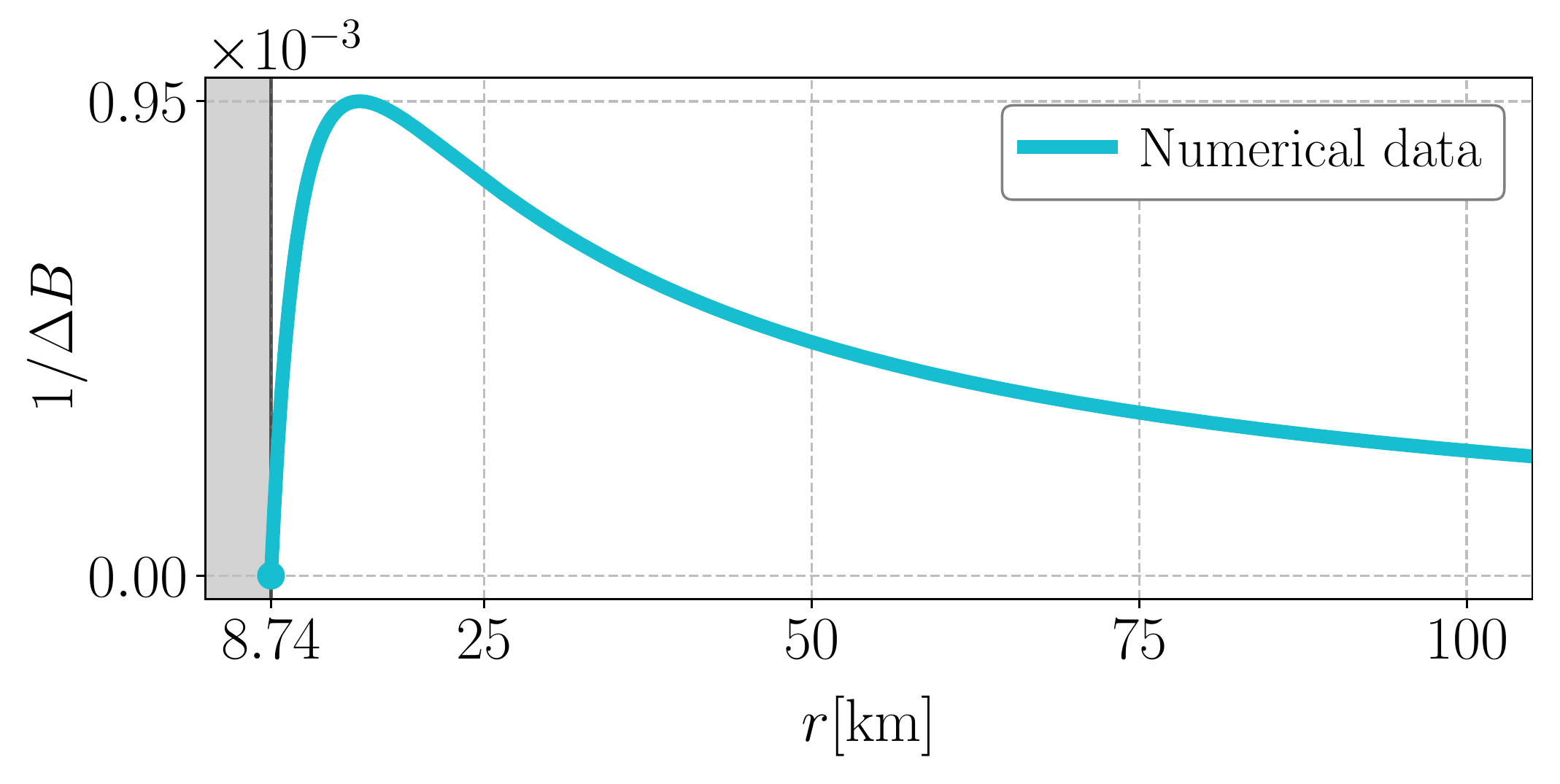}
	\end{tabular}
	
	\vspace{-2mm}
	
	\begin{tabular}{@{}c@{}}\hspace{-5mm}
		\includegraphics[width=.5\linewidth]{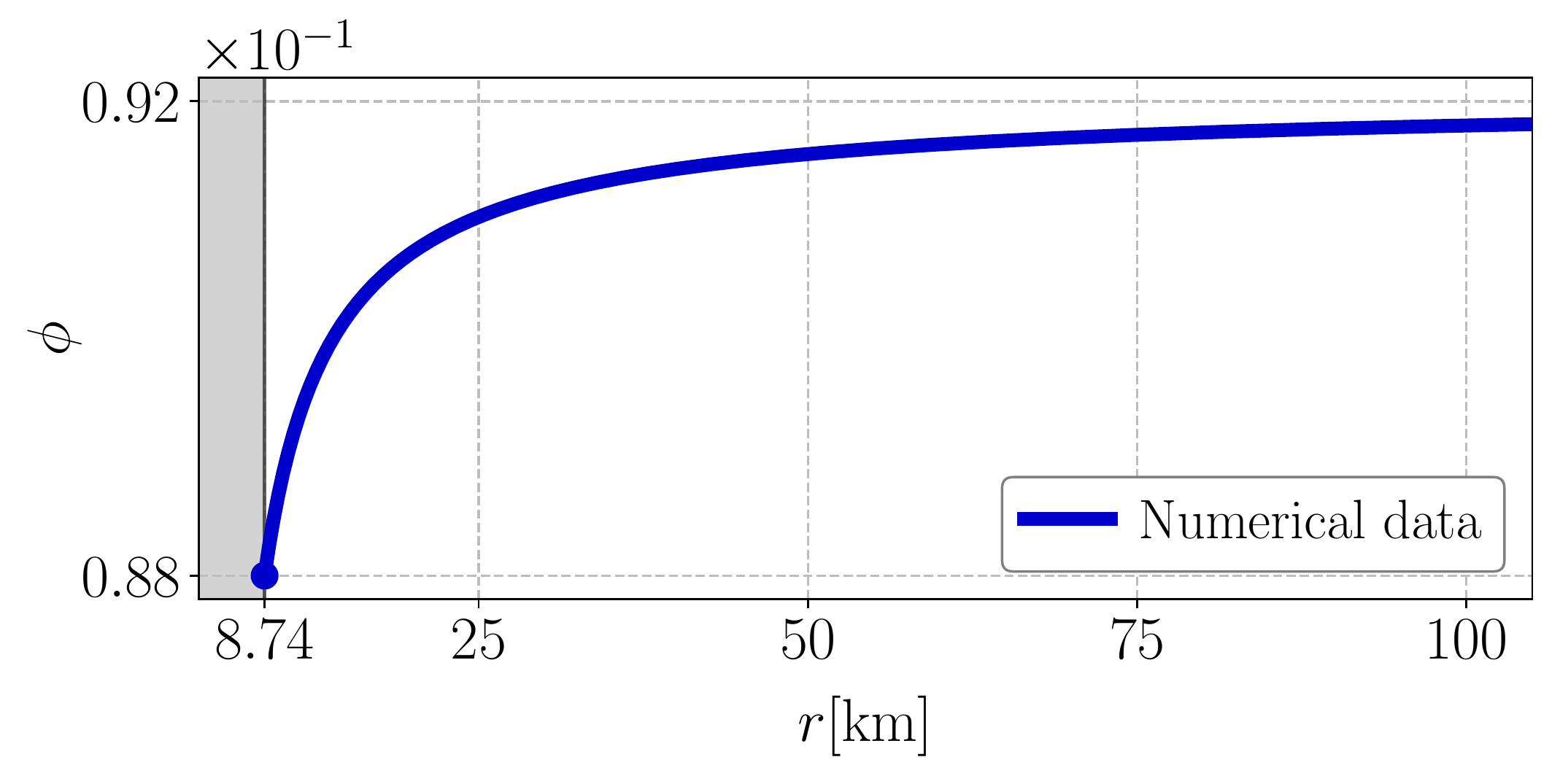}
	\end{tabular}\hspace{1mm}
	\begin{tabular}{@{}c@{}}
		\includegraphics[width=.5\linewidth]{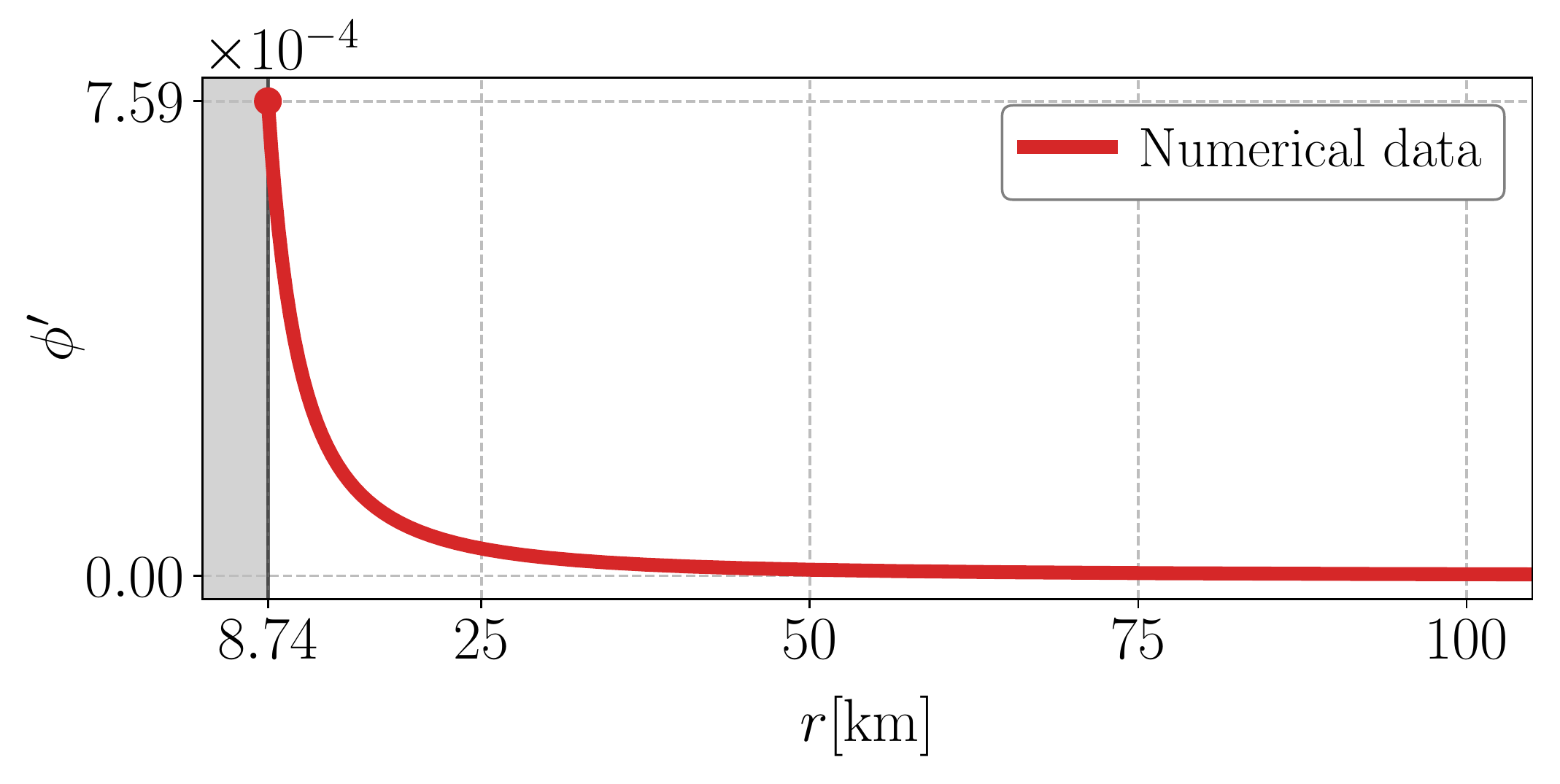}
	\end{tabular}
	
	\vspace{-2mm}
	
	\begin{tabular}{@{}c@{}}\hspace{-5mm}
		\includegraphics[width=.5\linewidth]{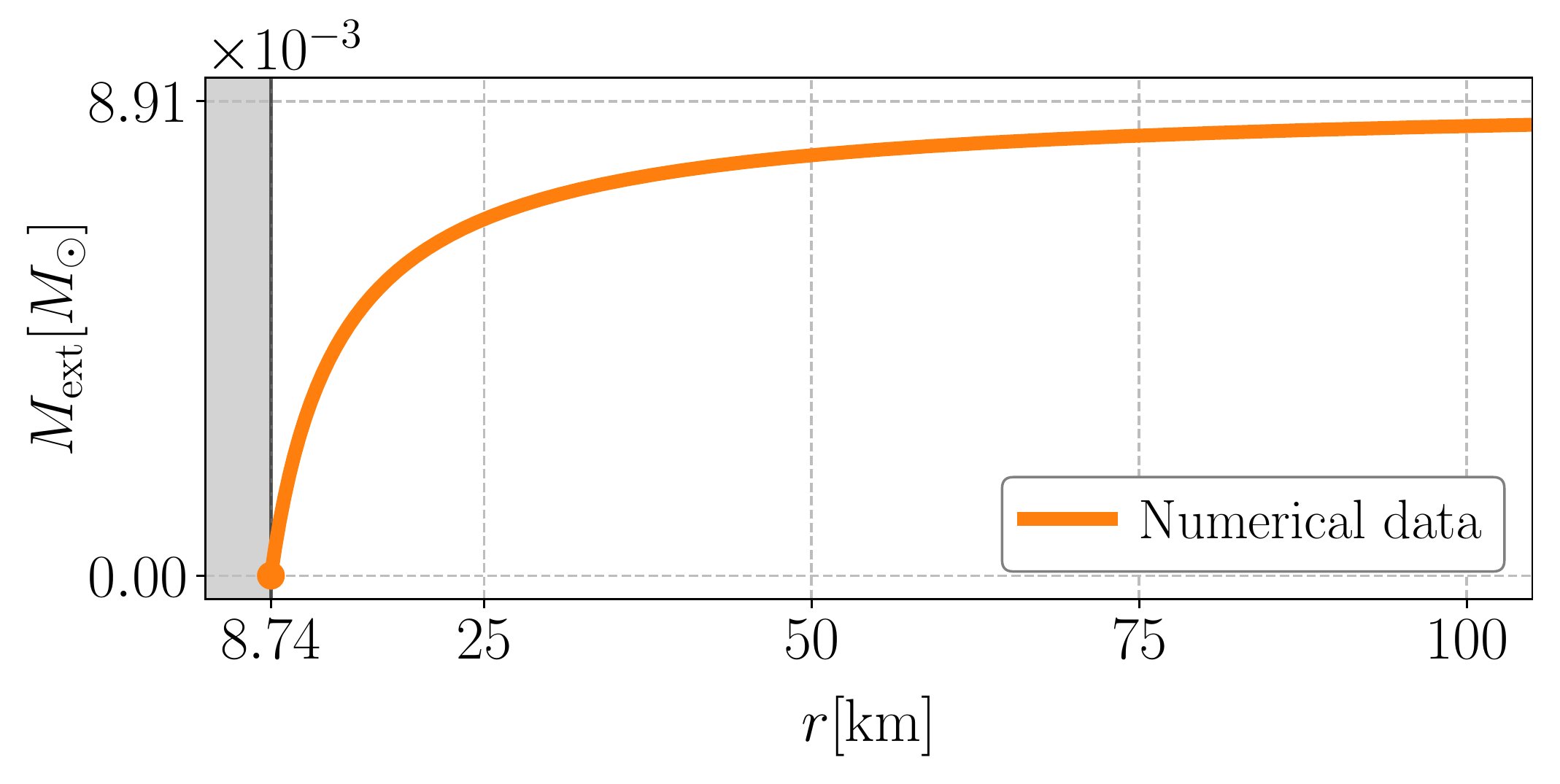}
	\end{tabular}\hspace{1mm}
	
	\vspace{-2mm}
	
	\caption{Radial evolution for $\Delta A(r)$, $\Delta B(r)$, $\phi(r)$, $\phi'(r)$, and $M_{\rm{ext}}(r)$ for a specific configuration with $\xi=-0.5$ and $\phi_{\rm{c}}=0.1\,$. The mass and the radius values are calculated as $M_{\rm{s}}=1.87$ and $R=8.74$ by using SLy dataset for the EOS with the central pressure value $P_{\rm{c}}=2.25 \times 10^{-3} \: \mbox{km}^{-2}$. Horizontal lines indicate the minimum and the maximum values for each data.}
\label{fig:funcs_eval}
\end{figure*}

\subsection{Radial Evolution of the Numerical Solutions}

\begin{figure*}[!ht]
	\centering
	
	\begin{tabular}{@{}c@{}}\hspace{-3mm}
		\includegraphics[width=.57\linewidth]{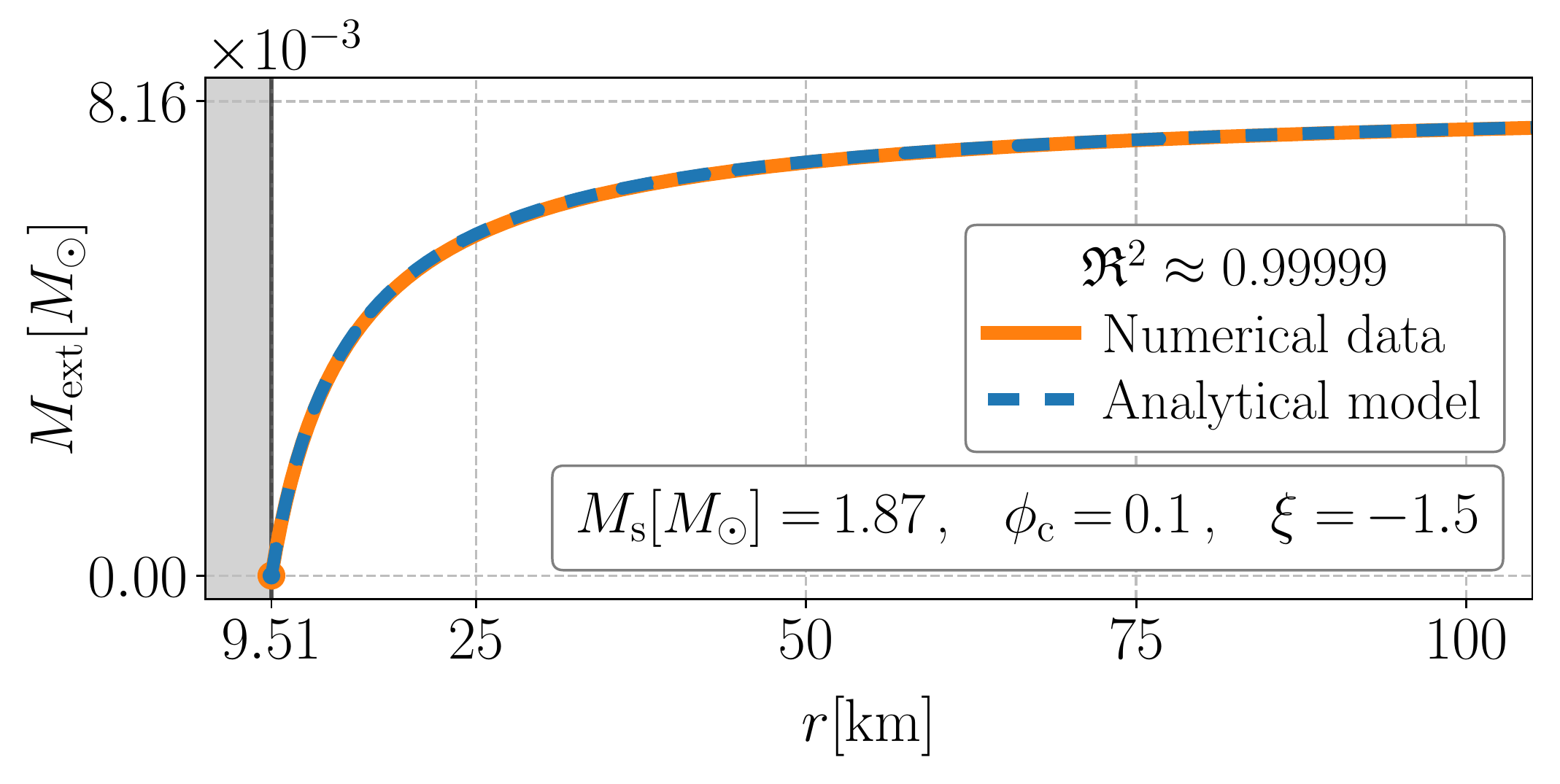}
	\end{tabular}
	\begin{tabular}{@{}c@{}}
		\includegraphics[width=.43\linewidth]{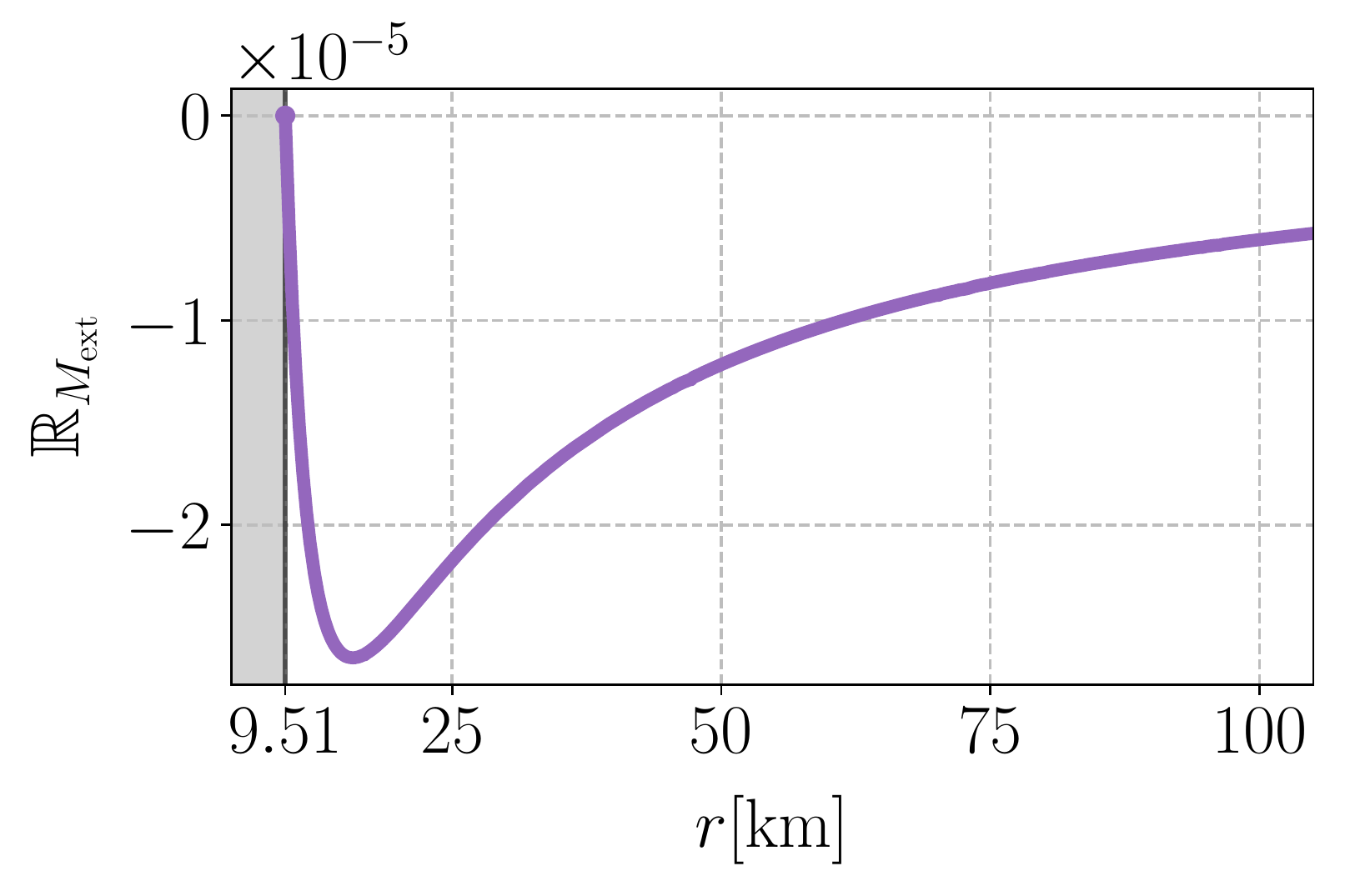}
	\end{tabular}
	
	\vspace{-4mm}
	
	\begin{tabular}{@{}c@{}}\hspace{-3mm}
		\includegraphics[width=.57\linewidth]{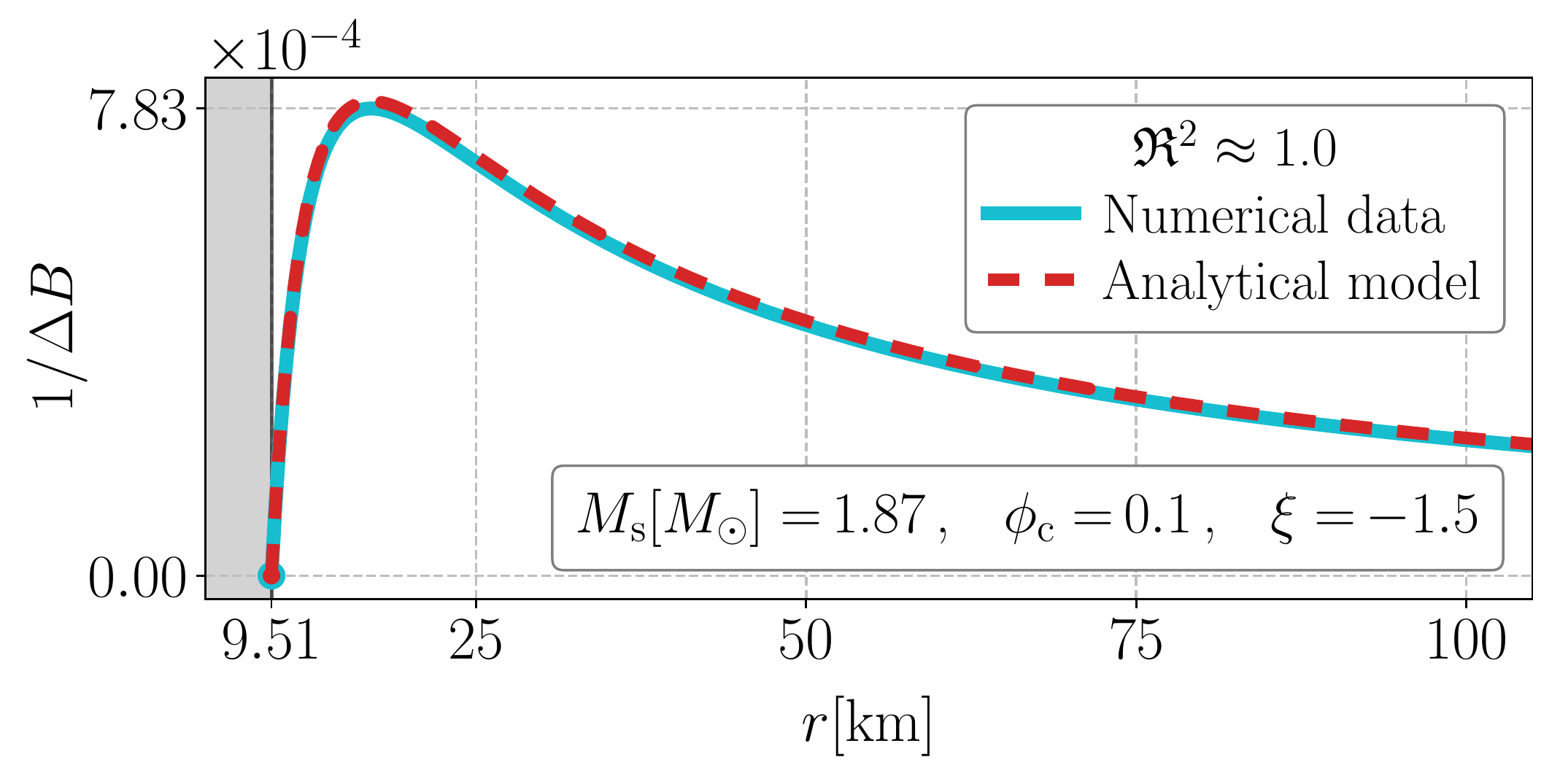}
	\end{tabular}
	\begin{tabular}{@{}c@{}}
		\includegraphics[width=.43\linewidth]{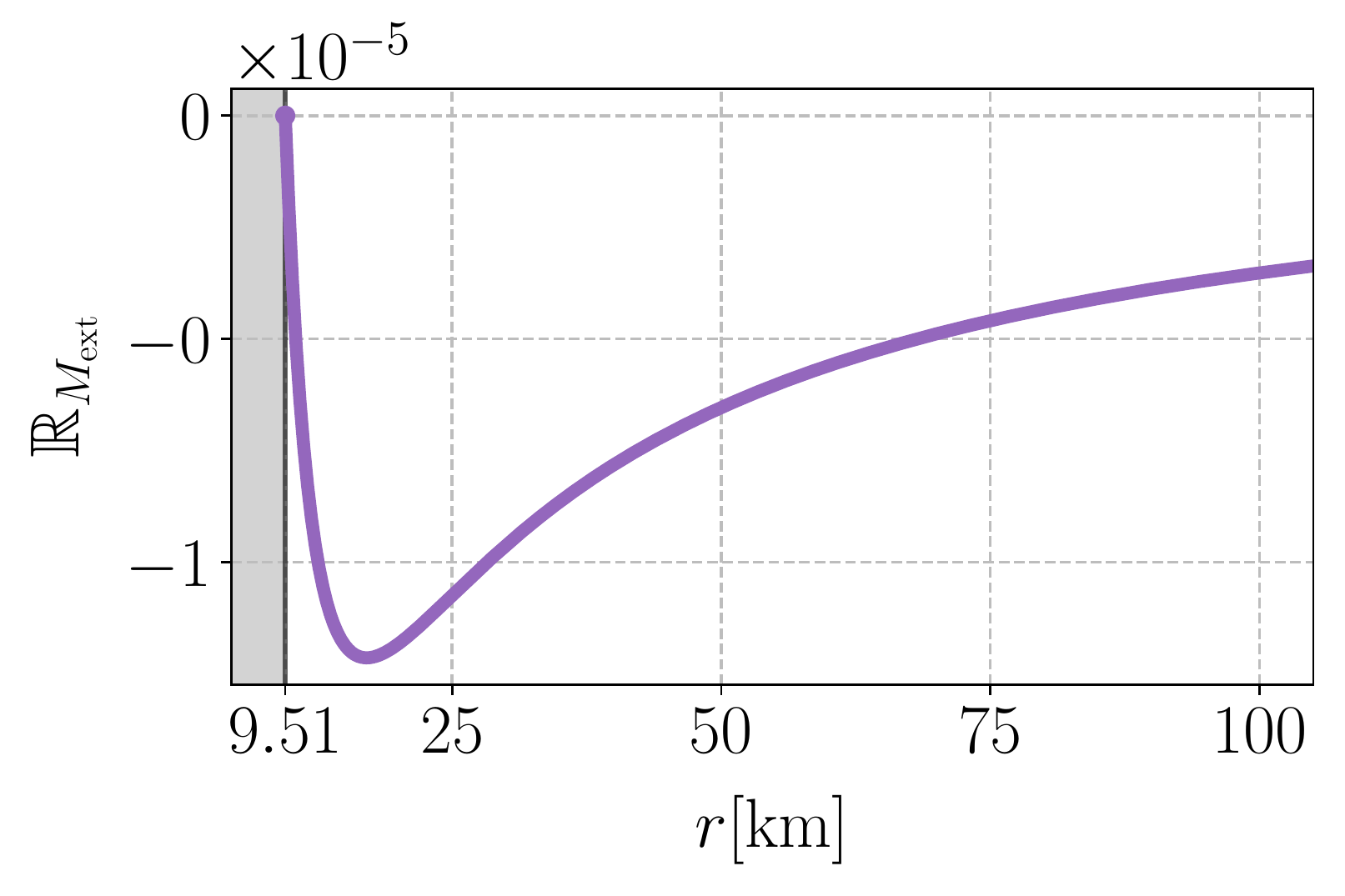}
	\end{tabular}
	
	\vspace{-4mm}
	
	\begin{tabular}{@{}c@{}}\hspace{-3mm}
		\includegraphics[width=.57\linewidth]{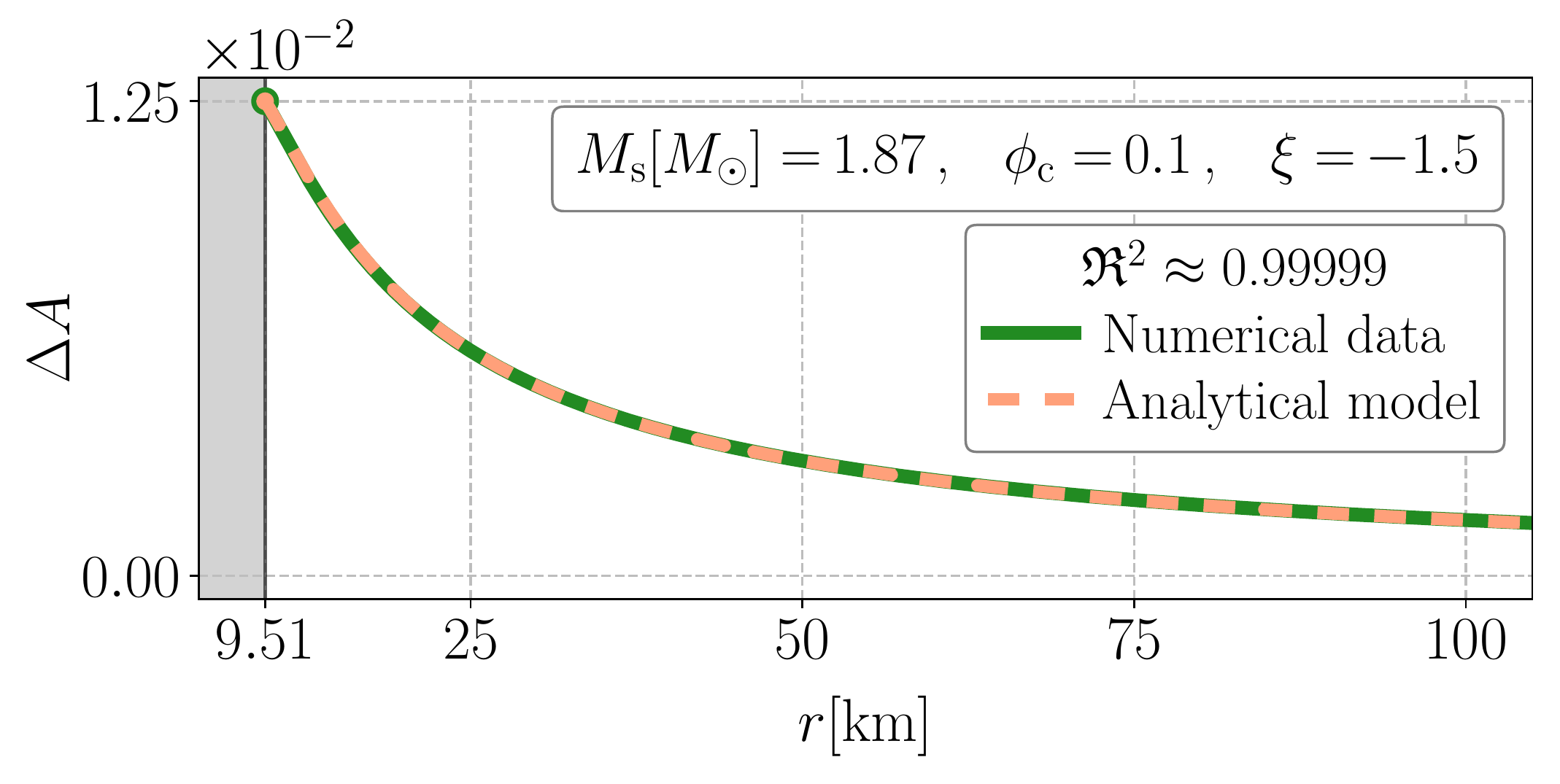}
	\end{tabular}
	\begin{tabular}{@{}c@{}}
		\includegraphics[width=.43\linewidth]{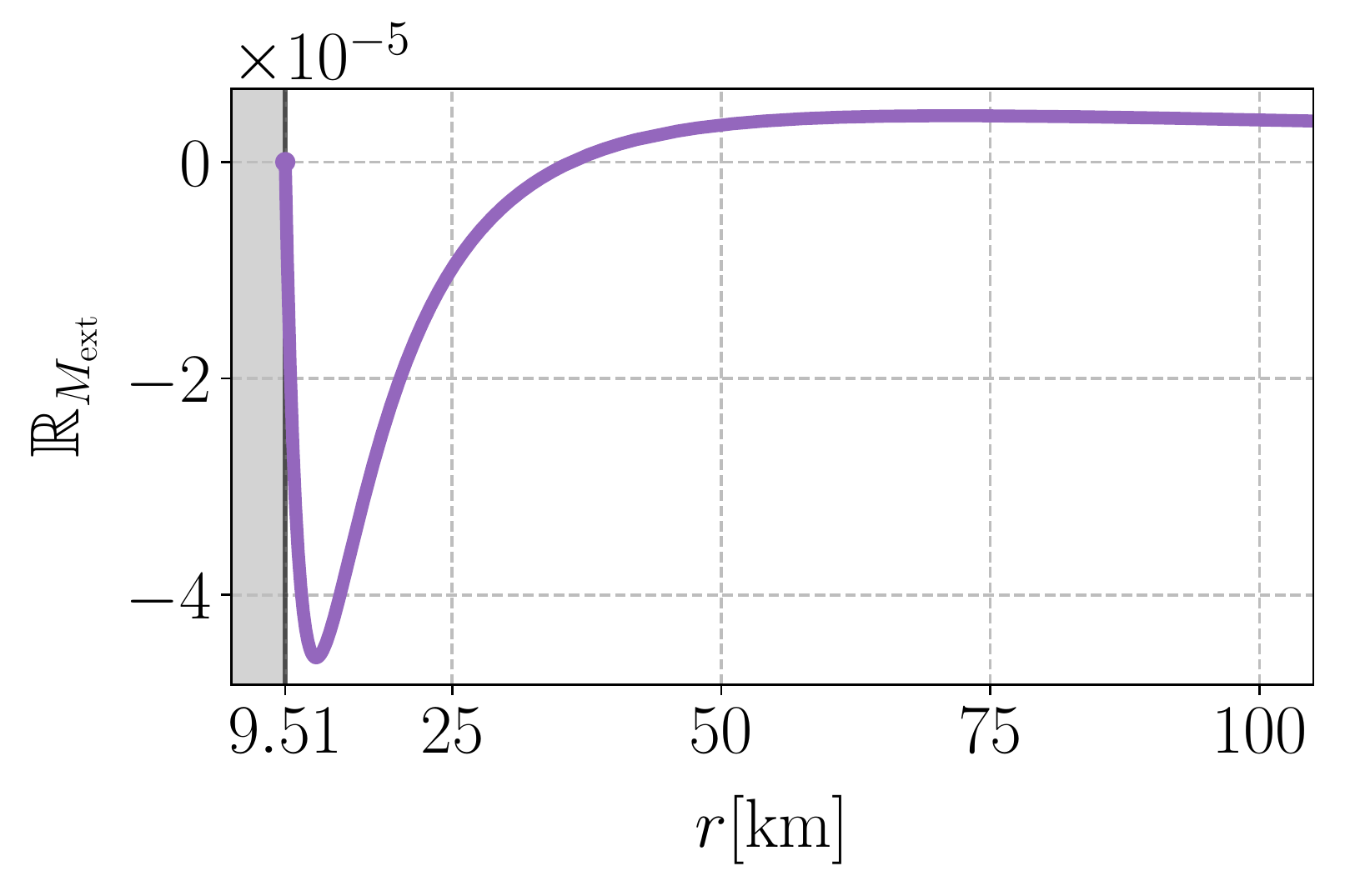}
	\end{tabular}
	
	\vspace{-4mm}
	
	\begin{tabular}{@{}c@{}}\hspace{-3mm}
		\includegraphics[width=.57\linewidth]{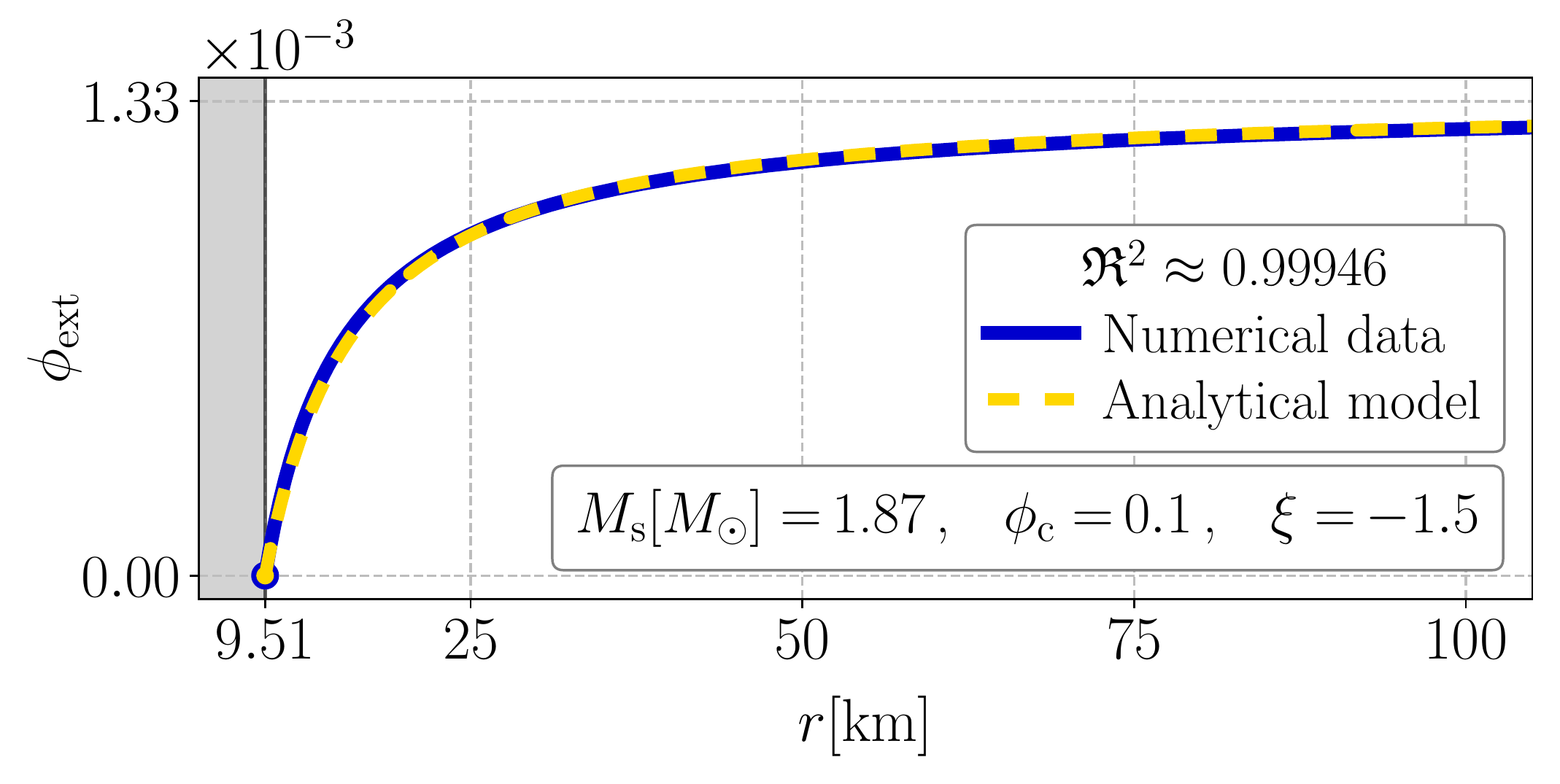}
	\end{tabular}
	\begin{tabular}{@{}c@{}}
		\includegraphics[width=.43\linewidth]{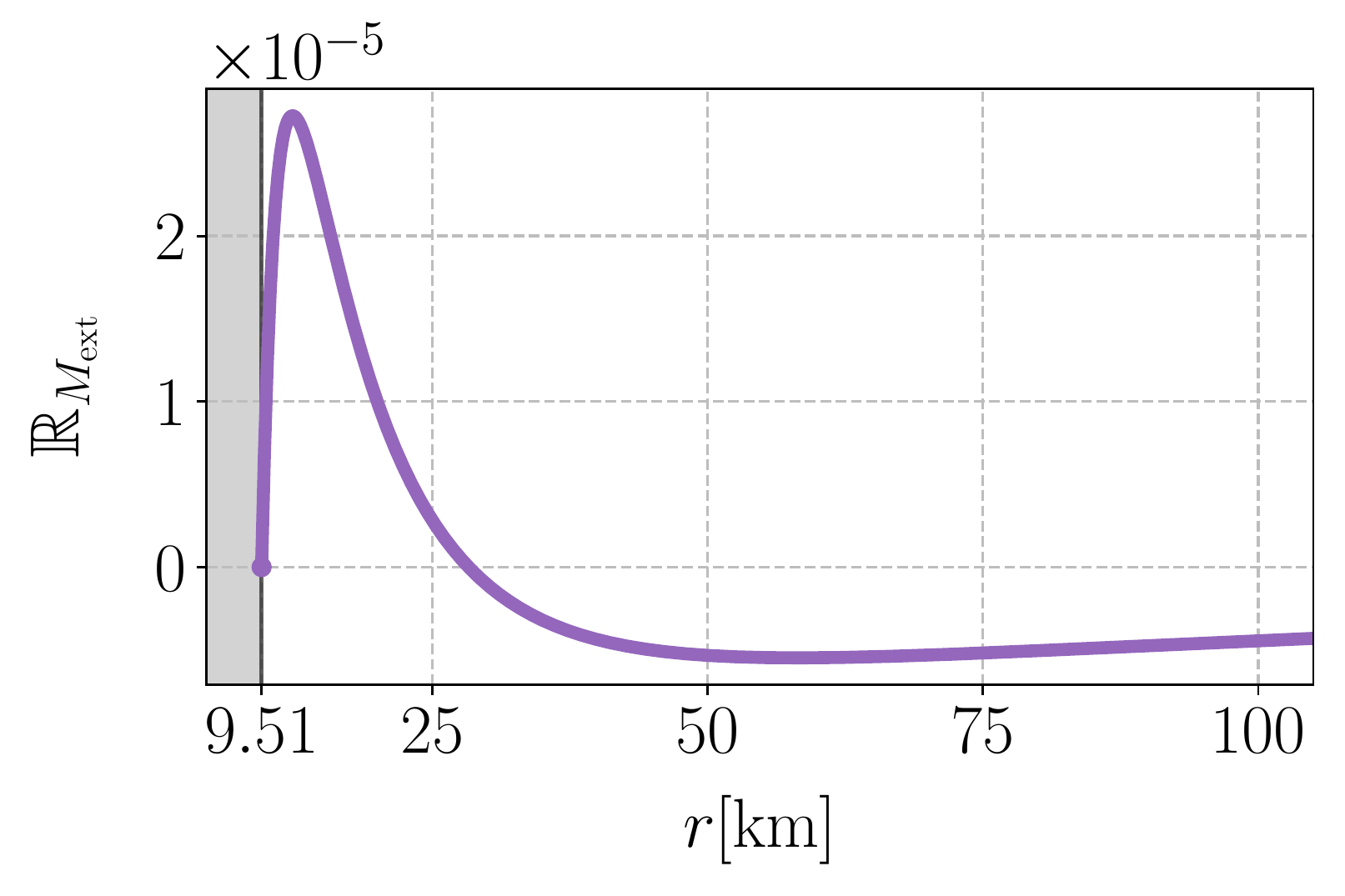}
	\end{tabular}
	
	\vspace{-4mm}
	
	\caption{Results generated through the analytical models given in Eqs.\ \autoref{eq:Mext_fit}, \autoref{eq:DeltaB_fit}, \autoref{eq:DeltaA_fit}, \autoref{eq:phi_fit} from top to bottom respectively. The surface values $M_{\rm{s}}$, $\phi_{\rm{s}}$, $\Delta A_{\rm{s}}$ are the outcomes of the interior solution obtained by using MS1 EoS.}
\label{fig:MS1_results}
\end{figure*}

In spite of the fact that the interior solutions are not needed for our analysis, nevertheless, we solve the equations beginning from the center of the star to determine the mass and the radius values properly as it is the usual method in the literature. To do that, we use SLy \cite{eos_sly} dataset for the EOS throughout the paper although we provide one particular set for MS1 EOS \cite{eos_ms1} as well in the Fig. \autoref{fig:MS1_results}. The order of the differences between two curves are the same with the configuration that has the same approximate mass value for SLy EOS although we have used the different set of $\xi$ and $\phi_{c}$. The distinction in the radius values should be taken into account for the slight differences in the residuals. The only free parameters in our model are the nonminimal coupling constant, $\xi$, and the central value for the scalar field, $\phi_{\rm{c}}$, because choice of these two parameters cause the mass ($M_{\rm{s}}$) and the radius ($R$) values to change, therefore, by applying this approach, we also cover investigation of different mass-radius (MR) configurations which can be achieved by altering the central pressure as well. But, this way is much more compatible with our purpose that is to determine analytical models independent of the free parameters of the model as much as possible.

We give a sample of the radial evolution for $\Delta A(r)$, $\Delta B(r)$, $\phi_{\rm{ext}}(r)$, $\phi'(r)$, and $M_{\rm{ext}}(r)$ in Fig.\ \autoref{fig:funcs_eval} for a specific configuration with $\xi=-0.5$ and $\phi_{\rm{c}}=0.1\,$. The first thing to notice from there is that $\phi_{\rm{ext}}(r)$ and $M_{\rm{ext}}(r)$ show very similar behavior. This observation suggests that the main contribution to the mass function outside the star should be proportional to the scalar field itself. The derivative of the scalar field, on the other hand, quickly vanishes outside indicating that the scalar field takes a constant value at infinity, that is $\phi_\infty$ as defined above. The deviations for the metric functions, namely $\Delta A$ and $\Delta B$, demonstrate that the difference in the vicinity of the star is much bigger as expected and tends to disappear at infinity since $\phi'(r \rightarrow \infty) \rightarrow 0$ which reduces Eq.\ \autoref{eq:components_eq} to Eq.\ \autoref{eq:numerical_sch} far away from the star. In light of these observations we will guess analytical models to fit the data in the following section.

For the parameters we use combinations of values $\xi = -0.5, -1.5$ and $\phi_{\rm{c}} = 0.1, 0.2$ for the representative purposes although we have mainly worked on intervals $\xi \in [-1.5,-0.1]$ and $\phi_{\rm{c}} \in [0.1,0.5]$ in our analysis taking into account the range $\xi \phi_{\rm{c}}^2 \in (-0.01,0)$ found in Ref.\ \cite{Arapoglu2019} and considering a little beyond that limitation as well.

\section{NUMERICAL DATA FITTING} \label{sec:fitting}

When fitting the numerical data, we guess an appropriate function based on the radial evolution in Fig.\ \autoref{fig:funcs_eval} and then check the accuracy through R-squared ($\Rsq$) analysis. Since this test alone does not guarantee the correctness of the fit, we also provide the residuals to see the bias in our analytical models.

\subsection{The Mass Function}

\begin{figure*}[!t]
	\centering

	\begin{tabular}{@{}c@{}}\hspace{-3mm}
		\includegraphics[width=.57\linewidth]{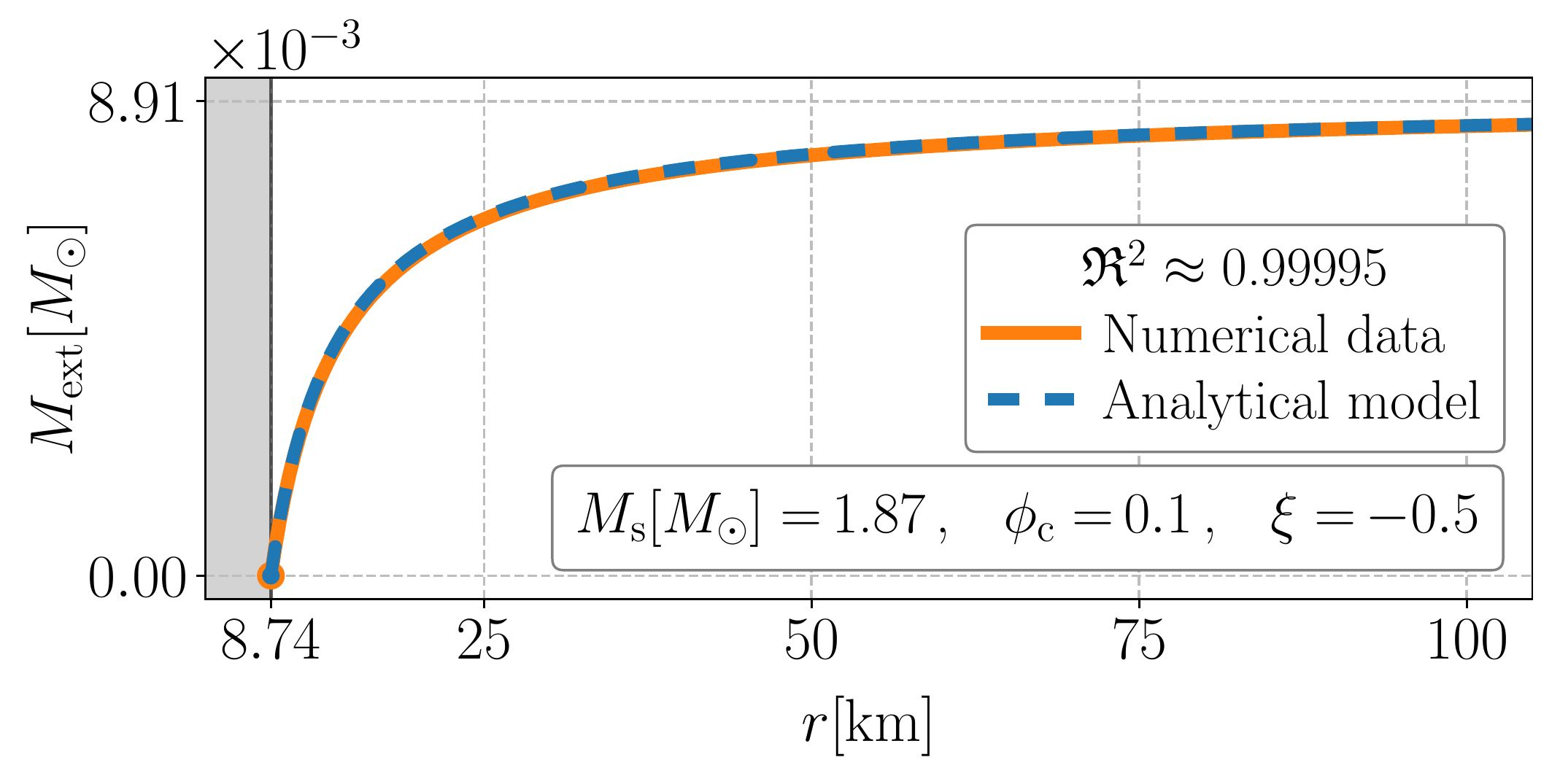}
	\end{tabular}
	\begin{tabular}{@{}c@{}}
		\includegraphics[width=.43\linewidth]{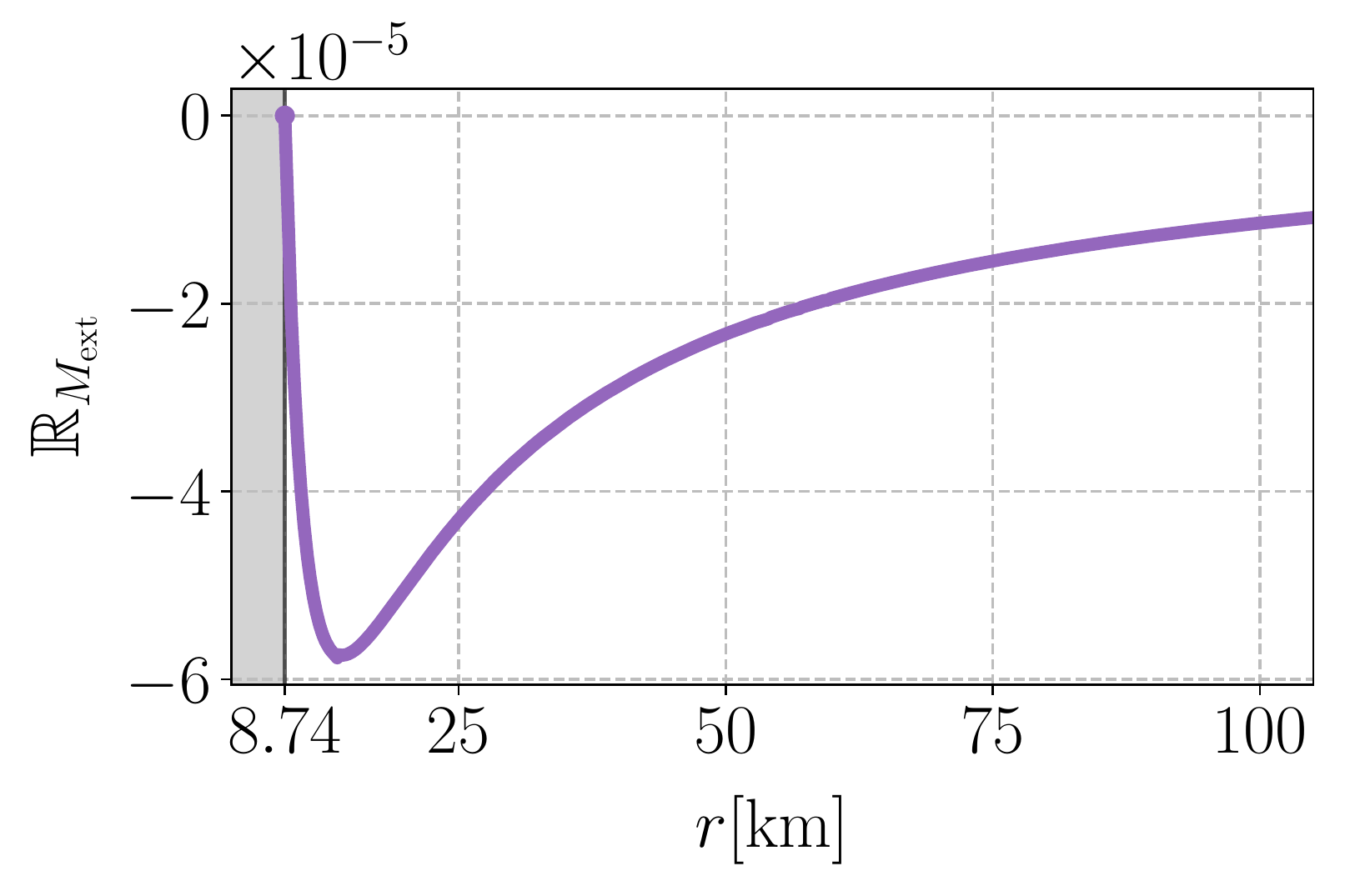}
	\end{tabular}
	
	\vspace{-2mm}
	
	\begin{tabular}{@{}c@{}}\hspace{-3mm}
		\includegraphics[width=.57\linewidth]{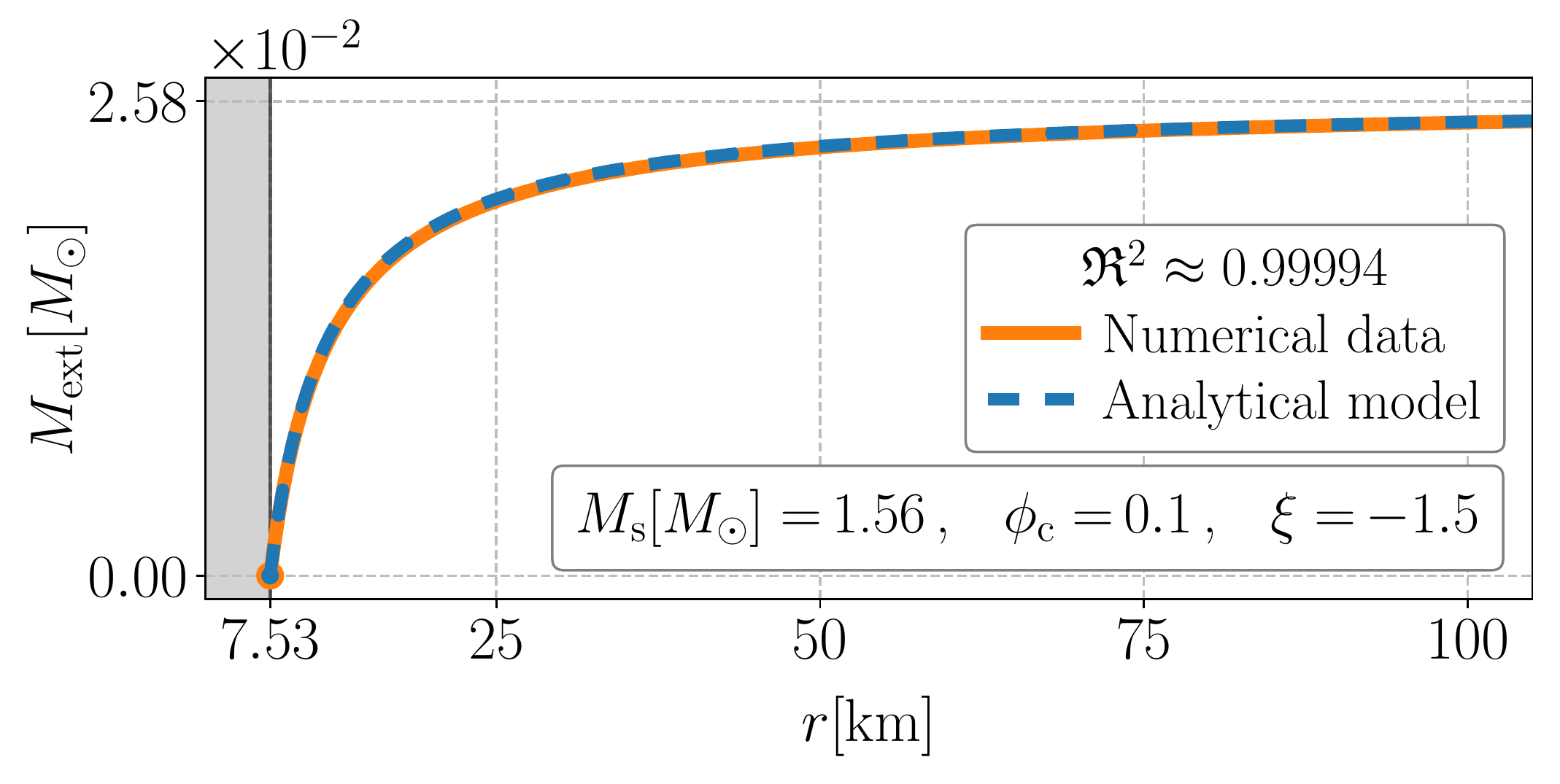}
	\end{tabular}
	\begin{tabular}{@{}c@{}}
		\includegraphics[width=.43\linewidth]{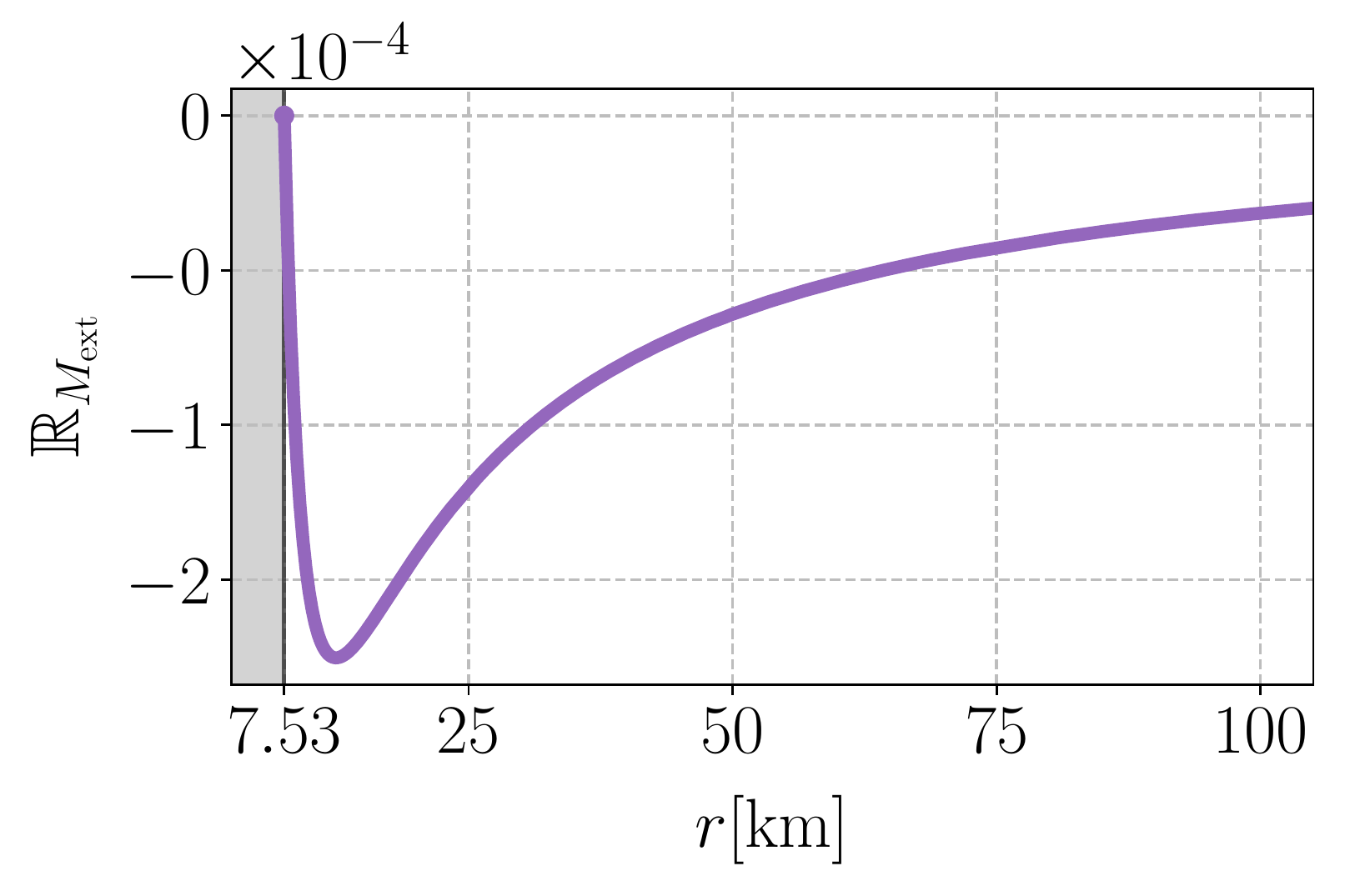}
	\end{tabular}
	
	\vspace{-2mm}
	
	\begin{tabular}{@{}c@{}}\hspace{-3mm}
		\includegraphics[width=.57\linewidth]{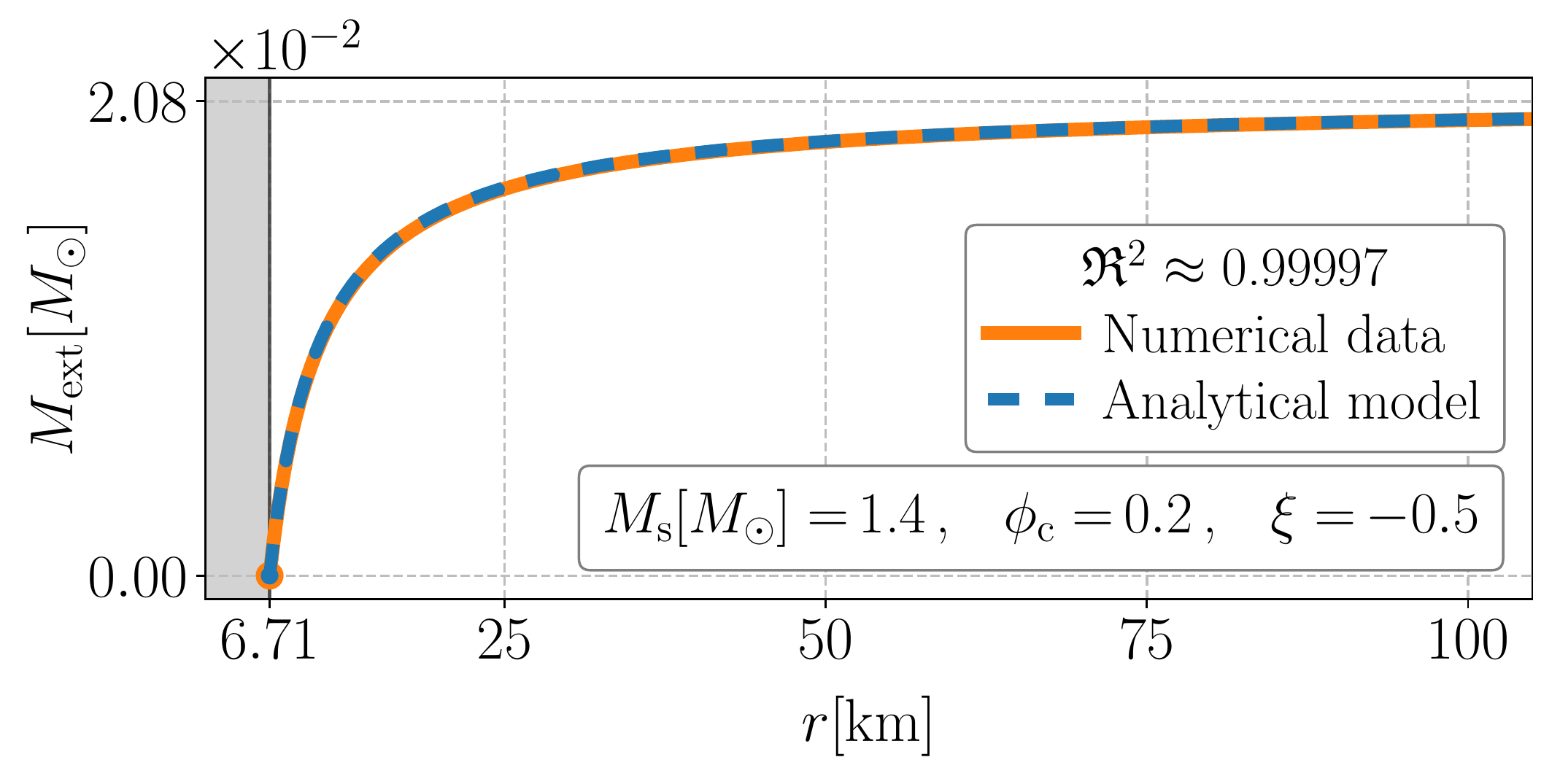}
	\end{tabular}
	\begin{tabular}{@{}c@{}}
		\includegraphics[width=.43\linewidth]{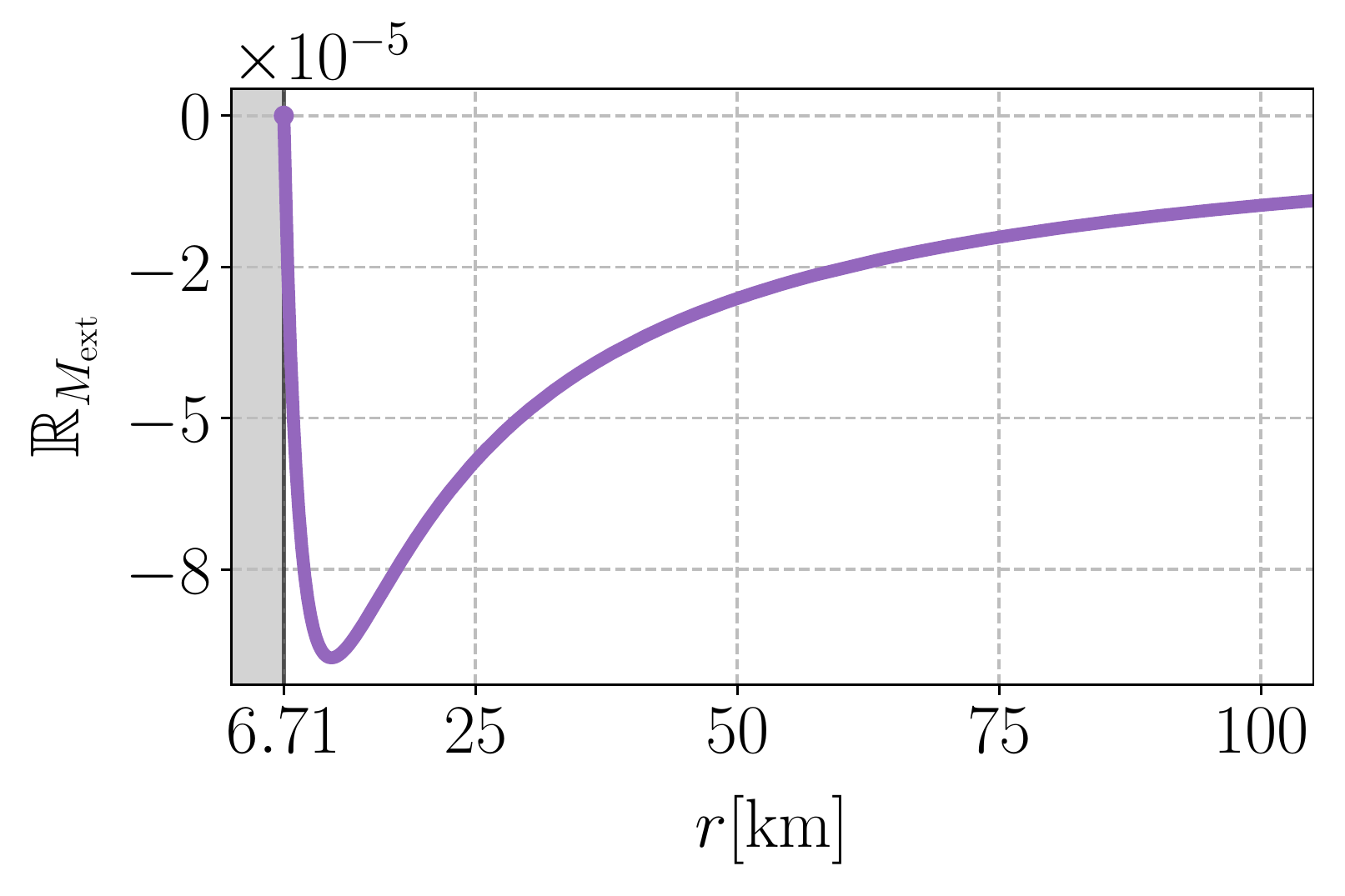}
	\end{tabular}
	
	\vspace{-2mm}
	
	\caption{Comparison of the numerical data and the analytical model given in Eq.\ \autoref{eq:Mext_fit} for $M_{\rm{ext}}(r)$. $M_{\rm{s}}$, $\phi_{\rm{c}}$, and $\xi$ indicate the mass value at the surface, central value of the scalar field and the coupling constant, respectively. Radius of the star is the leftmost number in the horizontal axes. The central pressure value is $P_{\rm{c}}=2.25 \times 10^{-3} \: \mbox{km}^{-2}$. Residuals in the right panel show the difference between two curves plotted in the left column as explained in the text. The result of $\Rsq$ analysis is also given inside the plots for each case.}
	\label{fig:M_ext_fits_and_residuals}
\end{figure*}

As mentioned before, the radial evolution given in Fig.\ \autoref{fig:funcs_eval} for the mass function and the scalar field in the exterior region suggests that these two functions should be directly proportional to some extent. This relation can be sought through the following form
\begin{equation}
    M_{\rm{ext}}(r) = \mu \, \phi_{\rm{ext}}(r) + \Rsdl{M_{\rm{ext}}}
\label{eq:Mext_fit}
\end{equation}
where $\mu$ is a constant and $\Rsdl{M_{\rm{ext}}}$ designates the residual that is the difference between the numerical data and the analytical model. In order to find the proportionality constant, $\mu$, we normalize the functions on both sides as a straightforward attempt, in other words, we define $\mu \equiv (M_{\infty} - M_{\rm{s}})/(\phi_{\infty} - \phi_{\rm{s}})$, where, as indicated before, the subscripts ``$\infty$" and ``s" stand for the values of the functions at infinity and at the surface of the star, respectively. We represent the results in Fig.\ \autoref{fig:M_ext_fits_and_residuals} for different choices of the free parameters, thus, creating various MR configurations. In the same figure we also provide the residuals, i.e. $\Rsdl{M_{\rm{ext}}}$, for each case. It seems that the expression for the external mass function written in Eq.\ \autoref{eq:Mext_fit} fits well to the data for different initial conditions to the scalar field value and different nonminimal coupling parameter, $\xi$. Deviation of the R-squared value from unity is of the order, $\mathcal{O}(\Rsq - 1) \simeq 10^{-5}$ which shows the accuracy of our analytical model in all cases. However, there is an evident deviation in a form which seems to be related to $1/\Delta B$ at first glance. Our analysis shows that this is not the case entirely. One may reduce this difference in the vicinity of the star by using $1/\Delta B$, but this becomes problematic for the further points in the radial evolution.

\begin{figure*}[!h]
	\centering

	\begin{tabular}{@{}c@{}}\hspace{-3mm}
		\includegraphics[width=.57\linewidth]{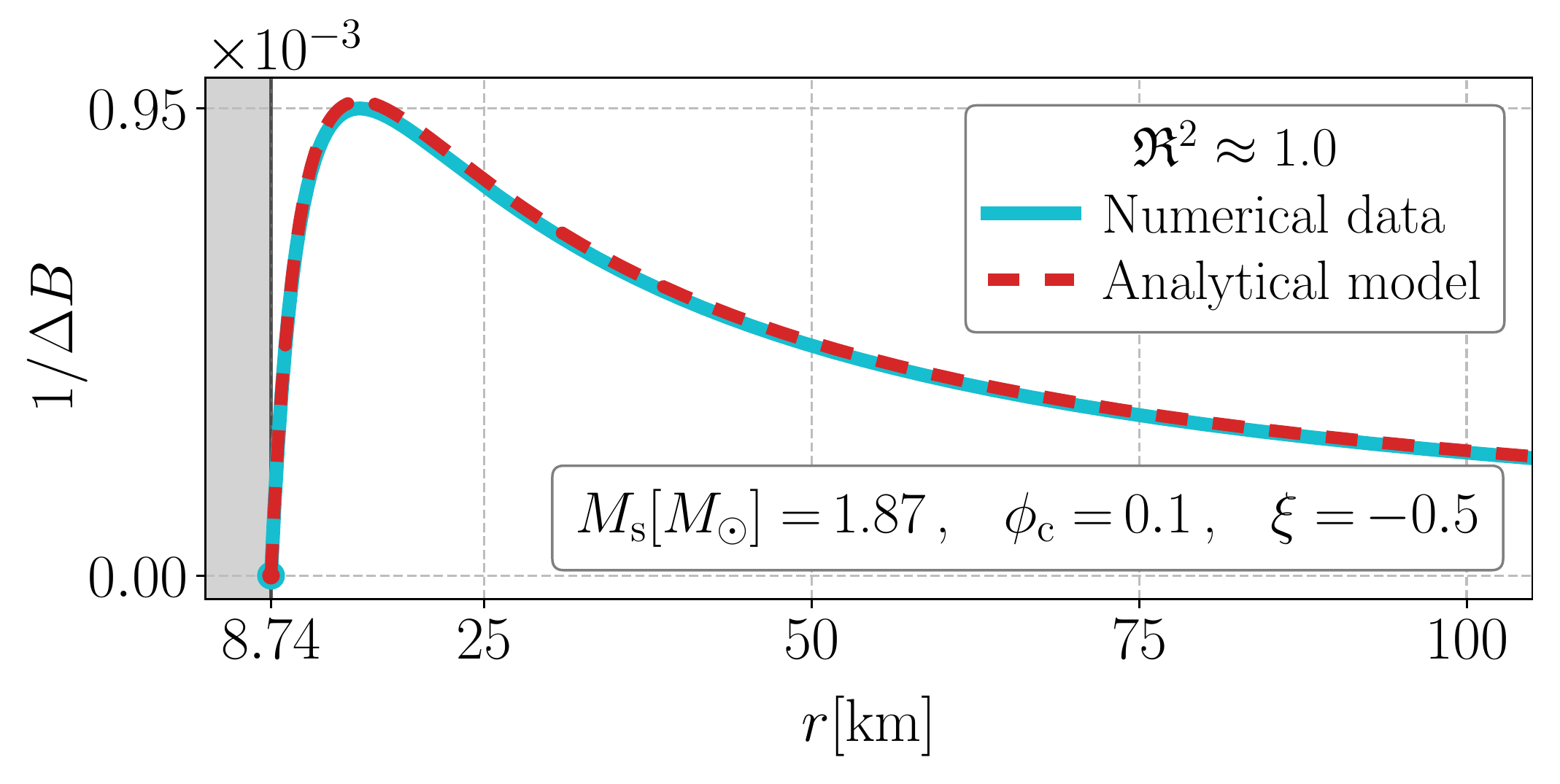}
	\end{tabular}
	\begin{tabular}{@{}c@{}}
		\includegraphics[width=.43\linewidth]{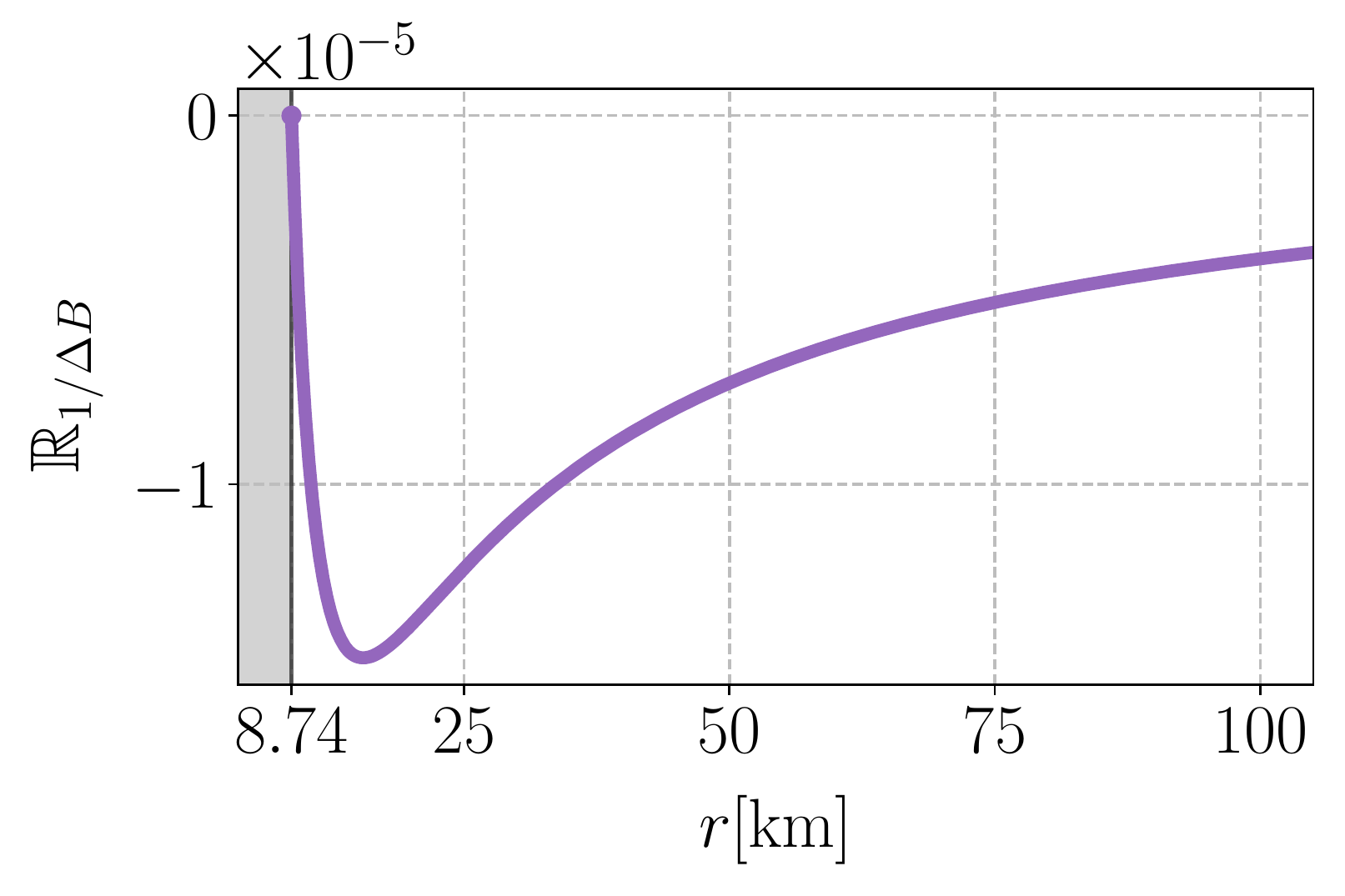}
	\end{tabular}
	
	\vspace{-4mm}
	
	\begin{tabular}{@{}c@{}}\hspace{-3mm}
		\includegraphics[width=.57\linewidth]{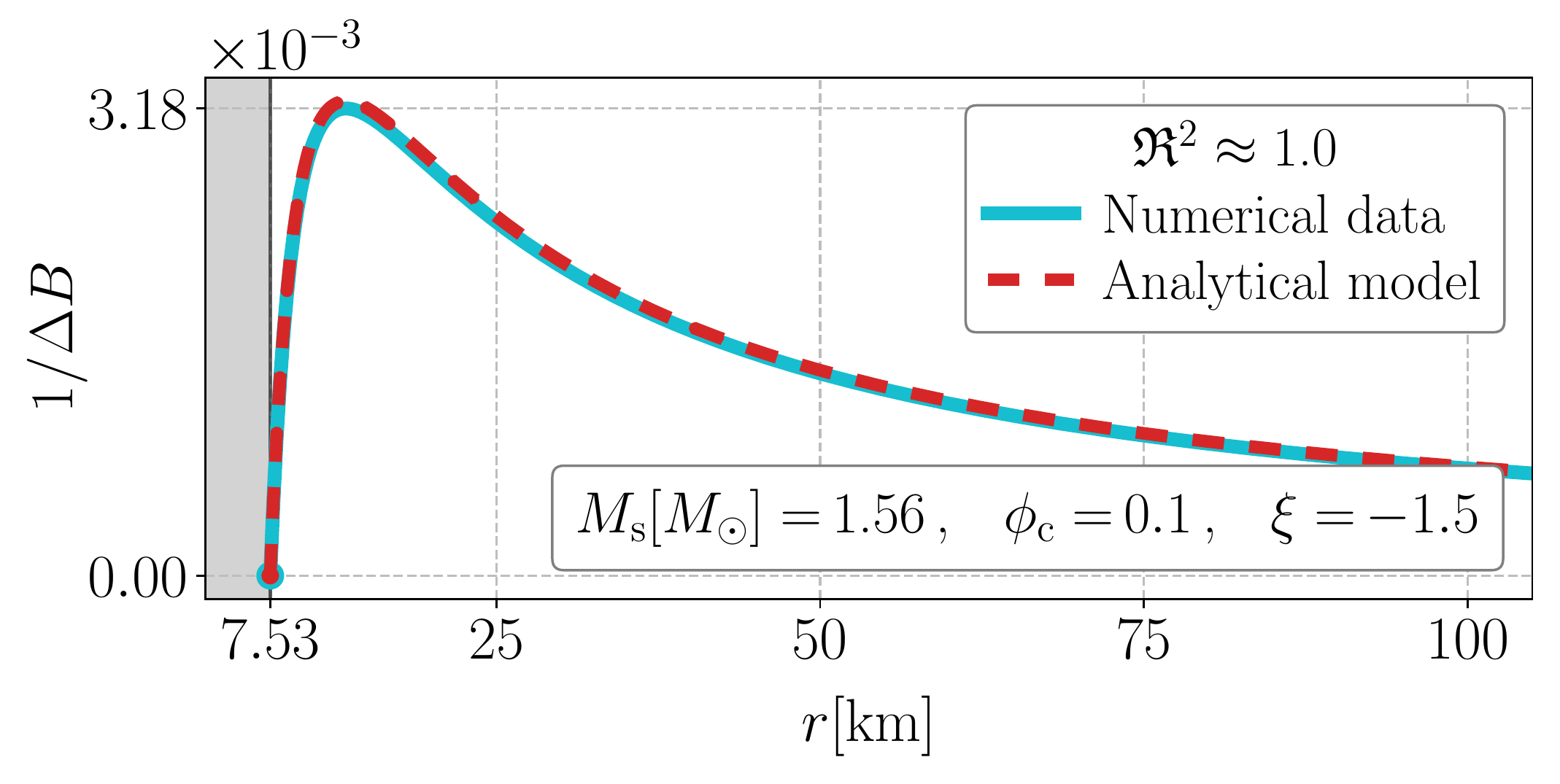}
	\end{tabular}
	\begin{tabular}{@{}c@{}}
		\includegraphics[width=.43\linewidth]{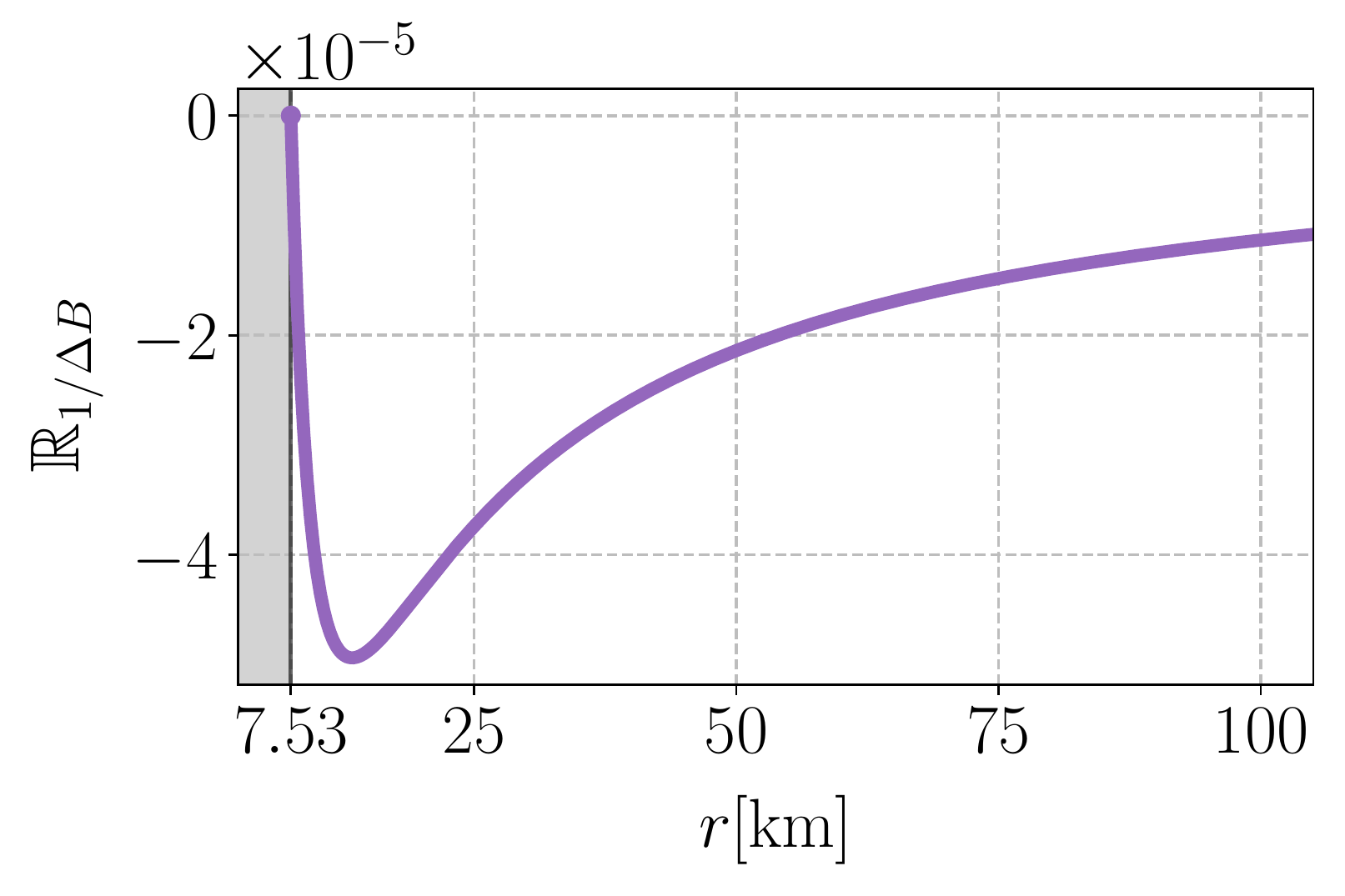}
	\end{tabular}
	
	\vspace{-4mm}
	
	\begin{tabular}{@{}c@{}}\hspace{-3mm}
		\includegraphics[width=.57\linewidth]{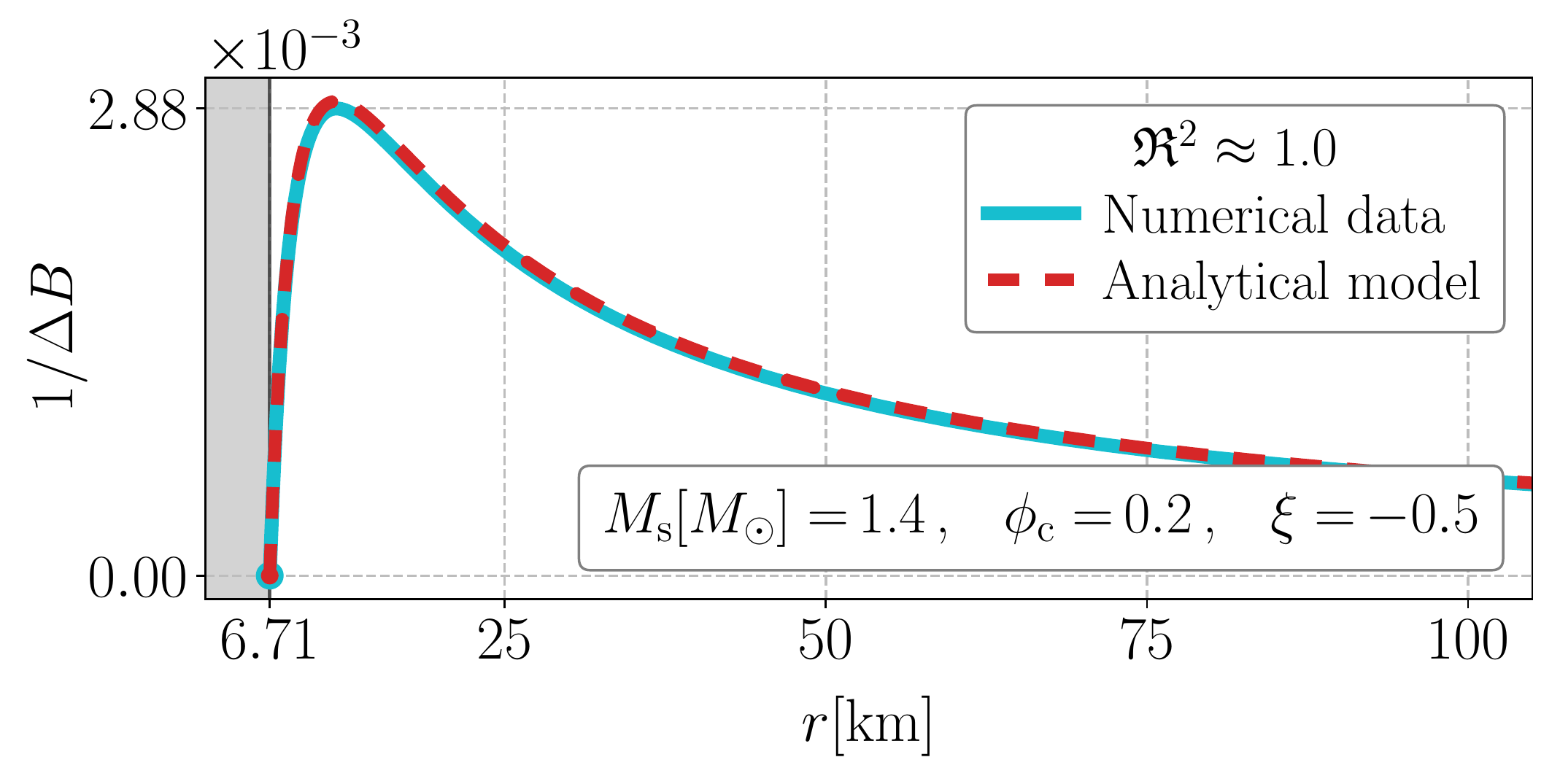}
	\end{tabular}
	\begin{tabular}{@{}c@{}}
		\includegraphics[width=.43\linewidth]{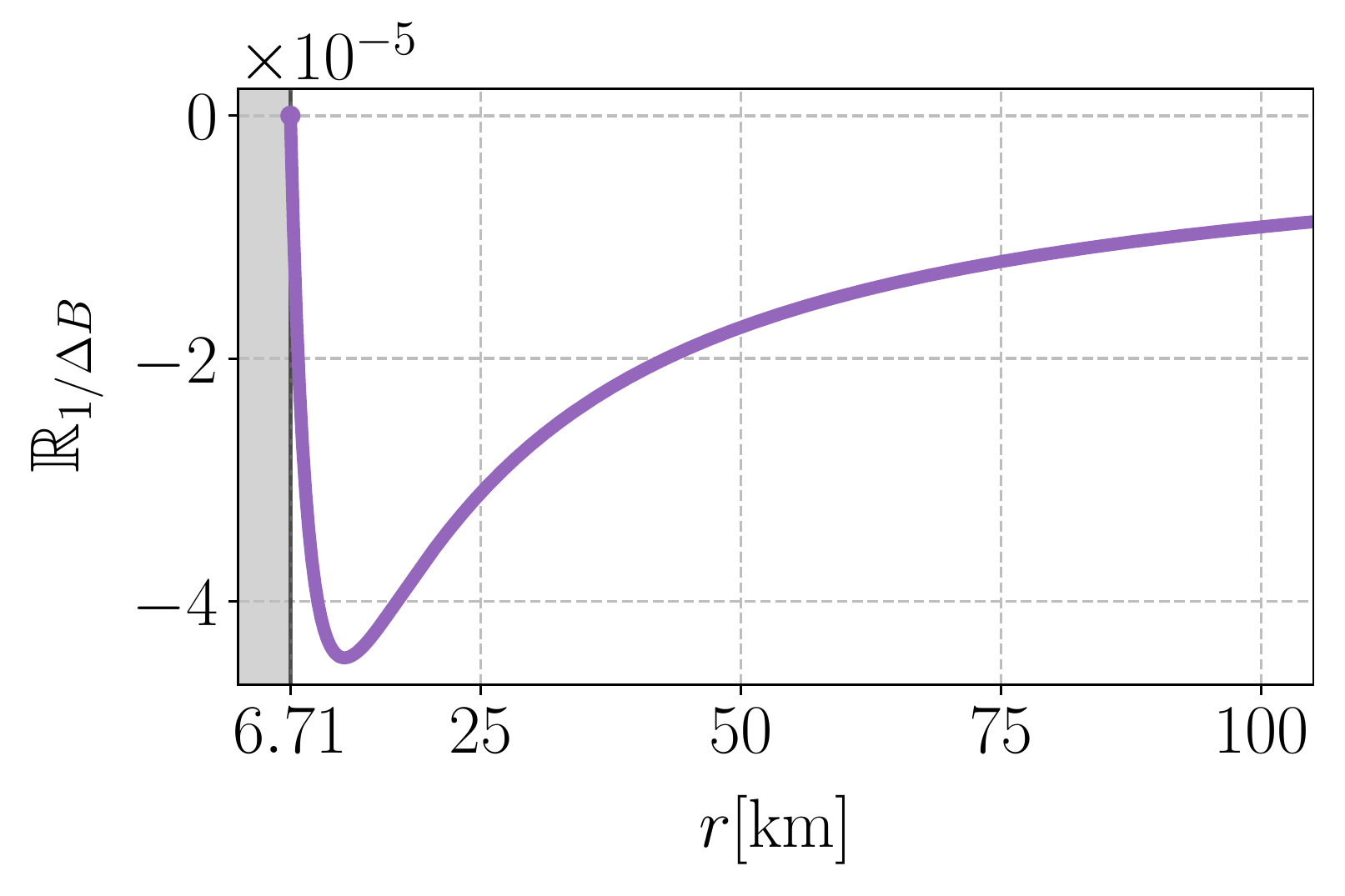}
	\end{tabular}
	
	\vspace{-2mm}
	
	\caption{Comparison of the numerical data and the analytical model given in Eq.\ \autoref{eq:DeltaB_fit} for $\Delta B(r)$. (For details see the caption of Fig.\ \autoref{fig:M_ext_fits_and_residuals}.)}
	\label{fig:DeltaB_fits_and_residuals}
\end{figure*}

\subsection{The Metric Functions}

\paragraph*{Radial part ($\Delta B$) :}
An educated guess for an analytical form of the correction to the radial part of the metric function, $1/\Delta B(r)$, can be made in the form of the second term of the Schwarzchild solution, i.e., mass to radial coordinate ratio up to a scaling constant. However, this time the mass function is not constant outside the configuration. Hence, we should replace the constant mass with its dynamical counterpart, $M_{\rm{ext}}(r)$. Analyzing various configurations to determine the scaling constant we have found that $1/\Delta B(r)$ can be written in terms of the external mass function and the radial coordinate as follows
\begin{equation}
    \sfrac{1}{\Delta B(r)} = \sfrac{3 M_{\rm{ext}}(r)}{r} + \Rsdl{1/\Delta B}
\label{eq:DeltaB_fit}
\end{equation}
the results of which are shown in Fig.\ \autoref{fig:DeltaB_fits_and_residuals} along with the residuals $\Rsdl{1/\Delta B}$ on the right-hand side of the same figure. It seems from the graphs that this model shows very compatible behavior with the numerical solution up to the residual order of $10^{-5}$ which has no major changes against different parameter values. Moreover, it is important to note that the above \textit{ansatz} does not explicitly depend on neither the nonminimal coupling parameter, $\xi$, nor the central value of the scalar field, $\phi_{\rm{c}}$. As a result, our initial attempt of writing the metric function deviation, $1/\Delta B(r)$, has $10^{-5}$ orders of magnitude deviation from the data with a perfect $\Rsq$ value.

\begin{figure*}[!t]
	\centering

	\begin{tabular}{@{}c@{}}\hspace{-3mm}
		\includegraphics[width=.57\linewidth]{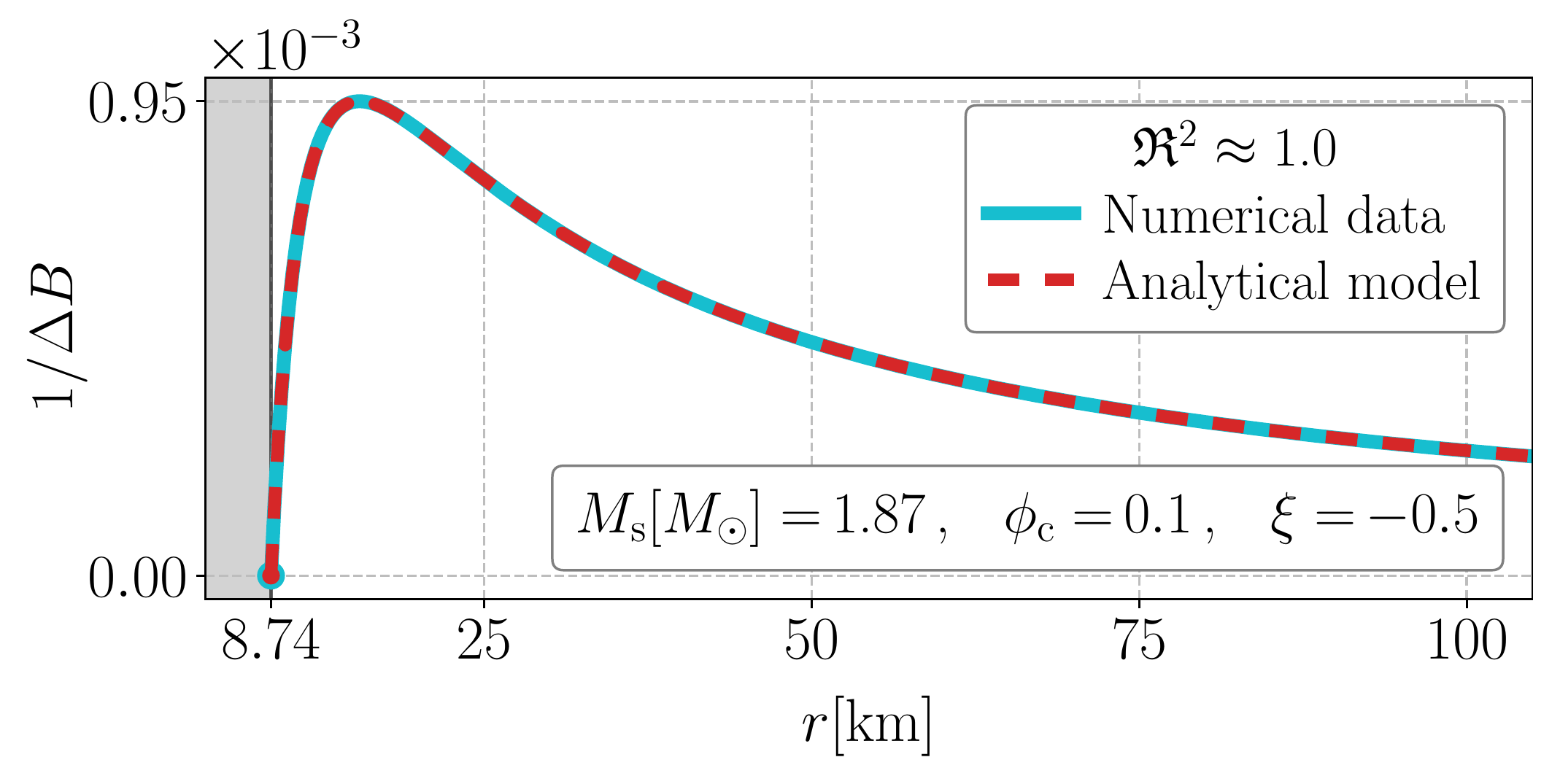}
	\end{tabular}
	\begin{tabular}{@{}c@{}}
		\includegraphics[width=.43\linewidth]{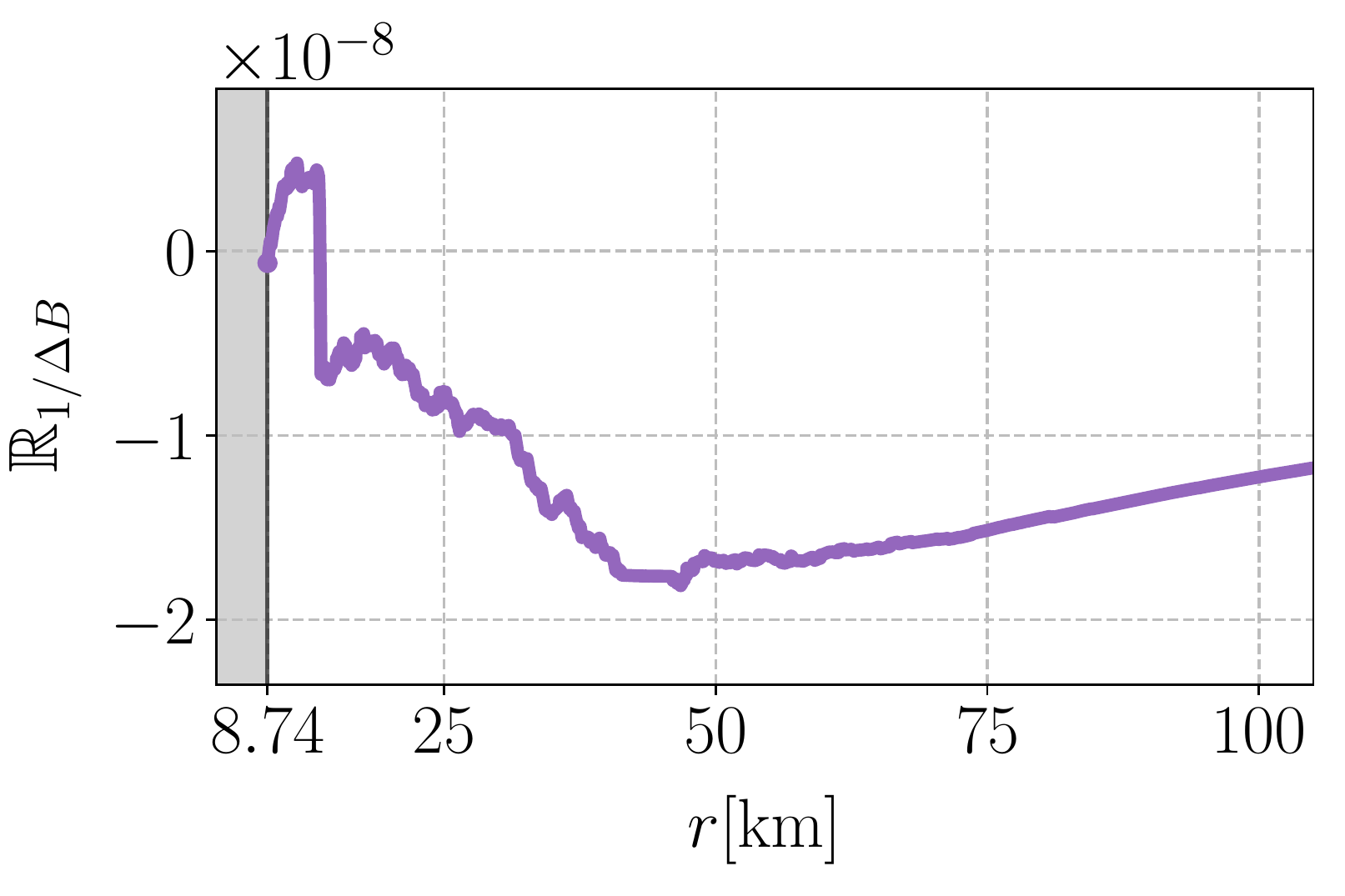}
	\end{tabular}
	
	\vspace{-2mm}
	
	\begin{tabular}{@{}c@{}}\hspace{-3mm}
		\includegraphics[width=.57\linewidth]{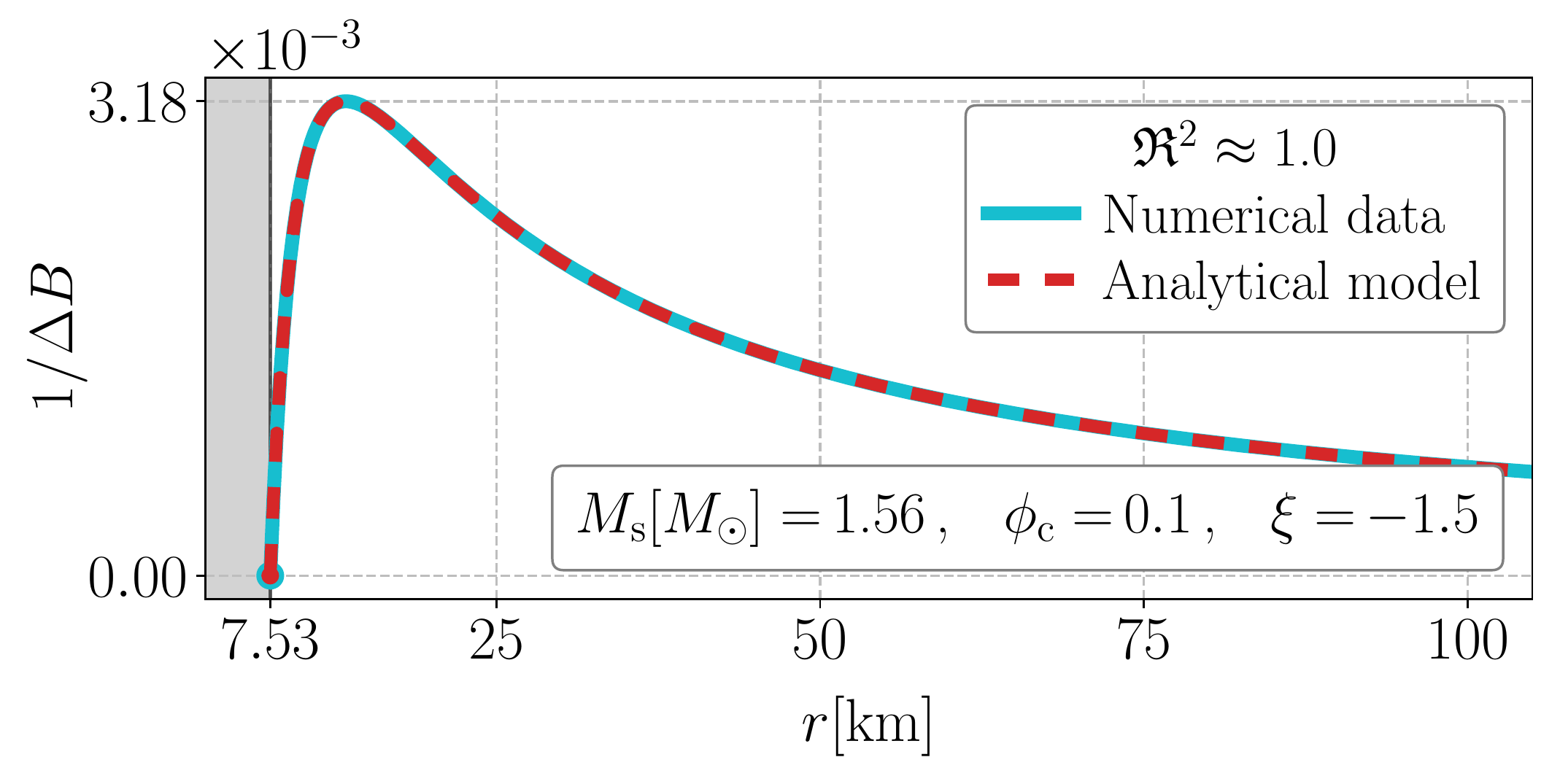}
	\end{tabular}
	\begin{tabular}{@{}c@{}}
		\includegraphics[width=.43\linewidth]{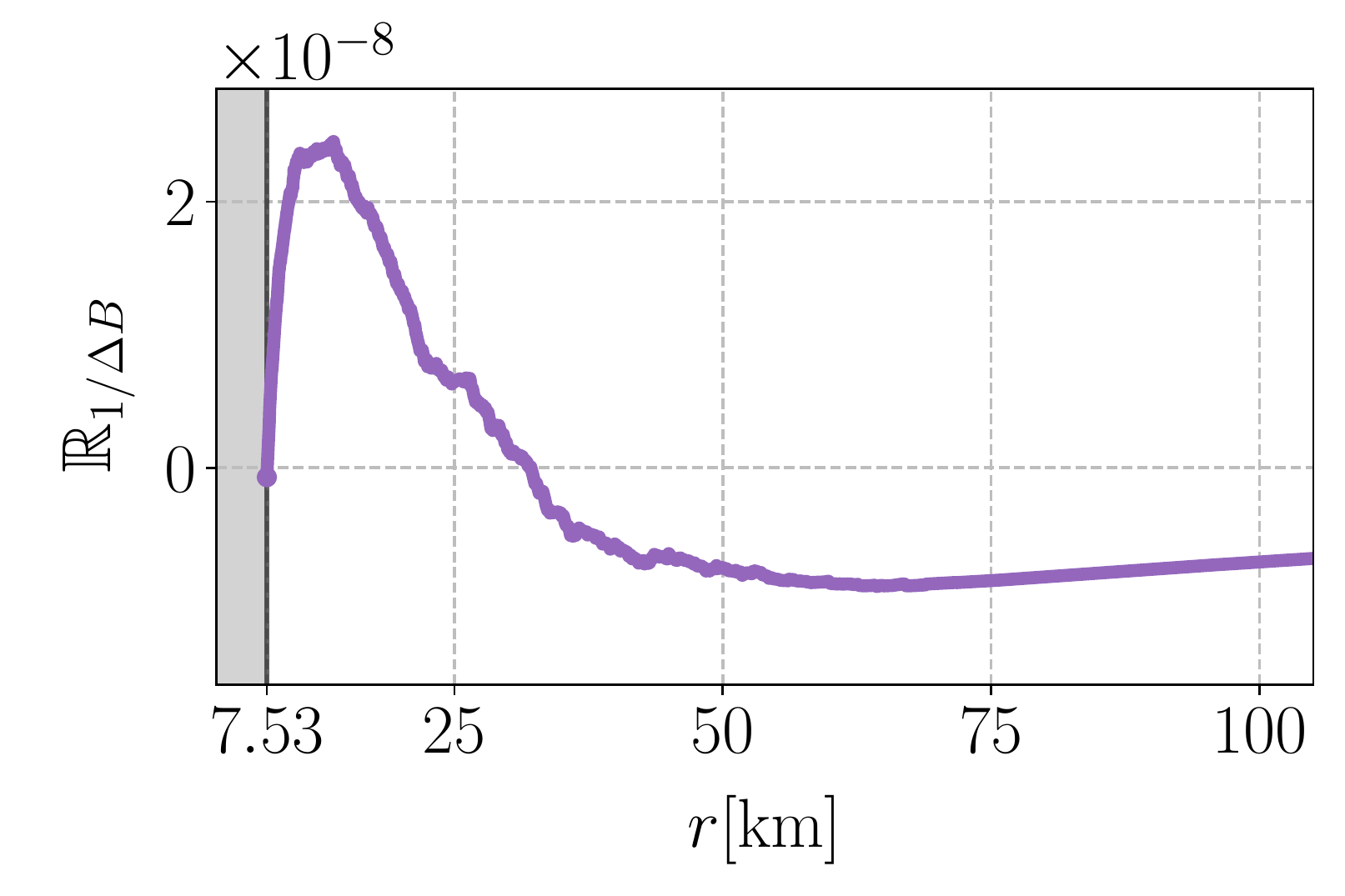}
	\end{tabular}
	
	\vspace{-2mm}
	
	\begin{tabular}{@{}c@{}}\hspace{-3mm}
		\includegraphics[width=.57\linewidth]{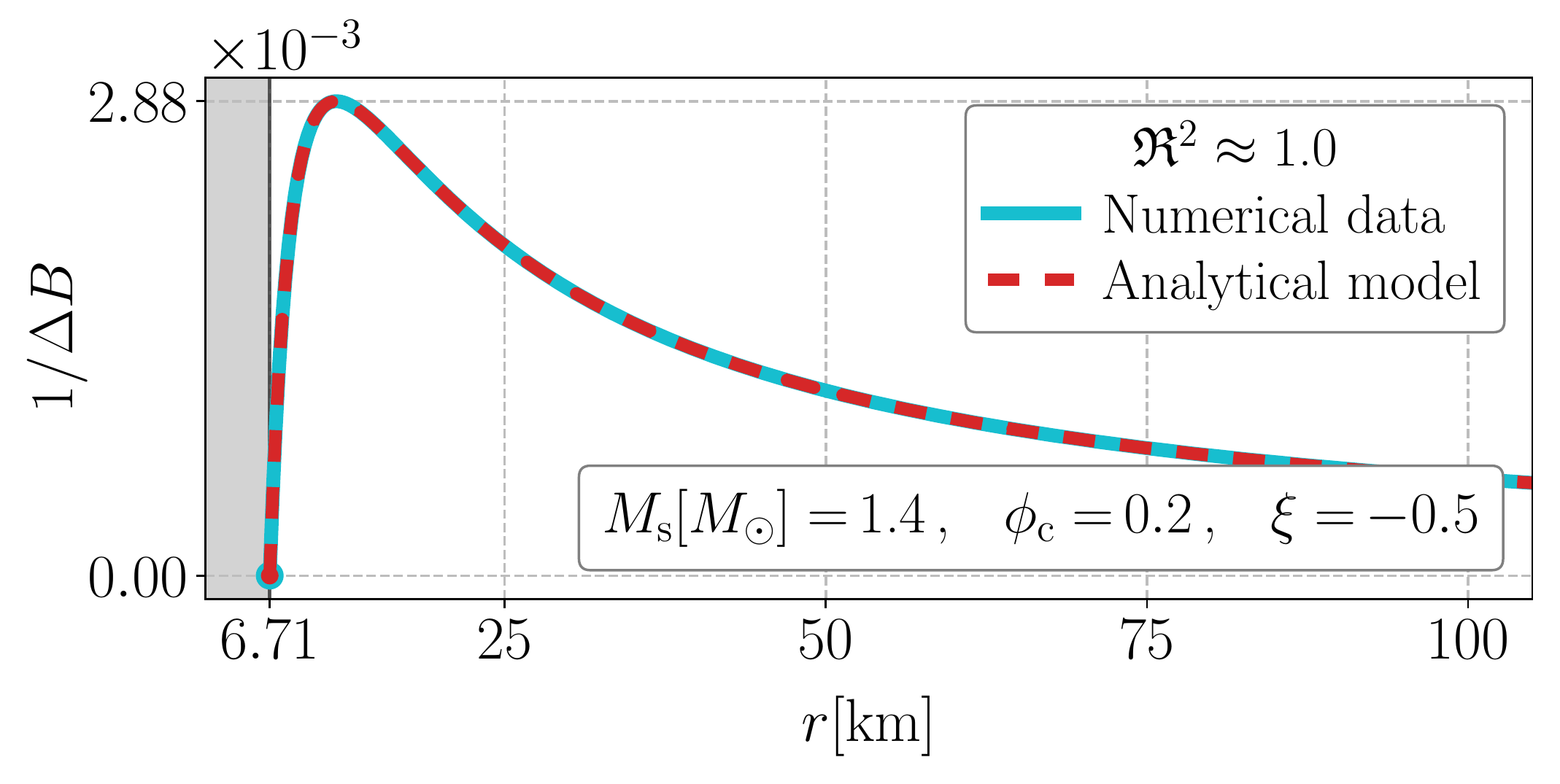}
	\end{tabular}
	\begin{tabular}{@{}c@{}}
		\includegraphics[width=.43\linewidth]{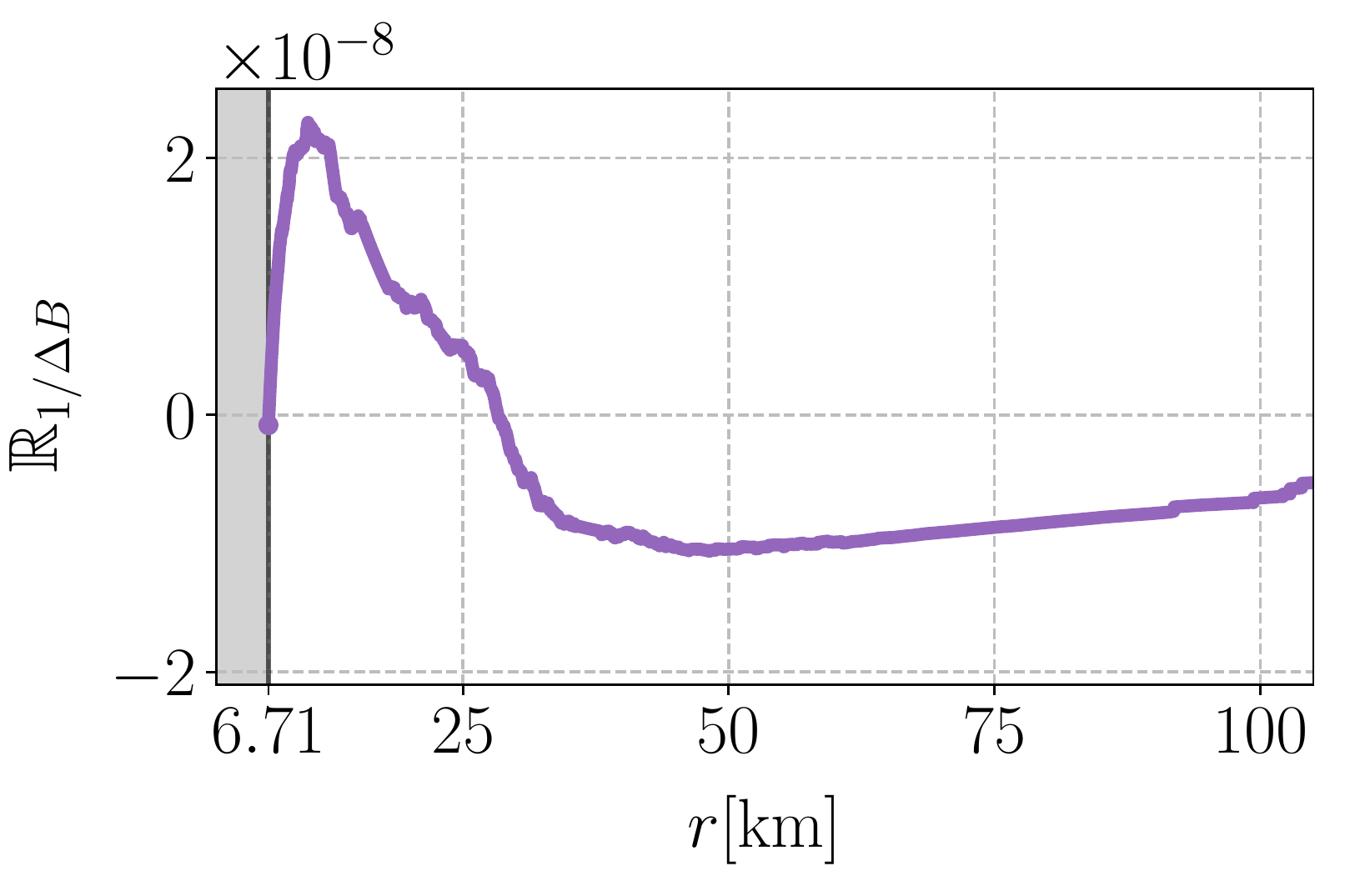}
	\end{tabular}
	
	\vspace{-2mm}
	
	\caption{The same configuration represented in Fig.\ \autoref{fig:DeltaB_fits_and_residuals} with the analytical model given in Eq.\ \autoref{eq:DeltaB_fit_00155} where $\al=0.01552\,$.}
	\label{fig:DeltaB_fits_and_residuals_00155}
\end{figure*}

Examination of the residuals indicates that its form is showing similar behavior as negative of $1/\Delta B$ itself. Therefore, in order to improve our analytical model we subtract a term that is proportional to the $1/\Delta B$ from the initial expression in Eq.\ \autoref{eq:DeltaB_fit} and we arrive at
\begin{equation}
    \sfrac{1}{\Delta B(r)} = \sfrac{3}{1+\al} \sfrac{M_{\rm{ext}}(r)}{r} + \Rsdl{1/\Delta B}
\label{eq:DeltaB_fit_00155}
\end{equation}
for which the results can be seen in Fig.\ \autoref{fig:DeltaB_fits_and_residuals_00155} where we set the proportionality constant $\alpha$ to $0.01552$ as a result of a series of analysis with different configurations. The order of magnitude of the residuals with the introduction of the $\alpha$ constant is now $\mathcal{O}\left(\Rsdl{1/\Delta B}\right) = 10^{-8}$ at the immediate outside the star and becomes around $\mathcal{O}\left(\Rsdl{1/\Delta B}\right)=10^{-7}$ at most at far outside the star. These residual values are in the same and even mostly beyond the order of our error calculations in our code as represented in Fig.\ \autoref{fig:GR_sol_abs_err} and this also explains the random oscillations in the plots.

\begin{figure*}[!t]
	\centering

	\begin{tabular}{@{}c@{}}\hspace{-3mm}
		\includegraphics[width=.57\linewidth]{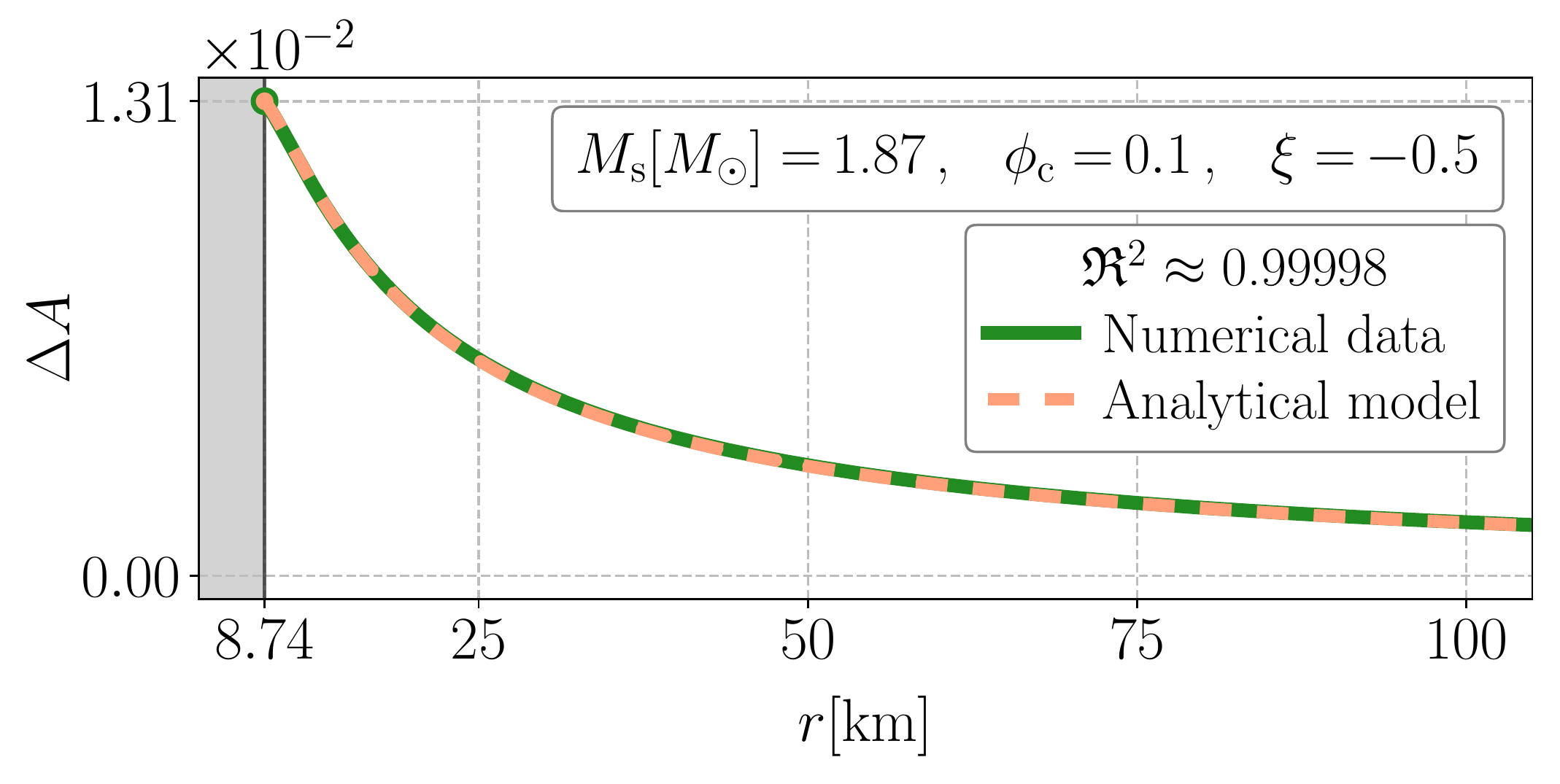}
	\end{tabular}
	\begin{tabular}{@{}c@{}}
		\includegraphics[width=.43\linewidth]{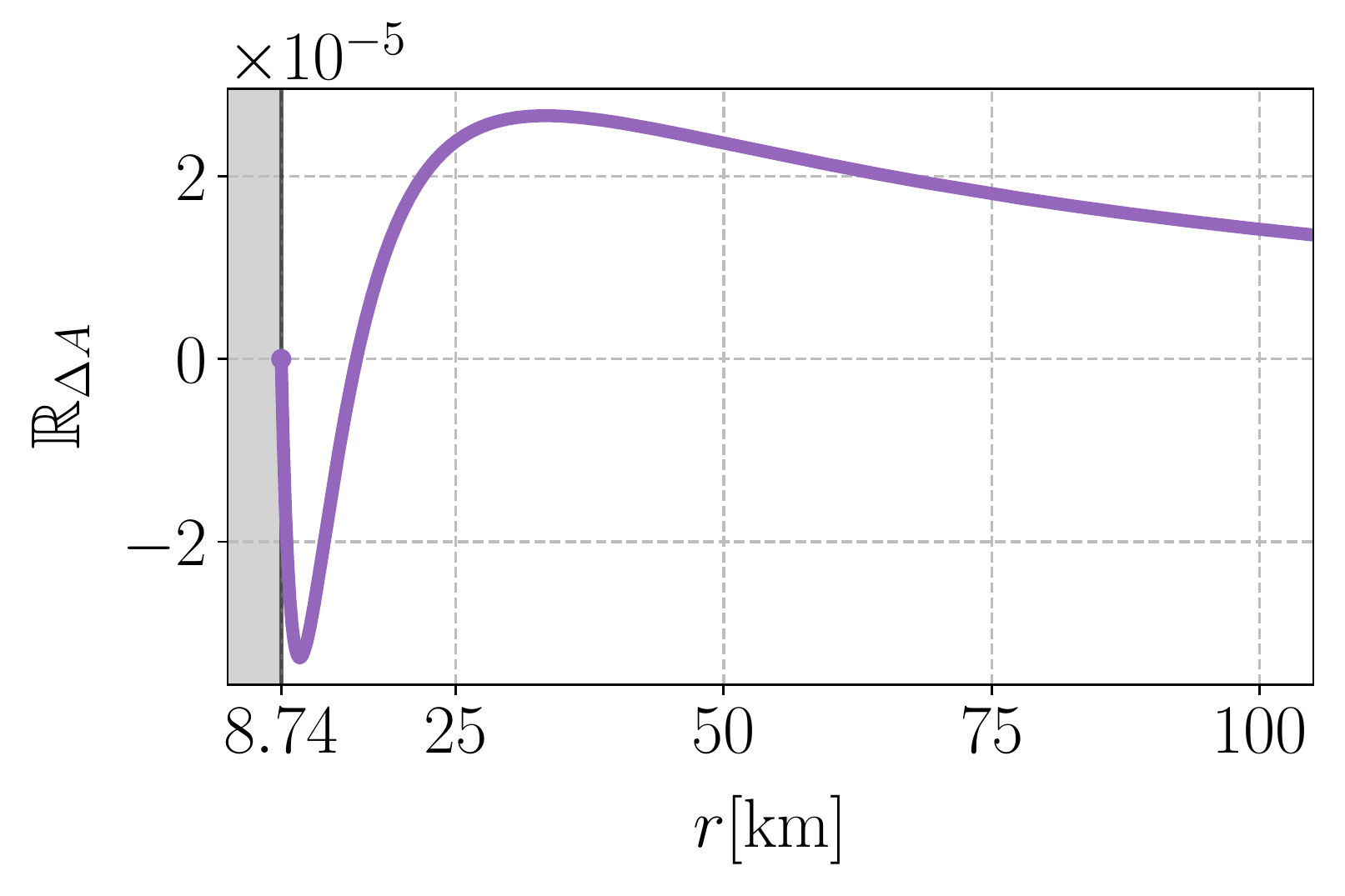}
	\end{tabular}
	
	\vspace{-2mm}
	
	\begin{tabular}{@{}c@{}}\hspace{-3mm}
		\includegraphics[width=.57\linewidth]{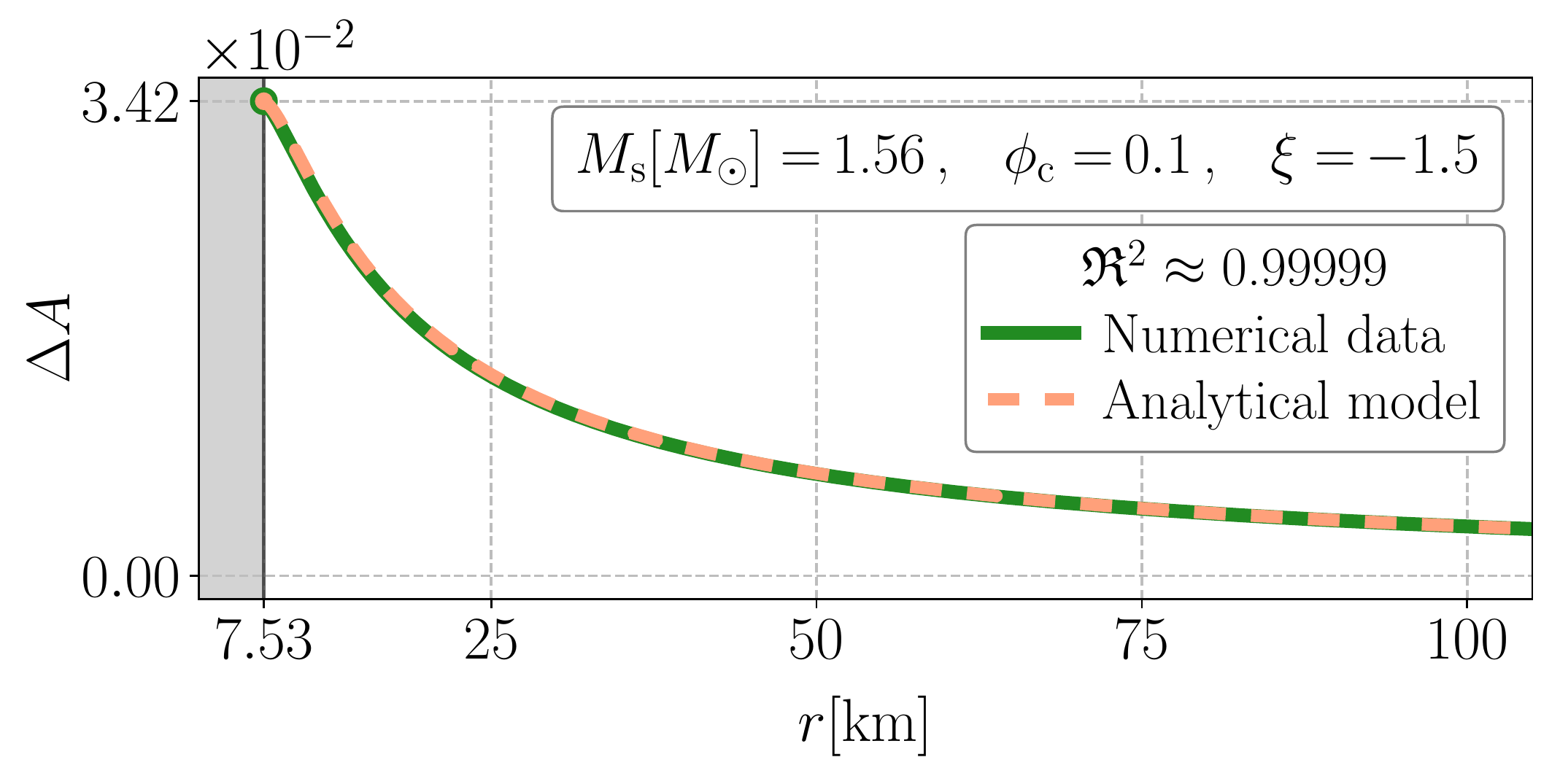}
	\end{tabular}
	\begin{tabular}{@{}c@{}}
		\includegraphics[width=.43\linewidth]{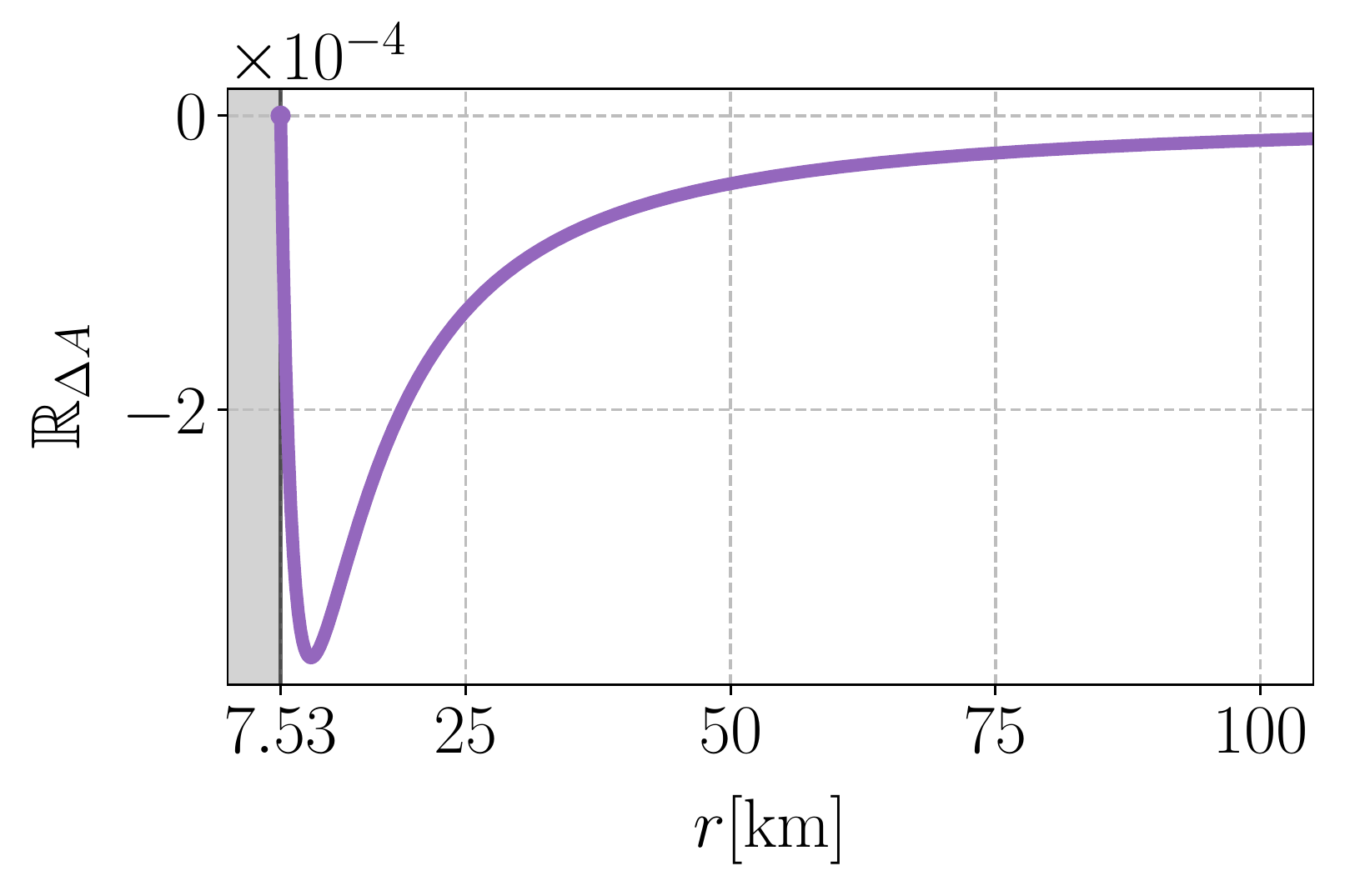}
	\end{tabular}
	
	\vspace{-2mm}
	
	\begin{tabular}{@{}c@{}}\hspace{-3mm}
		\includegraphics[width=.57\linewidth]{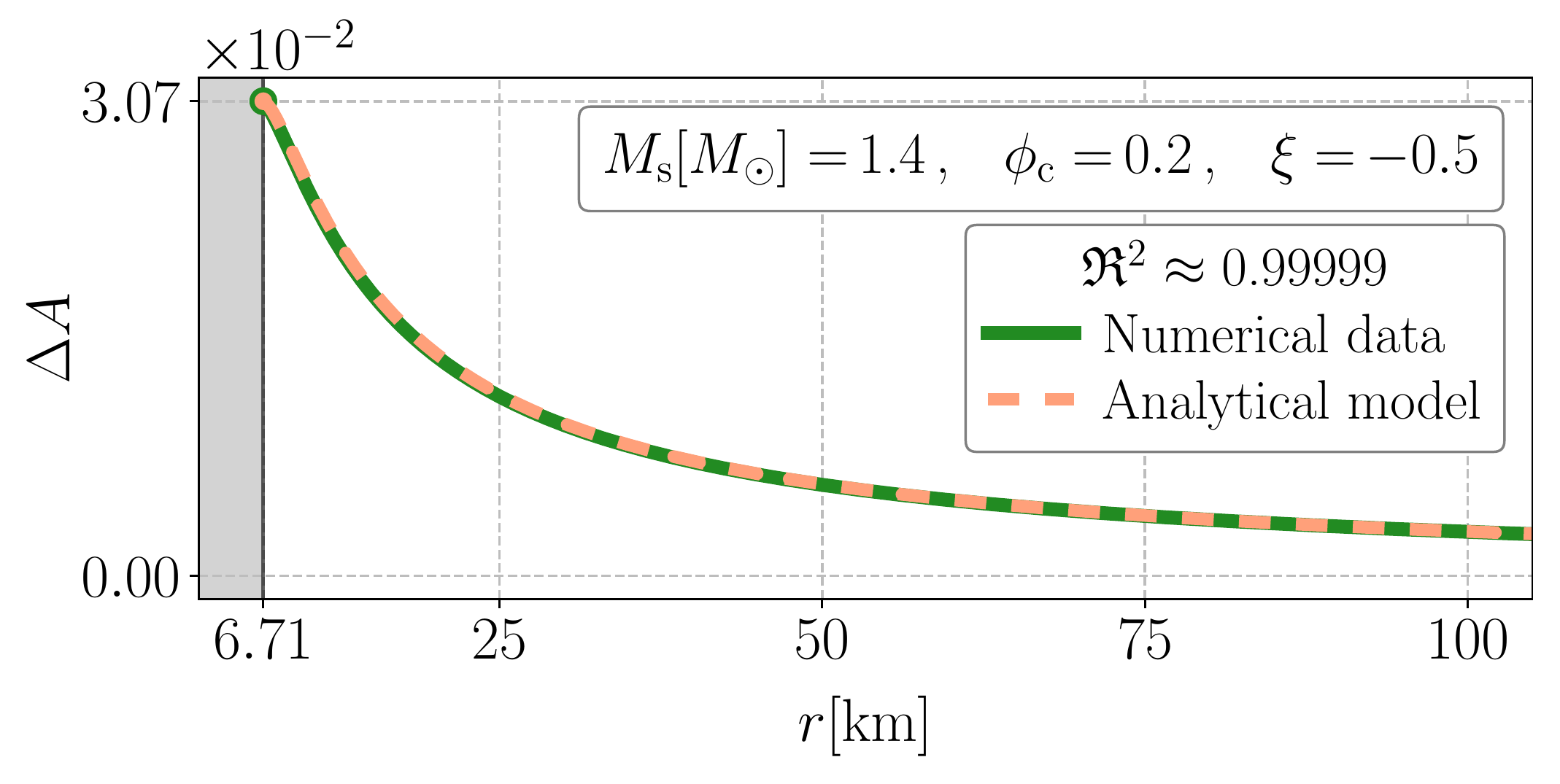}
	\end{tabular}
	\begin{tabular}{@{}c@{}}
		\includegraphics[width=.43\linewidth]{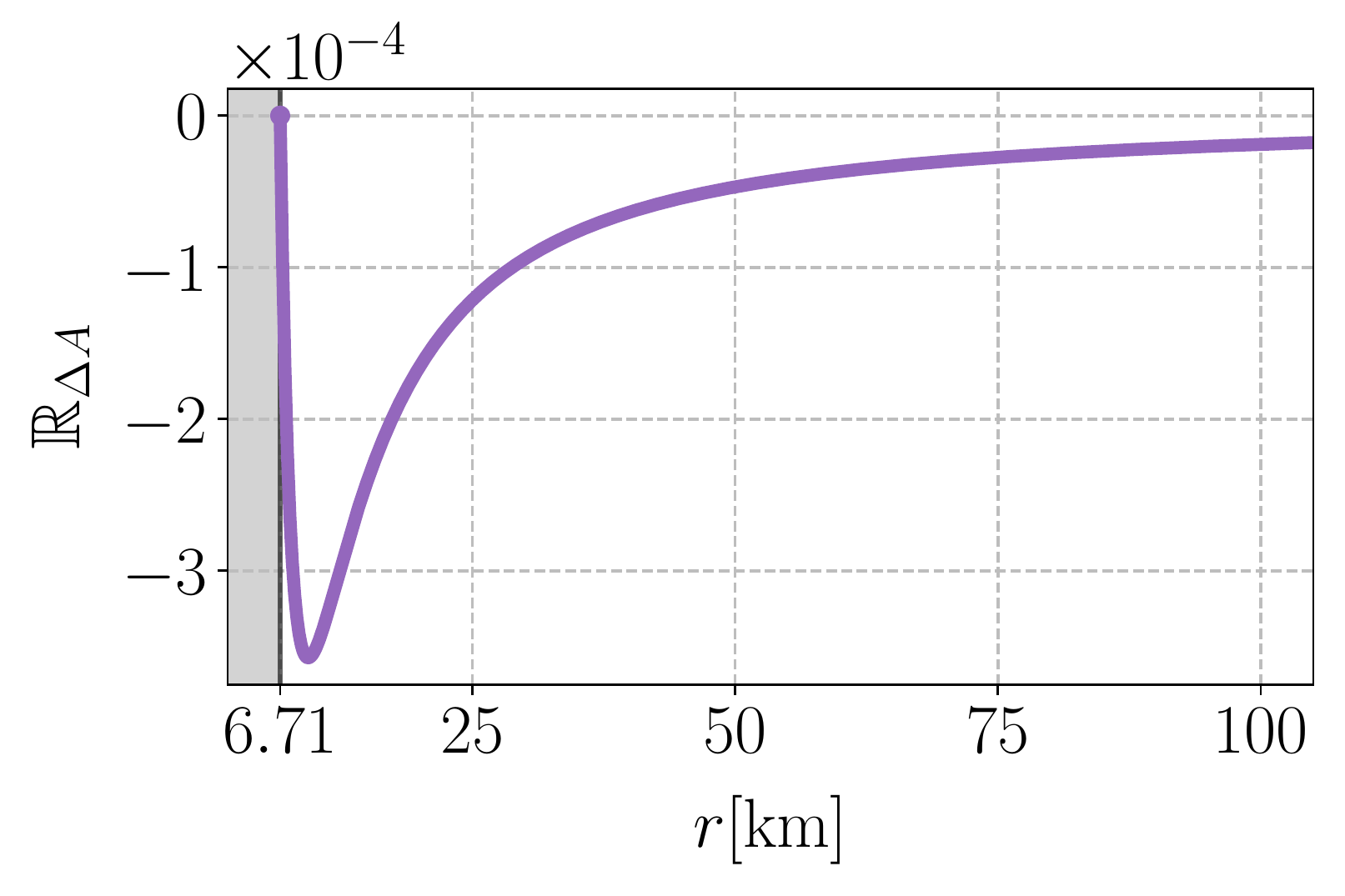}
	\end{tabular}
	
	\vspace{-2mm}
	
	\caption{Comparison of the numerical data and the analytical model given in Eq.\ \autoref{eq:DeltaA_fit} for $\Delta A(r)$. (For details see the caption of Fig.\ \autoref{fig:M_ext_fits_and_residuals}.)}
	\label{fig:DeltaA_fits_and_residuals}
\end{figure*}

\paragraph*{Temporal part ($\Delta A$) :}
Regarding to the metric function deviation of the temporal part, we need to emphasize that the function $A(r)$ has no effect on the dynamics of the system, but it is the one who is affected by the others. This can be clearly seen from the equation set \autoref{eq:components_eq} [$f(r)$ corresponds to $A(r)$] since it can be eliminated from the equations completely in favor of the scalar field and the other metric function. Nevertheless, we have found that the following form fits well with the data
\begin{equation}
    \Delta A(r) = \Delta A_{\rm{s}} \sfrac{R}{r} + \bt \sfrac{M_{\rm{ext}}(r)}{r} + \Rsdl{\Delta A}
\label{eq:DeltaA_fit}
\end{equation}
where $\bt$ is a constant and $\Delta A_{\rm{s}}$ is the value of the metric function $\Delta A(r)$ calculated at the surface of the star. Our investigations have shown that the value of $\bt$ actually changes in a range $5\le \bt \le 6$ and the outcomes shown in Fig.\ \autoref{fig:DeltaA_fits_and_residuals} along with the residuals, $\Rsdl{\Delta A}$, are generated simply by using the average value $\bt = 5.5\,$ without introducing large errors in the fit. Although it may seem a little strange to use the surface value of the deviation, namely $\Delta A_{\rm{s}}$, in the analytical model, we need to remind that it is necessary to find an initial value for $A(r)$ as explained in Sec.\ \autoref{sec:setup} for the numerical solution as well.

\subsection{The Scalar Field}

\begin{figure*}[!t]
	\centering

	\begin{tabular}{@{}c@{}}\hspace{-3mm}
		\includegraphics[width=.57\linewidth]{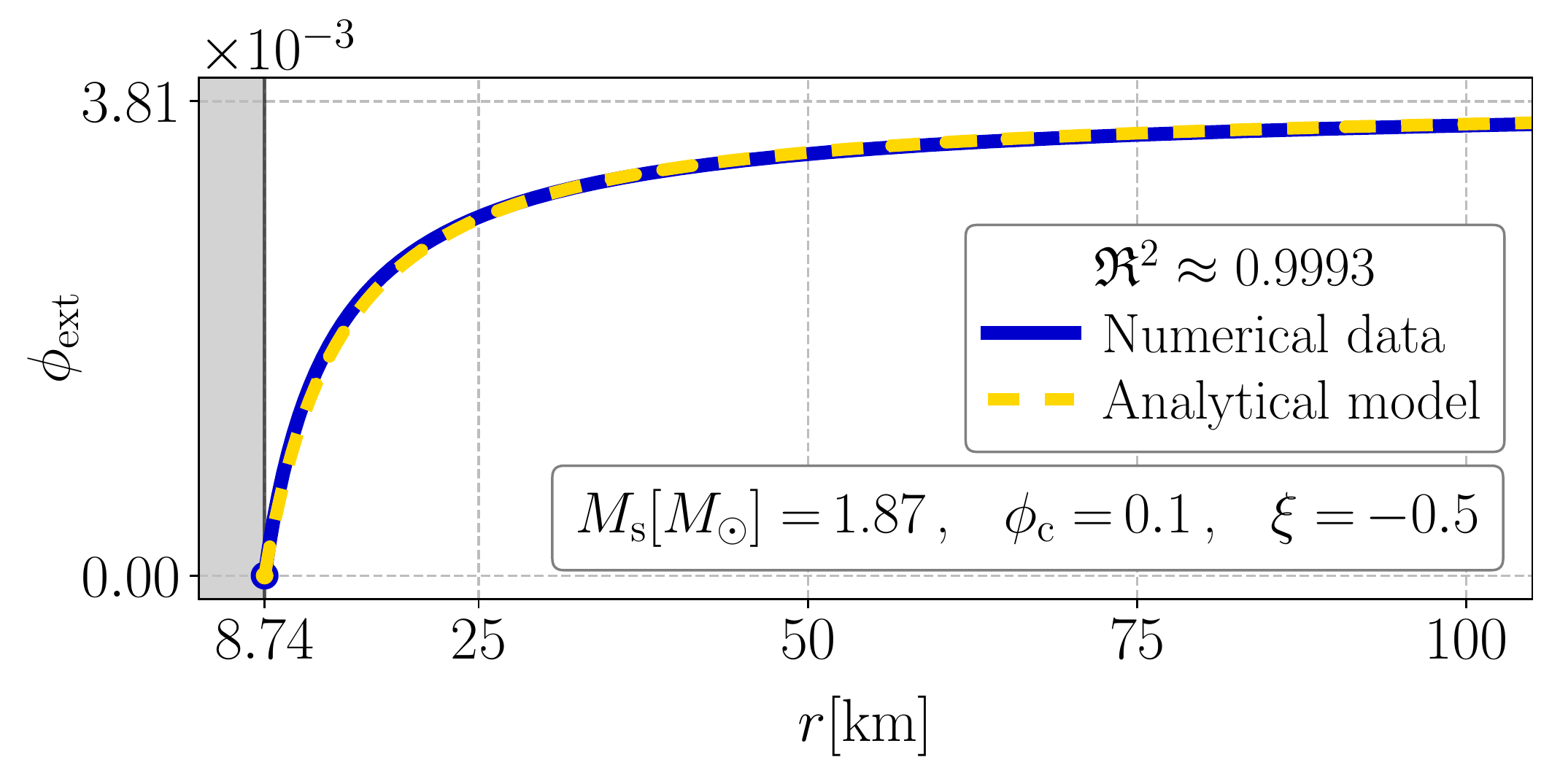}
	\end{tabular}
	\begin{tabular}{@{}c@{}}
		\includegraphics[width=.43\linewidth]{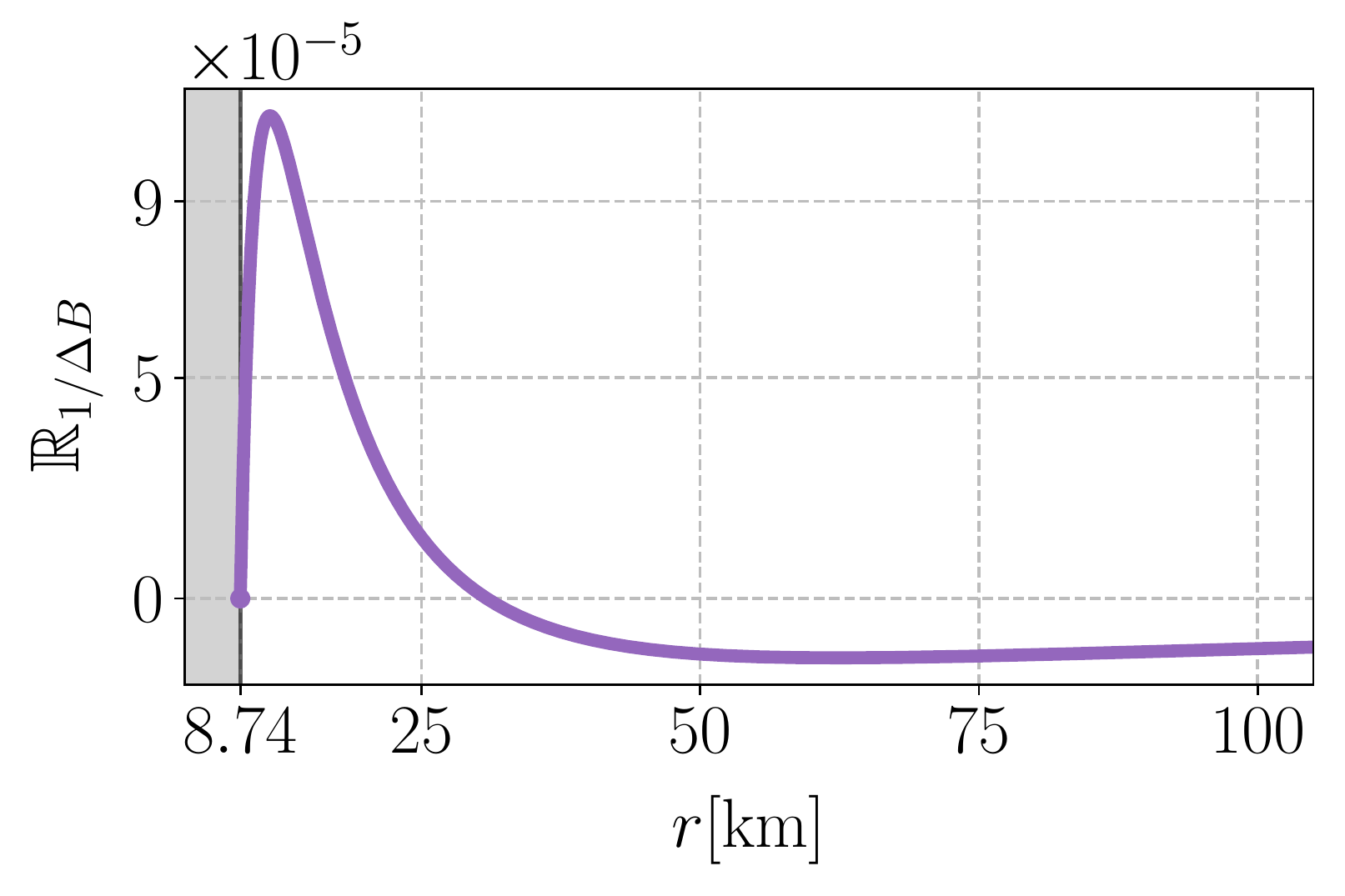}
	\end{tabular}
	
	\vspace{-2mm}
	
	\begin{tabular}{@{}c@{}}\hspace{-3mm}
		\includegraphics[width=.57\linewidth]{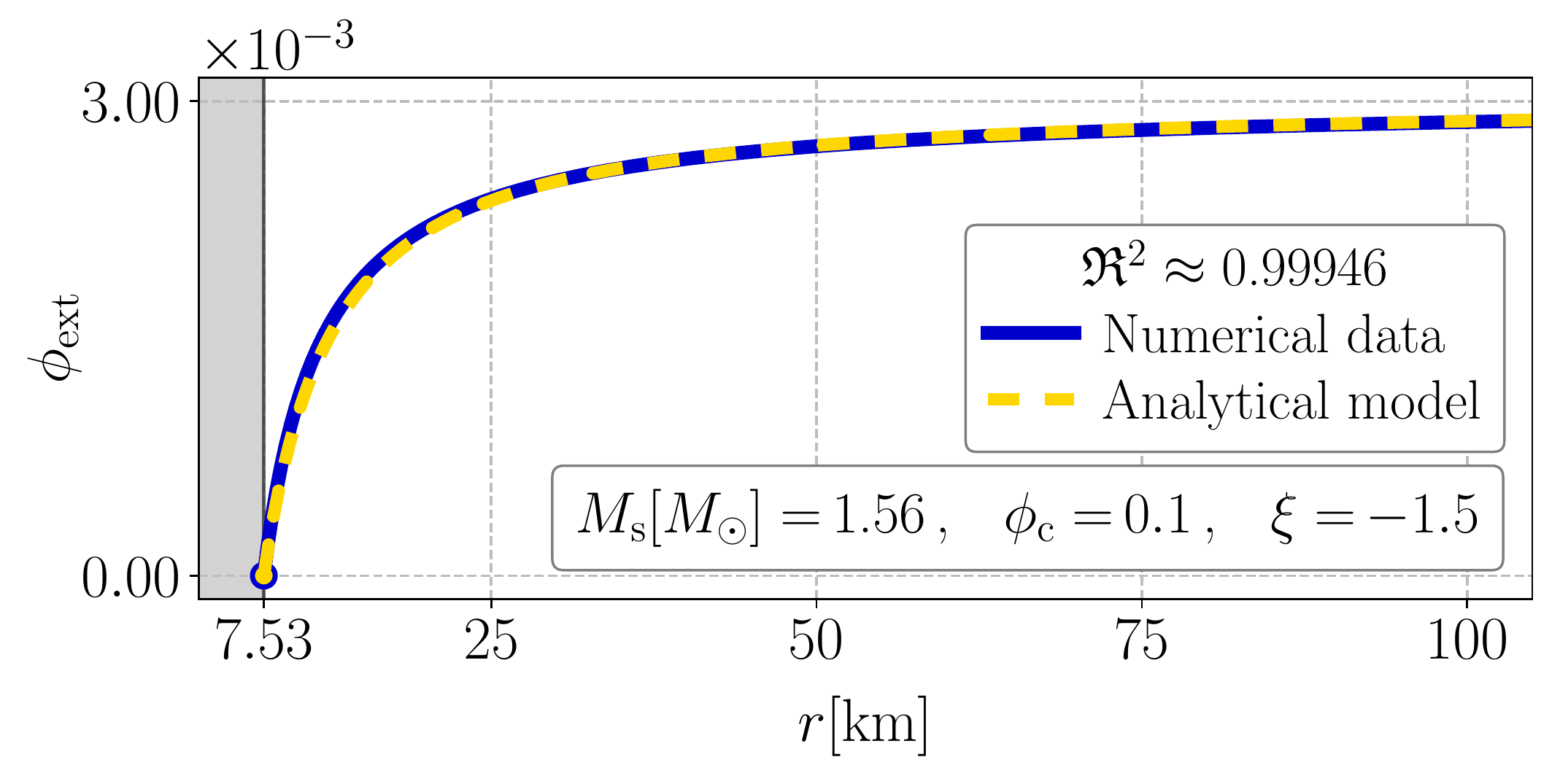}
	\end{tabular}
	\begin{tabular}{@{}c@{}}
		\includegraphics[width=.43\linewidth]{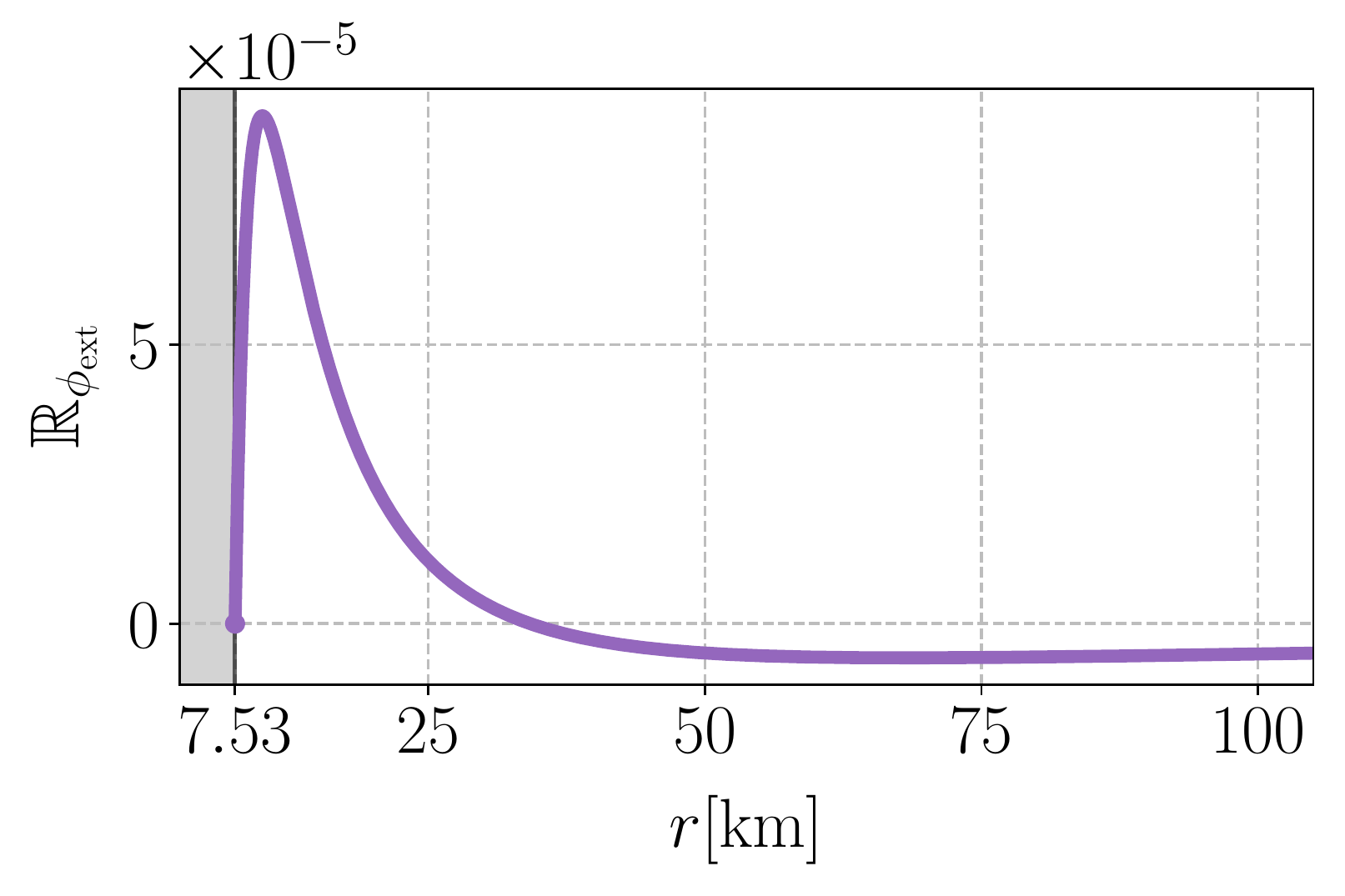}
	\end{tabular}
	
	\vspace{-2mm}
	
	\begin{tabular}{@{}c@{}}\hspace{-3mm}
		\includegraphics[width=.57\linewidth]{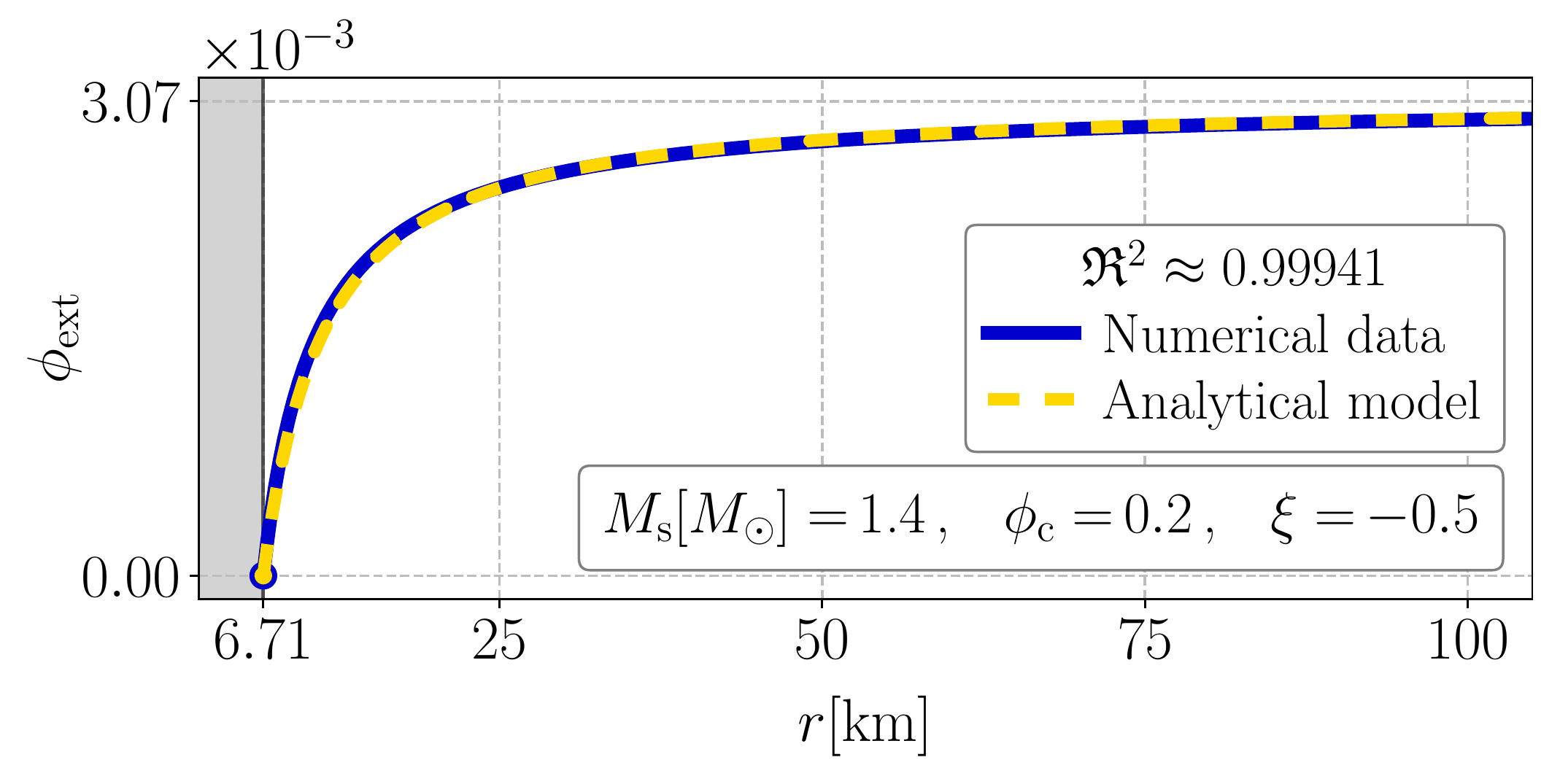}
	\end{tabular}
	\begin{tabular}{@{}c@{}}
		\includegraphics[width=.43\linewidth]{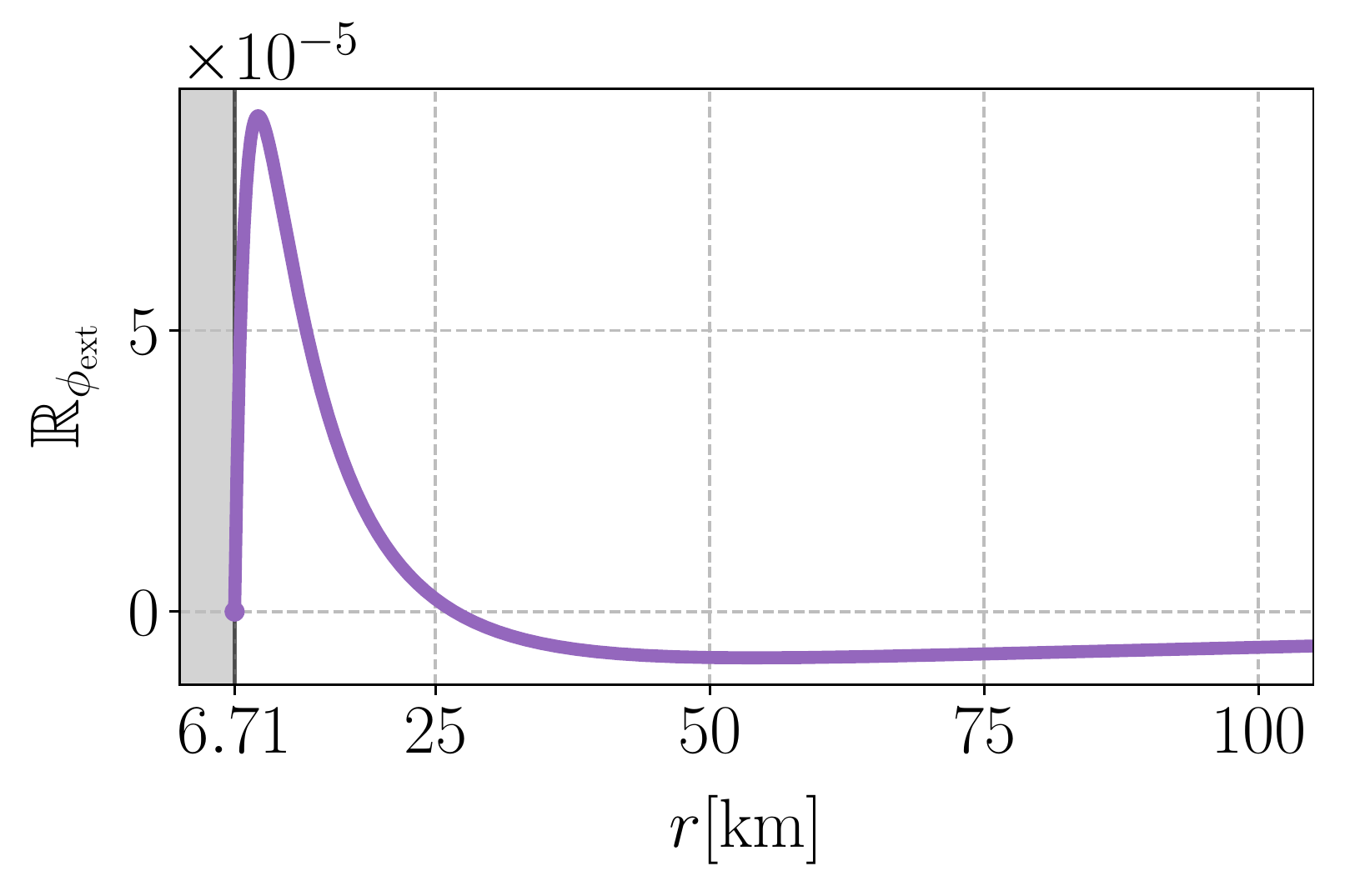}
	\end{tabular}
	
	\vspace{-2mm}
	
	\caption{Comparison of the numerical data and the analytical model given in Eq.\ \autoref{eq:phi_fit} for $\phi_{\rm{ext}}(r)$. (For details see the caption of Fig.\ \autoref{fig:M_ext_fits_and_residuals}.)}
	\label{fig:phi_fits_and_residuals}
\end{figure*}

In order to find an analytic form for the corrections to the metric functions we have used the exterior part of the mass function, $M_{\rm{ext}}$, which has turned out to be proportional to the exterior part of the scalar field, $\phi_{\rm{ext}}$. Therefore, to complete the definition, we need to give an analytic expression for the scalar field outside the star in terms of the radial coordinate, $r$, and the other parameters such as the mass, $M_{\rm{s}}$, and the radius, $R$, of the star. To this end, similar to what we have done in scaling for the mass function, we look for the solutions in the form of
\begin{equation}
    \phi_{\rm{ext}}(r) \approx (\phi_\infty - \phi_{\rm{s}}) \, \Phi(r)
\end{equation}
where we define the following \textit{ansatz}
\begin{equation}
    \Phi(r) \equiv a_1 \left( 1 - \sfrac{R}{r} \right) + a_2 \left( 1 - \sfrac{R^2}{r^2} \right) + a_3 \left( 1 - \sfrac{R^3}{r^3} \right) + \cdots
\label{eq:phi_ansatz}
\end{equation}
with constant coefficients $a_1$, $a_2$, $a_3$, ... and it satisfies the scalar field value at the surface $\phi_{\rm{ext}}(r=R)=0$ since $\Phi(r=R)=0$. On the other hand, $\Phi(r)$ is required to be equal to unity at infinity since we have $\phi_{\rm{total}}(r) = \phi_{\rm{s}} + \phi_{\rm{ext}}(r)$ and $\phi_{\rm{total}}(r \rightarrow \infty) = \phi_\infty$. Applying this condition for the first three terms we have
\begin{equation}
    a_1 = 1 - \left( a_2 + a_3 \right) \:.
\end{equation}
Since the form of $\Phi$ have been specifically chosen in accordance with the numerical data and, although it is not necessary, we seek a solution that the main contribution comes from the term with coefficient $a_1$ we constrain it as $a_1>0$ which, then, together with the above relation necessitates that $a_2+a_3 < 1$.

Our analysis shows that the values $a_2=2M_{\rm{s}}/R$ and $a_3=0$ give very compatible results with the data as shown in Fig.\ \autoref{fig:phi_fits_and_residuals} for different configurations. It seems that we do not need the cubic term at all since even without that factor we have an accuracy at the order around $\mathcal{O}(10^{-5})$ for all cases with different choices of the free parameters of the model, namely $\xi$ and $\phi_{\rm{c}}$. Then, using these values and rearranging the result we get
\begin{equation}
    \phi_{\rm{ext}}(r) = (\phi_\infty - \phi_{\rm{s}}) \left( 1 - \sfrac{R}{r} \right) \! \left( 1 + \sfrac{2M_{\rm{s}}}{r} \right) + \Rsdl{\phi_{\rm{ext}}} \:.
\label{eq:phi_fit}
\end{equation}
We need to mention that we have also worked on the higher order corrections in Eq.\ \autoref{eq:phi_ansatz}, but the result did not show any improvement compared to the one we have presented here.

\section{CONCLUSION} \label{sec:conclusion}
In this work we have examined the exterior solution of spherically symmetric and static configuration in scalar-tensor theories by fitting the data obtained from the numerical solutions of the differential equations. Our main goal was to determine analytical expressions for the exterior solution of the scalar field, the mass and the metric functions that are independent of the parameters of a particular model. To this end, we have taken the nonminimally coupled free scalar field as our model and calculated the mass and the metric functions in terms of the scalar field. Furthermore, in order to complete the definitions, we have introduced an analytical expression for the scalar field that only has a dependency on the mass and the radius of the configuration as its parameters.

We have found the expressions in Eqs.\ \autoref{eq:Mext_fit}, \autoref{eq:DeltaB_fit}, \autoref{eq:DeltaA_fit} and \autoref{eq:phi_fit} that agree with the numerical data of the external mass $M_{\rm{ext}}(r)$, the deviation of the metric functions $\Delta B(r)$, $\Delta A(r)$ and the external scalar field $\phi_{\rm{ext}}(r)$ respectively at around $\mathcal{O}(10^{-5})$ orders of magnitude accuracy. Furthermore, we went one step further to improve our results on the deviation $\Delta B(r)$ as in Eq.\ \autoref{eq:DeltaB_fit_00155} that reduces the discrepancy between the data and the model to $\mathcal{O}(10^{-8})$ orders of magnitude for which at points that are far outside the star it goes up to $\mathcal{O}(10^{-7})$; yet this will not affect the validity of the expression in the mainly interested region that is near outside the star.

As can be seen in the analytical models mentioned above, it seems that the nonminimal coupling constant, $\xi$, has no explicit effect on the external mass function and the deviation of the metric functions. Moreover, the expression for the external scalar field also does not contain any free parameters of the model. However, considering the accuracy we have obtained it is expected that its explicit presence, possibly with additional parameters that are dependent on the coupling parameter, will be necessary to find a more accurate model than ours, an example of which is presented for the deviation $\Delta B$ that shows $\mathcal{O}(10^{-8})$ orders of magnitude agreement with the data as a consequence of introduction of an additional parameter $\alpha$ in Eq.\ \autoref{eq:DeltaB_fit_00155}. Therefore, we have shown that the expressions written by including higher order terms and/or introducing additional free parameters, one can achieve higher orders of accuracy. Nevertheless in any case, our set of expressions fits the numerical data to an acceptable degree such that the analytical form of the exterior region characteristics can be clearly understood, while omitting the model parameter dependency. 

We have mentioned before that the central value of the scalar field can be restricted via its asymptotic value by using the observational constraints such as the PPN parameters or even the asymptotic value can be taken as zero in practice which certainly determines the central value of the scalar field and leaves the nonminimal coupling constant as the only free parameter of the model. However, even in this case, our results show that for the same central values of the scalar field the exterior solutions are sensitive to $\xi$ values only around $\mathcal{O}(10^{-5})$ in the vicinity of the star and even less in the farther distances. This indicates that it may not be possible to constrain the parameters of this model, and possibly the corresponding transformable models as well, at least up to $\mathcal{O}(10^{-5})$ order of accuracy.

\clearpage
\section*{Acknowledgement}
The authors thank Kazım Yavuz Ek\c{s}i in particular for his idea, comments and insightful suggestions.

\bibliographystyle{apsrev4-2}
\bibliography{references}

\begin{thebibliography}{48}%
\makeatletter
\providecommand \@ifxundefined [1]{%
 \@ifx{#1\undefined}
}%
\providecommand \@ifnum [1]{%
 \ifnum #1\expandafter \@firstoftwo
 \else \expandafter \@secondoftwo
 \fi
}%
\providecommand \@ifx [1]{%
 \ifx #1\expandafter \@firstoftwo
 \else \expandafter \@secondoftwo
 \fi
}%
\providecommand \natexlab [1]{#1}%
\providecommand \enquote  [1]{``#1''}%
\providecommand \bibnamefont  [1]{#1}%
\providecommand \bibfnamefont [1]{#1}%
\providecommand \citenamefont [1]{#1}%
\providecommand \href@noop [0]{\@secondoftwo}%
\providecommand \href [0]{\begingroup \@sanitize@url \@href}%
\providecommand \@href[1]{\@@startlink{#1}\@@href}%
\providecommand \@@href[1]{\endgroup#1\@@endlink}%
\providecommand \@sanitize@url [0]{\catcode `\\12\catcode `\$12\catcode
  `\&12\catcode `\#12\catcode `\^12\catcode `\_12\catcode `\%12\relax}%
\providecommand \@@startlink[1]{}%
\providecommand \@@endlink[0]{}%
\providecommand \url  [0]{\begingroup\@sanitize@url \@url }%
\providecommand \@url [1]{\endgroup\@href {#1}{\urlprefix }}%
\providecommand \urlprefix  [0]{URL }%
\providecommand \Eprint [0]{\href }%
\providecommand \doibase [0]{https://doi.org/}%
\providecommand \selectlanguage [0]{\@gobble}%
\providecommand \bibinfo  [0]{\@secondoftwo}%
\providecommand \bibfield  [0]{\@secondoftwo}%
\providecommand \translation [1]{[#1]}%
\providecommand \BibitemOpen [0]{}%
\providecommand \bibitemStop [0]{}%
\providecommand \bibitemNoStop [0]{.\EOS\space}%
\providecommand \EOS [0]{\spacefactor3000\relax}%
\providecommand \BibitemShut  [1]{\csname bibitem#1\endcsname}%
\let\auto@bib@innerbib\@empty
\bibitem [{\citenamefont {Psaltis}(2008)}]{psaltis-rw-2008}%
  \BibitemOpen
  \bibfield  {author} {\bibinfo {author} {\bibfnamefont {D.}~\bibnamefont
  {Psaltis}},\ }\href {https://doi.org/10.12942/lrr-2008-9} {\bibfield
  {journal} {\bibinfo  {journal} {Living Reviews in Relativity}\ }\textbf
  {\bibinfo {volume} {11}},\ \bibinfo {pages} {9} (\bibinfo {year} {2008})},\
  \Eprint {https://arxiv.org/abs/0806.1531} {\color{red}arXiv:0806.1531}
  \BibitemShut {NoStop}%
\bibitem [{\citenamefont {Schwarzschild}(1916)}]{Schwarzschild1916}%
  \BibitemOpen
  \bibfield  {author} {\bibinfo {author} {\bibfnamefont {K.}~\bibnamefont
  {Schwarzschild}},\ }\href@noop {} {\bibfield  {journal} {\bibinfo  {journal}
  {Sitzungsber. Preuss. Akad. Wiss. Berlin (Math. Phys.)}\ }\textbf {\bibinfo
  {volume} {1916}},\ \bibinfo {pages} {189} (\bibinfo {year} {1916})},\ \Eprint
  {https://arxiv.org/abs/physics/9905030} {\color{red}arXiv:physics/9905030}
  \BibitemShut {NoStop}%
\bibitem [{\citenamefont {McGruder}\ and\ \citenamefont
  {VanDerMeer}(2018)}]{Droste1916}%
  \BibitemOpen
  \bibfield  {author} {\bibinfo {author} {\bibfnamefont {C.~H.}\ \bibnamefont
  {McGruder}}\ and\ \bibinfo {author} {\bibfnamefont {B.~W.}\ \bibnamefont
  {VanDerMeer}},\ }\href@noop {} {\  (\bibinfo {year} {2018})},\ \Eprint
  {https://arxiv.org/abs/1801.07592} {\color{red}arXiv:1801.07592} \BibitemShut
  {NoStop}%
\bibitem [{\citenamefont {Misner}\ \emph {et~al.}(1973)\citenamefont {Misner},
  \citenamefont {Thorne},\ and\ \citenamefont {Wheeler}}]{Misner1973}%
  \BibitemOpen
  \bibfield  {author} {\bibinfo {author} {\bibfnamefont {C.~W.}\ \bibnamefont
  {Misner}}, \bibinfo {author} {\bibfnamefont {K.~S.}\ \bibnamefont {Thorne}},\
  and\ \bibinfo {author} {\bibfnamefont {J.~A.}\ \bibnamefont {Wheeler}},\
  }\href@noop {} {\emph {\bibinfo {title} {{Gravitation}}}}\ (\bibinfo
  {publisher} {W. H. Freeman},\ \bibinfo {address} {San Francisco},\ \bibinfo
  {year} {1973})\BibitemShut {NoStop}%
\bibitem [{\citenamefont {Landau}\ and\ \citenamefont
  {Lifschits}(1975)}]{Landau1975}%
  \BibitemOpen
  \bibfield  {author} {\bibinfo {author} {\bibfnamefont {L.~D.}\ \bibnamefont
  {Landau}}\ and\ \bibinfo {author} {\bibfnamefont {E.~M.}\ \bibnamefont
  {Lifschits}},\ }\href@noop {} {\emph {\bibinfo {title} {{The Classical Theory
  of Fields}}}},\ \bibinfo {series} {Course of Theoretical Physics}, Vol.\
  \bibinfo {volume} {Volume 2}\ (\bibinfo  {publisher} {Pergamon Press},\
  \bibinfo {address} {Oxford},\ \bibinfo {year} {1975})\BibitemShut {NoStop}%
\bibitem [{\citenamefont {{Zaglauer}}(1992)}]{zag92}%
  \BibitemOpen
  \bibfield  {author} {\bibinfo {author} {\bibfnamefont {H.~W.}\ \bibnamefont
  {{Zaglauer}}},\ }\href {https://doi.org/10.1086/171537} {\bibfield  {journal}
  {\bibinfo  {journal} {\apj}\ }\textbf {\bibinfo {volume} {393}},\ \bibinfo
  {pages} {685} (\bibinfo {year} {1992})}\BibitemShut {NoStop}%
\bibitem [{\citenamefont {{Harada}}(1998)}]{har98}%
  \BibitemOpen
  \bibfield  {author} {\bibinfo {author} {\bibfnamefont {T.}~\bibnamefont
  {{Harada}}},\ }\href {https://doi.org/10.1103/PhysRevD.57.4802} {\bibfield
  {journal} {\bibinfo  {journal} {\prd}\ }\textbf {\bibinfo {volume} {57}},\
  \bibinfo {pages} {4802} (\bibinfo {year} {1998})},\ \Eprint
  {https://arxiv.org/abs/gr-qc/9801049} {\color{red}arXiv:gr-qc/9801049}
  \BibitemShut {NoStop}%
\bibitem [{\citenamefont {Salgado}\ \emph {et~al.}(1998)\citenamefont
  {Salgado}, \citenamefont {Sudarsky},\ and\ \citenamefont
  {Nucamendi}}]{salgado1998}%
  \BibitemOpen
  \bibfield  {author} {\bibinfo {author} {\bibfnamefont {M.}~\bibnamefont
  {Salgado}}, \bibinfo {author} {\bibfnamefont {D.}~\bibnamefont {Sudarsky}},\
  and\ \bibinfo {author} {\bibfnamefont {U.}~\bibnamefont {Nucamendi}},\ }\href
  {https://doi.org/10.1103/PhysRevD.58.124003} {\bibfield  {journal} {\bibinfo
  {journal} {Phys. Rev. D}\ }\textbf {\bibinfo {volume} {58}},\ \bibinfo
  {pages} {124003} (\bibinfo {year} {1998})},\ \Eprint
  {https://arxiv.org/abs/gr-qc/9806070} {\color{red}arXiv:gr-qc/9806070}
  \BibitemShut {NoStop}%
\bibitem [{\citenamefont {{Horbatsch}}\ and\ \citenamefont
  {{Burgess}}(2011)}]{hor11}%
  \BibitemOpen
  \bibfield  {author} {\bibinfo {author} {\bibfnamefont {M.~W.}\ \bibnamefont
  {{Horbatsch}}}\ and\ \bibinfo {author} {\bibfnamefont {C.~P.}\ \bibnamefont
  {{Burgess}}},\ }\href {https://doi.org/10.1088/1475-7516/2011/08/027}
  {\bibfield  {journal} {\bibinfo  {journal} {JCAP}\ }\textbf {\bibinfo
  {volume} {8}},\ \bibinfo {eid} {027}},\ \Eprint
  {https://arxiv.org/abs/1006.4411} {\color{red}arXiv:1006.4411} \BibitemShut
  {NoStop}%
\bibitem [{\citenamefont {Sultana}\ \emph {et~al.}(2014)\citenamefont
  {Sultana}, \citenamefont {Bose},\ and\ \citenamefont
  {Kazanas}}]{kazanas2014}%
  \BibitemOpen
  \bibfield  {author} {\bibinfo {author} {\bibfnamefont {J.}~\bibnamefont
  {Sultana}}, \bibinfo {author} {\bibfnamefont {B.}~\bibnamefont {Bose}},\ and\
  \bibinfo {author} {\bibfnamefont {D.}~\bibnamefont {Kazanas}},\ }\href
  {https://doi.org/10.1142/S0218271814500904} {\bibfield  {journal} {\bibinfo
  {journal} {International Journal of Modern Physics D}\ }\textbf {\bibinfo
  {volume} {23}},\ \bibinfo {pages} {1450090} (\bibinfo {year}
  {2014})}\BibitemShut {NoStop}%
\bibitem [{\citenamefont {F\"uzfa}\ \emph {et~al.}(2013)\citenamefont {F\"uzfa}
  \emph {et~al.}}]{fuzfa1}%
  \BibitemOpen
  \bibfield  {author} {\bibinfo {author} {\bibfnamefont {A.}~\bibnamefont
  {F\"uzfa}} \emph {et~al.},\ }\href
  {https://doi.org/10.1103/PhysRevLett.111.121103} {\bibfield  {journal}
  {\bibinfo  {journal} {Phys. Rev. Lett.}\ }\textbf {\bibinfo {volume} {111}},\
  \bibinfo {pages} {121103} (\bibinfo {year} {2013})},\ \Eprint
  {https://arxiv.org/abs/1305.2640} {\color{red}arXiv:1305.2640} \BibitemShut
  {NoStop}%
\bibitem [{\citenamefont {{Sotani}}\ and\ \citenamefont
  {{Kokkotas}}(2018)}]{sot18}%
  \BibitemOpen
  \bibfield  {author} {\bibinfo {author} {\bibfnamefont {H.}~\bibnamefont
  {{Sotani}}}\ and\ \bibinfo {author} {\bibfnamefont {K.~D.}\ \bibnamefont
  {{Kokkotas}}},\ }\href {https://doi.org/10.1103/PhysRevD.97.124034}
  {\bibfield  {journal} {\bibinfo  {journal} {\prd}\ }\textbf {\bibinfo
  {volume} {97}},\ \bibinfo {eid} {124034} (\bibinfo {year} {2018})},\ \Eprint
  {https://arxiv.org/abs/1806.00568} {\color{red}arXiv:1806.00568} \BibitemShut
  {NoStop}%
\bibitem [{\citenamefont {Olmo}\ \emph {et~al.}(2020)\citenamefont {Olmo},
  \citenamefont {Rubiera-Garcia},\ and\ \citenamefont {Wojnar}}]{Olmo2019}%
  \BibitemOpen
  \bibfield  {author} {\bibinfo {author} {\bibfnamefont {G.~J.}\ \bibnamefont
  {Olmo}}, \bibinfo {author} {\bibfnamefont {D.}~\bibnamefont
  {Rubiera-Garcia}},\ and\ \bibinfo {author} {\bibfnamefont {A.}~\bibnamefont
  {Wojnar}},\ }\href {https://doi.org/10.1016/j.physrep.2020.07.001} {\bibfield
   {journal} {\bibinfo  {journal} {Phys. Rept.}\ }\textbf {\bibinfo {volume}
  {876}},\ \bibinfo {pages} {1} (\bibinfo {year} {2020})},\ \Eprint
  {https://arxiv.org/abs/1912.05202} {\color{red}arXiv:1912.05202} \BibitemShut
  {NoStop}%
\bibitem [{\citenamefont {S.~Arapo\u{g}lu}\ \emph {et~al.}(2019)\citenamefont
  {S.~Arapo\u{g}lu}, \citenamefont {Y.~Ek\c{s}i},\ and\ \citenamefont
  {E.~Y\"ukselci}}]{Arapoglu2019}%
  \BibitemOpen
  \bibfield  {author} {\bibinfo {author} {\bibfnamefont {A.}~\bibnamefont
  {S.~Arapo\u{g}lu}}, \bibinfo {author} {\bibfnamefont {K.}~\bibnamefont
  {Y.~Ek\c{s}i}},\ and\ \bibinfo {author} {\bibfnamefont {A.}~\bibnamefont
  {E.~Y\"ukselci}},\ }\href {https://doi.org/10.1103/PhysRevD.99.064055}
  {\bibfield  {journal} {\bibinfo  {journal} {Phys. Rev. D}\ }\textbf {\bibinfo
  {volume} {99}},\ \bibinfo {pages} {064055} (\bibinfo {year} {2019})},\
  \Eprint {https://arxiv.org/abs/1903.00391} {\color{red}arXiv:1903.00391}
  \BibitemShut {NoStop}%
\bibitem [{\citenamefont {Arapo\u{g}lu}\ and\ \citenamefont
  {Emrah~Y\"ukselci}(2020)}]{Arapoglu2020}%
  \BibitemOpen
  \bibfield  {author} {\bibinfo {author} {\bibfnamefont {A.~S.}\ \bibnamefont
  {Arapo\u{g}lu}}\ and\ \bibinfo {author} {\bibfnamefont {A.}~\bibnamefont
  {Emrah~Y\"ukselci}},\ }\href {https://doi.org/10.1016/j.dark.2020.100488}
  {\bibfield  {journal} {\bibinfo  {journal} {Phys. Dark Univ.}\ }\textbf
  {\bibinfo {volume} {28}},\ \bibinfo {pages} {100488} (\bibinfo {year}
  {2020})},\ \Eprint {https://arxiv.org/abs/2002.03025}
  {\color{red}arXiv:2002.03025} \BibitemShut {NoStop}%
\bibitem [{\citenamefont {Damour}\ and\ \citenamefont
  {Esposito-Farese}(1992)}]{Damour:1992}%
  \BibitemOpen
  \bibfield  {author} {\bibinfo {author} {\bibfnamefont {T.}~\bibnamefont
  {Damour}}\ and\ \bibinfo {author} {\bibfnamefont {G.}~\bibnamefont
  {Esposito-Farese}},\ }\href {https://doi.org/10.1088/0264-9381/9/9/015}
  {\bibfield  {journal} {\bibinfo  {journal} {Class. Quant. Grav.}\ }\textbf
  {\bibinfo {volume} {9}},\ \bibinfo {pages} {2093} (\bibinfo {year}
  {1992})}\BibitemShut {NoStop}%
\bibitem [{\citenamefont {Silva}\ and\ \citenamefont
  {Yunes}(2019)}]{yunes-silva2018}%
  \BibitemOpen
  \bibfield  {author} {\bibinfo {author} {\bibfnamefont {H.~O.}\ \bibnamefont
  {Silva}}\ and\ \bibinfo {author} {\bibfnamefont {N.}~\bibnamefont {Yunes}},\
  }\href {https://doi.org/10.1103/PhysRevD.99.044034} {\bibfield  {journal}
  {\bibinfo  {journal} {Phys. Rev.}\ }\textbf {\bibinfo {volume} {D99}},\
  \bibinfo {pages} {044034} (\bibinfo {year} {2019})},\ \Eprint
  {https://arxiv.org/abs/1808.04391} {\color{red}arXiv:1808.04391} \BibitemShut
  {NoStop}%
\bibitem [{\citenamefont {Yazadjiev}(2011)}]{Yazadjiev2011}%
  \BibitemOpen
  \bibfield  {author} {\bibinfo {author} {\bibfnamefont {S.~S.}\ \bibnamefont
  {Yazadjiev}},\ }\href {https://doi.org/10.1103/PhysRevD.83.127501} {\bibfield
   {journal} {\bibinfo  {journal} {Phys. Rev. D}\ }\textbf {\bibinfo {volume}
  {83}},\ \bibinfo {pages} {127501} (\bibinfo {year} {2011})},\ \Eprint
  {https://arxiv.org/abs/1104.1865} {\color{red}arXiv:1104.1865} \BibitemShut
  {NoStop}%
\bibitem [{\citenamefont {Saffer}\ \emph {et~al.}(2019)\citenamefont {Saffer},
  \citenamefont {Silva},\ and\ \citenamefont {Yunes}}]{Saffer2019}%
  \BibitemOpen
  \bibfield  {author} {\bibinfo {author} {\bibfnamefont {A.}~\bibnamefont
  {Saffer}}, \bibinfo {author} {\bibfnamefont {H.~O.}\ \bibnamefont {Silva}},\
  and\ \bibinfo {author} {\bibfnamefont {N.}~\bibnamefont {Yunes}},\ }\href
  {https://doi.org/10.1103/PhysRevD.100.044030} {\bibfield  {journal} {\bibinfo
   {journal} {Phys. Rev. D}\ }\textbf {\bibinfo {volume} {100}},\ \bibinfo
  {pages} {044030} (\bibinfo {year} {2019})},\ \Eprint
  {https://arxiv.org/abs/1903.07779} {\color{red}arXiv:1903.07779} \BibitemShut
  {NoStop}%
\bibitem [{\citenamefont {Tolman}(1934)}]{Tolman1934}%
  \BibitemOpen
  \bibfield  {author} {\bibinfo {author} {\bibfnamefont {R.~C.}\ \bibnamefont
  {Tolman}},\ }\href {https://doi.org/10.1073/pnas.20.3.169} {\bibfield
  {journal} {\bibinfo  {journal} {Proceedings of the National Academy of
  Sciences}\ }\textbf {\bibinfo {volume} {20}},\ \bibinfo {pages} {169}
  (\bibinfo {year} {1934})}\BibitemShut {NoStop}%
\bibitem [{\citenamefont {Tolman}(1939)}]{Tolman1939}%
  \BibitemOpen
  \bibfield  {author} {\bibinfo {author} {\bibfnamefont {R.~C.}\ \bibnamefont
  {Tolman}},\ }\href {https://doi.org/10.1103/PhysRev.55.364} {\bibfield
  {journal} {\bibinfo  {journal} {Phys. Rev.}\ }\textbf {\bibinfo {volume}
  {55}},\ \bibinfo {pages} {364} (\bibinfo {year} {1939})}\BibitemShut
  {NoStop}%
\bibitem [{\citenamefont {Oppenheimer}\ and\ \citenamefont
  {Volkoff}(1939)}]{Oppenheimer1939}%
  \BibitemOpen
  \bibfield  {author} {\bibinfo {author} {\bibfnamefont {J.~R.}\ \bibnamefont
  {Oppenheimer}}\ and\ \bibinfo {author} {\bibfnamefont {G.~M.}\ \bibnamefont
  {Volkoff}},\ }\href {https://doi.org/10.1103/PhysRev.55.374} {\bibfield
  {journal} {\bibinfo  {journal} {Phys. Rev.}\ }\textbf {\bibinfo {volume}
  {55}},\ \bibinfo {pages} {374} (\bibinfo {year} {1939})}\BibitemShut
  {NoStop}%
\bibitem [{\citenamefont {Lattimer}(2012)}]{EoS1}%
  \BibitemOpen
  \bibfield  {author} {\bibinfo {author} {\bibfnamefont {J.~M.}\ \bibnamefont
  {Lattimer}},\ }\href {https://doi.org/10.1146/annurev-nucl-102711-095018}
  {\bibfield  {journal} {\bibinfo  {journal} {Annual Review of Nuclear and
  Particle Science}\ }\textbf {\bibinfo {volume} {62}},\ \bibinfo {pages} {485}
  (\bibinfo {year} {2012})},\ \Eprint {https://arxiv.org/abs/1305.3510}
  {\color{red}arXiv:1305.3510} \BibitemShut {NoStop}%
\bibitem [{\citenamefont {Baym}\ and\ \citenamefont {the
  others}(2018)\citenamefont {Baym} \emph {et~al.}}]{EoS2}%
  \BibitemOpen
  \bibfield  {author} {\bibinfo {author} {\bibfnamefont {G.}~\bibnamefont
  {Baym}} \emph {et~al.},\ }\href {https://doi.org/10.1088/1361-6633/aaae14}
  {\bibfield  {journal} {\bibinfo  {journal} {Reports on Progress in Physics}\
  }\textbf {\bibinfo {volume} {81}},\ \bibinfo {eid} {056902} (\bibinfo {year}
  {2018})},\ \Eprint {https://arxiv.org/abs/1707.04966}
  {\color{red}arXiv:1707.04966} \BibitemShut {NoStop}%
\bibitem [{\citenamefont {Oertel}\ and\ \citenamefont {the
  others}(2017)\citenamefont {Oertel} \emph {et~al.}}]{EoS3}%
  \BibitemOpen
  \bibfield  {author} {\bibinfo {author} {\bibfnamefont {M.}~\bibnamefont
  {Oertel}} \emph {et~al.},\ }\href
  {https://doi.org/10.1103/RevModPhys.89.015007} {\bibfield  {journal}
  {\bibinfo  {journal} {Reviews of Modern Physics}\ }\textbf {\bibinfo {volume}
  {89}},\ \bibinfo {eid} {015007} (\bibinfo {year} {2017})},\ \Eprint
  {https://arxiv.org/abs/1610.03361} {\color{red}arXiv:1610.03361} \BibitemShut
  {NoStop}%
\bibitem [{\citenamefont {Horedt}(2004)}]{Horedt2004}%
  \BibitemOpen
  \bibfield  {author} {\bibinfo {author} {\bibfnamefont {G.}~\bibnamefont
  {Horedt}},\ }\bibinfo {title} {Polytropes - applications in astrophysics and
  related fields}\ (\bibinfo {year} {2004})\ pp.\ \bibinfo {pages}
  {1--724}\BibitemShut {NoStop}%
\bibitem [{\citenamefont {Potekhin}\ \emph {et~al.}(2013)\citenamefont
  {Potekhin}, \citenamefont {Fantina}, \citenamefont {Chamel}, \citenamefont
  {Pearson},\ and\ \citenamefont {Goriely}}]{Potekhin2013}%
  \BibitemOpen
  \bibfield  {author} {\bibinfo {author} {\bibfnamefont {A.~Y.}\ \bibnamefont
  {Potekhin}}, \bibinfo {author} {\bibfnamefont {A.~F.}\ \bibnamefont
  {Fantina}}, \bibinfo {author} {\bibfnamefont {N.}~\bibnamefont {Chamel}},
  \bibinfo {author} {\bibfnamefont {J.~M.}\ \bibnamefont {Pearson}},\ and\
  \bibinfo {author} {\bibfnamefont {S.}~\bibnamefont {Goriely}},\ }\href
  {https://doi.org/10.1051/0004-6361/201321697} {\bibfield  {journal} {\bibinfo
   {journal} {Astron. Astrophys.}\ }\textbf {\bibinfo {volume} {560}},\
  \bibinfo {pages} {A48} (\bibinfo {year} {2013})},\ \Eprint
  {https://arxiv.org/abs/1310.0049} {\color{red}arXiv:1310.0049} \BibitemShut
  {NoStop}%
\bibitem [{\citenamefont {{Ek{\c s}i}}\ \emph {et~al.}(2014)\citenamefont
  {{Ek{\c s}i}}, \citenamefont {{G{\"u}ng{\"o}r}},\ and\ \citenamefont
  {{T{\"u}rko{\v g}lu}}}]{degeneracy1}%
  \BibitemOpen
  \bibfield  {author} {\bibinfo {author} {\bibfnamefont {K.~Y.}\ \bibnamefont
  {{Ek{\c s}i}}}, \bibinfo {author} {\bibfnamefont {C.}~\bibnamefont
  {{G{\"u}ng{\"o}r}}},\ and\ \bibinfo {author} {\bibfnamefont {M.~M.}\
  \bibnamefont {{T{\"u}rko{\v g}lu}}},\ }\href
  {https://doi.org/10.1103/PhysRevD.89.063003} {\bibfield  {journal} {\bibinfo
  {journal} {\prd}\ }\textbf {\bibinfo {volume} {89}},\ \bibinfo {eid} {063003}
  (\bibinfo {year} {2014})},\ \Eprint {https://arxiv.org/abs/1402.0488}
  {\color{red}arXiv:1402.0488} \BibitemShut {NoStop}%
\bibitem [{\citenamefont {{DeDeo}}\ and\ \citenamefont
  {{Psaltis}}(2003)}]{degeneracy2}%
  \BibitemOpen
  \bibfield  {author} {\bibinfo {author} {\bibfnamefont {S.}~\bibnamefont
  {{DeDeo}}}\ and\ \bibinfo {author} {\bibfnamefont {D.}~\bibnamefont
  {{Psaltis}}},\ }\href {https://doi.org/10.1103/PhysRevLett.90.141101}
  {\bibfield  {journal} {\bibinfo  {journal} {Physical Review Letters}\
  }\textbf {\bibinfo {volume} {90}},\ \bibinfo {eid} {141101} (\bibinfo {year}
  {2003})},\ \Eprint {https://arxiv.org/abs/astro-ph/0302095}
  {\color{red}arXiv:astro-ph/0302095} \BibitemShut {NoStop}%
\bibitem [{\citenamefont {He}\ and\ \citenamefont {the
  others}(2015)\citenamefont {He} \emph {et~al.}}]{degeneracy3}%
  \BibitemOpen
  \bibfield  {author} {\bibinfo {author} {\bibfnamefont {X.-T.}\ \bibnamefont
  {He}} \emph {et~al.},\ }\href {https://doi.org/10.1103/PhysRevC.91.015810}
  {\bibfield  {journal} {\bibinfo  {journal} {\prc}\ }\textbf {\bibinfo
  {volume} {91}},\ \bibinfo {eid} {015810} (\bibinfo {year} {2015})},\ \Eprint
  {https://arxiv.org/abs/1408.0857} {\color{red}arXiv:1408.0857} \BibitemShut
  {NoStop}%
\bibitem [{\citenamefont {Doneva}\ and\ \citenamefont
  {Pappas}(2018)}]{dgnrcy4_Doneva:2017}%
  \BibitemOpen
  \bibfield  {author} {\bibinfo {author} {\bibfnamefont {D.~D.}\ \bibnamefont
  {Doneva}}\ and\ \bibinfo {author} {\bibfnamefont {G.}~\bibnamefont
  {Pappas}},\ }\href {https://doi.org/10.1007/978-3-319-97616-7_13} {\bibfield
  {journal} {\bibinfo  {journal} {Astrophys. Space Sci. Libr.}\ }\textbf
  {\bibinfo {volume} {457}},\ \bibinfo {pages} {737} (\bibinfo {year}
  {2018})},\ \Eprint {https://arxiv.org/abs/1709.08046}
  {\color{red}arXiv:1709.08046} \BibitemShut {NoStop}%
\bibitem [{\citenamefont {Shao}(2019)}]{dgnrcy5_Shao:2019}%
  \BibitemOpen
  \bibfield  {author} {\bibinfo {author} {\bibfnamefont {L.}~\bibnamefont
  {Shao}},\ }\href {https://doi.org/10.1063/1.5117806} {\bibfield  {journal}
  {\bibinfo  {journal} {AIP Conf. Proc.}\ }\textbf {\bibinfo {volume} {2127}},\
  \bibinfo {pages} {020016} (\bibinfo {year} {2019})},\ \Eprint
  {https://arxiv.org/abs/1901.07546} {\color{red}arXiv:1901.07546 [gr-qc]}
  \BibitemShut {NoStop}%
\bibitem [{\citenamefont {Chernikov}\ and\ \citenamefont
  {Tagirov}(1968)}]{chernikov1968}%
  \BibitemOpen
  \bibfield  {author} {\bibinfo {author} {\bibfnamefont {N.~A.}\ \bibnamefont
  {Chernikov}}\ and\ \bibinfo {author} {\bibfnamefont {E.~A.}\ \bibnamefont
  {Tagirov}},\ }\href {http://www.numdam.org/item/AIHPA_1968__9_2_109_0}
  {\bibfield  {journal} {\bibinfo  {journal} {Annales de l'I.H.P. Physique
  th\'eorique}\ }\textbf {\bibinfo {volume} {9}},\ \bibinfo {pages} {109}
  (\bibinfo {year} {1968})}\BibitemShut {NoStop}%
\bibitem [{\citenamefont {Callan}\ \emph {et~al.}(1970)\citenamefont {Callan},
  \citenamefont {Coleman},\ and\ \citenamefont {Jackiw}}]{callan-etal1970}%
  \BibitemOpen
  \bibfield  {author} {\bibinfo {author} {\bibfnamefont {C.~G.}\ \bibnamefont
  {Callan}}, \bibinfo {author} {\bibfnamefont {S.}~\bibnamefont {Coleman}},\
  and\ \bibinfo {author} {\bibfnamefont {R.}~\bibnamefont {Jackiw}},\ }\href
  {https://doi.org/10.1016/0003-4916(70)90394-5} {\bibfield  {journal}
  {\bibinfo  {journal} {Annals of Physics}\ }\textbf {\bibinfo {volume} {59}},\
  \bibinfo {pages} {42 } (\bibinfo {year} {1970})}\BibitemShut {NoStop}%
\bibitem [{\citenamefont {Birrell}\ and\ \citenamefont
  {Davies}(1980)}]{birrell-davies1980}%
  \BibitemOpen
  \bibfield  {author} {\bibinfo {author} {\bibfnamefont {N.~D.}\ \bibnamefont
  {Birrell}}\ and\ \bibinfo {author} {\bibfnamefont {P.~C.~W.}\ \bibnamefont
  {Davies}},\ }\href {https://doi.org/10.1103/PhysRevD.22.322} {\bibfield
  {journal} {\bibinfo  {journal} {Phys. Rev. D}\ }\textbf {\bibinfo {volume}
  {22}},\ \bibinfo {pages} {322} (\bibinfo {year} {1980})}\BibitemShut
  {NoStop}%
\bibitem [{\citenamefont {Birrell}\ and\ \citenamefont
  {Davies}(1982)}]{birrell-davies-1982}%
  \BibitemOpen
  \bibfield  {author} {\bibinfo {author} {\bibfnamefont {N.~D.}\ \bibnamefont
  {Birrell}}\ and\ \bibinfo {author} {\bibfnamefont {P.~C.~W.}\ \bibnamefont
  {Davies}},\ }\href {https://doi.org/10.1017/CBO9780511622632} {\emph
  {\bibinfo {title} {Quantum Fields in Curved Space}}}\ (\bibinfo  {publisher}
  {Cambridge University Press},\ \bibinfo {year} {1982})\BibitemShut {NoStop}%
\bibitem [{\citenamefont {{Douchin}}\ and\ \citenamefont
  {{Haensel}}(2001)}]{eos_sly}%
  \BibitemOpen
  \bibfield  {author} {\bibinfo {author} {\bibfnamefont {F.}~\bibnamefont
  {{Douchin}}}\ and\ \bibinfo {author} {\bibfnamefont {P.}~\bibnamefont
  {{Haensel}}},\ }\href {https://doi.org/10.1051/0004-6361:20011402} {\bibfield
   {journal} {\bibinfo  {journal} {\aap}\ }\textbf {\bibinfo {volume} {380}},\
  \bibinfo {pages} {151} (\bibinfo {year} {2001})},\ \Eprint
  {https://arxiv.org/abs/astro-ph/0111092} {\color{red}arXiv:astro-ph/0111092}
  \BibitemShut {NoStop}%
\bibitem [{\citenamefont {Arnowitt}\ \emph {et~al.}(1960)\citenamefont
  {Arnowitt}, \citenamefont {Deser},\ and\ \citenamefont {Misner}}]{adm_mass}%
  \BibitemOpen
  \bibfield  {author} {\bibinfo {author} {\bibfnamefont {R.}~\bibnamefont
  {Arnowitt}}, \bibinfo {author} {\bibfnamefont {S.}~\bibnamefont {Deser}},\
  and\ \bibinfo {author} {\bibfnamefont {C.~W.}\ \bibnamefont {Misner}},\
  }\href {https://doi.org/10.1103/PhysRev.118.1100} {\bibfield  {journal}
  {\bibinfo  {journal} {Phys. Rev.}\ }\textbf {\bibinfo {volume} {118}},\
  \bibinfo {pages} {1100} (\bibinfo {year} {1960})}\BibitemShut {NoStop}%
\bibitem [{\citenamefont {Komar}(1959)}]{komar_mass}%
  \BibitemOpen
  \bibfield  {author} {\bibinfo {author} {\bibfnamefont {A.}~\bibnamefont
  {Komar}},\ }\href {https://doi.org/10.1103/PhysRev.113.934} {\bibfield
  {journal} {\bibinfo  {journal} {Phys. Rev.}\ }\textbf {\bibinfo {volume}
  {113}},\ \bibinfo {pages} {934} (\bibinfo {year} {1959})}\BibitemShut
  {NoStop}%
\bibitem [{\citenamefont {Shibata}\ and\ \citenamefont
  {Kawaguchi}(2013)}]{mass_coincides}%
  \BibitemOpen
  \bibfield  {author} {\bibinfo {author} {\bibfnamefont {M.}~\bibnamefont
  {Shibata}}\ and\ \bibinfo {author} {\bibfnamefont {K.}~\bibnamefont
  {Kawaguchi}},\ }\href {https://doi.org/10.1103/PhysRevD.87.104031} {\bibfield
   {journal} {\bibinfo  {journal} {Phys. Rev. D}\ }\textbf {\bibinfo {volume}
  {87}},\ \bibinfo {pages} {104031} (\bibinfo {year} {2013})}\BibitemShut
  {NoStop}%
\bibitem [{\citenamefont {Bertotti}\ \emph {et~al.}(2003)\citenamefont
  {Bertotti}, \citenamefont {Iess},\ and\ \citenamefont {Tortora}}]{ppn_gamma}%
  \BibitemOpen
  \bibfield  {author} {\bibinfo {author} {\bibfnamefont {B.}~\bibnamefont
  {Bertotti}}, \bibinfo {author} {\bibfnamefont {L.}~\bibnamefont {Iess}},\
  and\ \bibinfo {author} {\bibfnamefont {P.}~\bibnamefont {Tortora}},\ }\href
  {https://doi.org/10.1038/nature01997} {\bibfield  {journal} {\bibinfo
  {journal} {Nature}\ }\textbf {\bibinfo {volume} {425}},\ \bibinfo {pages}
  {374} (\bibinfo {year} {2003})}\BibitemShut {NoStop}%
\bibitem [{\citenamefont {Dickey}\ \emph {et~al.}(1994)\citenamefont {Dickey}
  \emph {et~al.}}]{ppn_beta_1}%
  \BibitemOpen
  \bibfield  {author} {\bibinfo {author} {\bibfnamefont {J.~O.}\ \bibnamefont
  {Dickey}} \emph {et~al.},\ }\href
  {https://doi.org/10.1126/science.265.5171.482} {\bibfield  {journal}
  {\bibinfo  {journal} {Science}\ }\textbf {\bibinfo {volume} {265}},\ \bibinfo
  {pages} {482} (\bibinfo {year} {1994})}\BibitemShut {NoStop}%
\bibitem [{\citenamefont {Williams}\ \emph {et~al.}(1996)\citenamefont
  {Williams} \emph {et~al.}}]{ppn_beta_2}%
  \BibitemOpen
  \bibfield  {author} {\bibinfo {author} {\bibfnamefont {J.~G.}\ \bibnamefont
  {Williams}} \emph {et~al.},\ }\href
  {https://doi.org/10.1103/PhysRevD.53.6730} {\bibfield  {journal} {\bibinfo
  {journal} {Phys. Rev. D}\ }\textbf {\bibinfo {volume} {53}},\ \bibinfo
  {pages} {6730} (\bibinfo {year} {1996})}\BibitemShut {NoStop}%
\bibitem [{\citenamefont {Will}(2014)}]{ppn_review}%
  \BibitemOpen
  \bibfield  {author} {\bibinfo {author} {\bibfnamefont {C.~M.}\ \bibnamefont
  {Will}},\ }\href {https://doi.org/10.12942/lrr-2014-4} {\bibfield  {journal}
  {\bibinfo  {journal} {Living Reviews in Relativity}\ }\textbf {\bibinfo
  {volume} {17}},\ \bibinfo {pages} {4} (\bibinfo {year} {2014})},\ \Eprint
  {https://arxiv.org/abs/1403.7377} {\color{red}arXiv:1403.7377} \BibitemShut
  {NoStop}%
\bibitem [{\citenamefont {Freire}\ and\ \citenamefont {the
  others}(2012)\citenamefont {Freire} \emph {et~al.}}]{dipole_rad_Freire:2012}%
  \BibitemOpen
  \bibfield  {author} {\bibinfo {author} {\bibfnamefont {P.~C.~C.}\
  \bibnamefont {Freire}} \emph {et~al.},\ }\href
  {https://doi.org/10.1111/j.1365-2966.2012.21253.x} {\bibfield  {journal}
  {\bibinfo  {journal} {Mon. Not. Roy. Astron. Soc.}\ }\textbf {\bibinfo
  {volume} {423}},\ \bibinfo {pages} {3328} (\bibinfo {year} {2012})},\ \Eprint
  {https://arxiv.org/abs/1205.1450} {\color{red}arXiv:1205.1450} \BibitemShut
  {NoStop}%
\bibitem [{\citenamefont {Shao}\ and\ \citenamefont {the
  others}(2017)\citenamefont {Shao} \emph {et~al.}}]{dipole_rad_Shao:2017}%
  \BibitemOpen
  \bibfield  {author} {\bibinfo {author} {\bibfnamefont {L.}~\bibnamefont
  {Shao}} \emph {et~al.},\ }\href {https://doi.org/10.1103/PhysRevX.7.041025}
  {\bibfield  {journal} {\bibinfo  {journal} {Phys. Rev.}\ }\textbf {\bibinfo
  {volume} {X7}},\ \bibinfo {pages} {041025} (\bibinfo {year} {2017})},\
  \Eprint {https://arxiv.org/abs/1704.07561} {\color{red}arXiv:1704.07561}
  \BibitemShut {NoStop}%
\bibitem [{\citenamefont {Virtanen}\ \emph {et~al.}(2020)\citenamefont
  {Virtanen}, \citenamefont {Gommers}, \citenamefont {Oliphant} \emph
  {et~al.}}]{2020SciPy-NMeth}%
  \BibitemOpen
  \bibfield  {author} {\bibinfo {author} {\bibfnamefont {P.}~\bibnamefont
  {Virtanen}}, \bibinfo {author} {\bibfnamefont {R.}~\bibnamefont {Gommers}},
  \bibinfo {author} {\bibfnamefont {T.}~\bibnamefont {Oliphant}}, \emph
  {et~al.},\ }\href {https://doi.org/10.1038/s41592-019-0686-2} {\bibfield
  {journal} {\bibinfo  {journal} {Nature Methods}\ }\textbf {\bibinfo {volume}
  {17}},\ \bibinfo {pages} {261} (\bibinfo {year} {2020})}\BibitemShut
  {NoStop}%
\bibitem [{\citenamefont {Mueller}\ and\ \citenamefont
  {Serot}(1996)}]{eos_ms1}%
  \BibitemOpen
  \bibfield  {author} {\bibinfo {author} {\bibfnamefont {H.}~\bibnamefont
  {Mueller}}\ and\ \bibinfo {author} {\bibfnamefont {B.~D.}\ \bibnamefont
  {Serot}},\ }\href
  {https://doi.org/https://doi.org/10.1016/0375-9474(96)00187-X} {\bibfield
  {journal} {\bibinfo  {journal} {Nuclear Physics A}\ }\textbf {\bibinfo
  {volume} {606}},\ \bibinfo {pages} {508 } (\bibinfo {year} {1996})},\ \Eprint
  {https://arxiv.org/abs/nucl-th/9603037} {\color{red}arXiv:nucl-th/9603037}
  \BibitemShut {NoStop}%
\end{thebibliography}%


%

\end{document}